\documentclass[opre,nonblindrev]{informs3} 

\OneAndAHalfSpacedXI 

\usepackage{xr}
\usepackage{hyperref}       
\usepackage{url}            
\usepackage{booktabs}       
\usepackage{amsfonts}       
\usepackage{nicefrac}       
\usepackage{microtype}      

\usepackage{amsmath}
\usepackage{subcaption}
\usepackage{graphicx}
\usepackage[ruled,vlined]{algorithm2e}

\usepackage{floatrow}
\newfloatcommand{capbtabbox}{table}[][\FBwidth]
\usepackage{blindtext}
\usepackage{enumitem}
\usepackage{bbold}

\usepackage{natbib}
 \bibpunct[, ]{(}{)}{,}{a}{}{,}%
 \def\bibfont{\small}%
\TheoremsNumberedThrough     
\ECRepeatTheorems

\EquationsNumberedThrough    

\MANUSCRIPTNO{}

\begin{document}



\RUNTITLE{Certifiable Deep Importance Sampling}

\TITLE{Certifiable Deep Importance Sampling for Rare-Event Simulation of Black-Box Systems}

\ARTICLEAUTHORS{%
\AUTHOR{Mansur Arief}
\AFF{Carnegie Mellon University, \EMAIL{marief@andrew.cmu.edu}} 
\AUTHOR{Yuanlu Bai}
\AFF{Columbia University, \EMAIL{yb2436@columbia.edu}}
\AUTHOR{Wenhao Ding}
\AFF{Carnegie Mellon University, \EMAIL{wenhaod@andrew.cmu.edu}}
\AUTHOR{Shengyi He}
\AFF{Columbia University, \EMAIL{sh3972@columbia.edu}}
\AUTHOR{Zhiyuan Huang}
\AFF{Tongji University, \EMAIL{huangzy@tongji.edu.cn}}
\AUTHOR{Henry Lam}
\AFF{Columbia University, \EMAIL{henry.lam@columbia.edu}}
\AUTHOR{Ding Zhao}
\AFF{Carnegie Mellon University, \EMAIL{dingzhao@cmu.edu}}
} 

\ABSTRACT{%
Rare-event simulation techniques, such as importance sampling (IS), constitute powerful tools to speed up challenging estimation of rare catastrophic events. These techniques often leverage the knowledge and analysis on underlying system structures to endow desirable efficiency guarantees. However, black-box problems, especially those arising from recent safety-critical applications of AI-driven physical systems, can fundamentally undermine their efficiency guarantees and lead to dangerous under-estimation without diagnostically detected. We propose a framework called Deep Probabilistic Accelerated Evaluation (Deep-PrAE) to design statistically guaranteed IS, by converting black-box samplers that are versatile but could lack guarantees, into one with what we call a relaxed efficiency certificate that allows accurate estimation of bounds on the rare-event probability. We present the theory of Deep-PrAE that combines the dominating point concept with rare-event set learning via deep neural network classifiers, and demonstrate its effectiveness in numerical examples including the safety-testing of intelligent driving algorithms.
}



\maketitle

%

\newpage
\section{Introduction}

Rare-event simulation techniques have constituted a powerful toolbox to speed up challenging estimation of rare catastrophic events. From a Monte Carlo standpoint, rare-event estimation is difficult because, by definition, the probability of landing such an event is tiny, which necessitates an enormous amount of simulation runs to observe one ``hit'' and obtain meaningful information. Statistically, this difficulty manifests as a high variance-to-squared-mean ratio (i.e., the so-called relative error; \citealt{l2010asymptotic}), and rare-event simulation techniques, cast under the umbrella of so-called variance reduction \citep{bucklew2013introduction}, have been substantially studied over the years to reduce these errors. These techniques range from importance sampling (IS) \citep{juneja2006rare,blanchet2012state} to multi-level splitting \citep{glasserman1999multilevel,villen1994restart}. In many interesting problems, they can be shown to improve estimation efficiency by orders of magnitudes, thus forming a powerhouse to handle these challenging Monte Carlo problems.

While powerful, it is also well-known that, to endow desirable theoretical efficiency guarantees from variance reduction, one often requires knowledge and careful analysis on the underlying model structure \citep{juneja2006rare,dean2009splitting}. This is particularly so in rare-event estimation, because powerful methods that give substantial theoretical improvements, such as IS, could also greatly harm performances if not properly configured to suit the considered system process (a ``double-edged sword" so to speak) \citep{glasserman1997counterexamples}. The high-level issue is that just increasing the frequency of hits on the rare-event set alone does not necessarily produce a better estimate than naive Monte Carlo. In order to be statistically valid, one also needs to enforce the output estimator to be unbiased which, in turn, could blow up the variance if the estimator is poorly constructed. On the other hand, general-purpose methods, such as cross-entropy \citep{de2005tutorial} and subset sampling \citep{au2001estimation}, are advantageously versatile and do not require detailed model knowledge or analysis, but with the price that full theoretical guarantee is not provided and efficiency is empirically verified. 

Our goal in this paper is to study an approach that gives \emph{theoretical efficiency guarantees for rare-event estimation driven by black-box models}. Let us qualify our claim. By ``black-box" here we mean the system logic is not known or too complicated to support analytical tractability, but the input distribution in the simulation model is known and can be modified to observe outputs (thus allowing for techniques such as IS). This relaxes the traditional requirement on precise system knowledge in using provably efficient variance reduction methods. To this end, while one may attempt to use existing general-purpose algorithms, we show that, when applied in such black-box settings, these algorithms could result in dangerous \emph{under-estimation of rare-event probabilities without diagnostically detected}. We highlight that this is an issue that goes beyond theoretical interest - By not bearing a theoretical efficiency guarantee, the large statistical error could be hidden from empirical results and standard diagnostic checks, thus potentially trapping the user into making big mistakes without knowing it. Our approach, in some sense, aims to convert these general-purpose methods that lack guarantees and risk undetected under-estimation, into ones that have rigorous guarantees and subsequently avoid the perils. 

\subsection{Motivation}
Rare-event estimation has appeared ubiquitously in risk analysis and management across many disciplines such as queueing systems \citep{Sadowsky91,KN99,BGL10,blanchet2014rare,BMRT09,SP02,Ridder09,dupuis2009importance},
finance \citep{Glasserman04,glasserman2008fast,glasserman2005importance,nyquist2017moderate}, insurance \citep{ASM00Ruin,pASM85a,collamore2002importance} and reliability \citep{Heidelberger95,RT09,Tuffin04QEST,nicola1993fast,nicola2001techniques}. While efficiency-guaranteed black-box variance reduction algorithms investigated in this paper are expected to be applicable in many of these applications, our main motivation comes from the recent vigorous surge of safety-critical intelligent physical systems. The unprecedented deployment of these systems on many real-world applications comes with the need for safety validation and certification \citep{kalra2016driving,koopman2017autonomous, uesato2018rigorous}. For systems that interact with humans and are potentially safety-critical -- which can range from medical systems to self-driving cars and personal assistive robots -- it is imperative to rigorously assess their risks before their full-scale deployments. The challenge, however, is that these risks are often associated precisely with how AI reacts in rare and catastrophic scenarios, which presents rare-event problems that involve sophisticated AI-driven system dynamics.

To react to this challenge, one could attempt to employ traditional test methods which, unfortunately, fall short of addressing the black-box nature of these problems. In the self-driving context, for instance, the goal of validation is to ensure the AI-enabled system reduces human-level accident rate (in the order of 1.5 per $10^{8}$ miles of driving), thus delivering enhanced safety promise to the public \citep{Evan2016FatalPerfect,kalra2016driving, PreliminaryHWY16FH018}. Formal verification, which mathematically analyzes and verifies autonomous design,  lacks the analytic tractability to formulate failure cases or consider all execution trajectories when applied to complex models \citep{clarke2018handbook}. Automated scenario selection approaches generate test cases based on domain knowledge \citep{wegener2004evaluation} or adaptive searching algorithms (such as adaptive stress testing; \citealt{koren2018adaptive}), which is more implementable but falls short of rigor. Test matrix approaches, such as Euro NCAP \citep{national2007new}, use prepopulated test cases extracted from crash databases, but they only contain historical human-driver information. The closest analog to the latter for self-driving vehicles is ``naturalistic tests'', which means placing them in real-world environments and gathering observations. This method, however, is economically prohibitive because of the rarity of the target conflict events \citep{zhao2017accelerated, arief2018accelerated, claybrook2018autonomous,o2018scalable}.

Because of all these limitations, two potential lines of approaches to test intelligent physical systems have been recently proposed. First is simulation-based tests to validate complex black-box designs \citep{corso2020survey}. This approach operates by integrating the target intelligent algorithm into an interacting virtual simulation platform that models the surrounding environment. By running enough Monte Carlo sampling of this (stochastic) environment, one hopes to observe catastrophic conflict events and subsequently conduct statistical analyses. This approach is flexible and scalable, as it hinges on building a virtual environment instead of physical systems, and provides a probabilistic assessment on the occurrences and behaviors of safety-critical events  \citep{ koopman2018toward}. Second is selective experimental scenario testing on real physical grounds, where the scenarios are suitably randomized to capture the uncertainties in actual road driving. Experiments have been designed to be conducted on the proving ground \citep{zhao2018lab}, in the public streets  \citep{arief2018accelerated}, and with mixed reality settings \citep{8500545}.

 Nonetheless, similar to the challenge encountered by naturalistic tests, because of their rarity, safety-critical events are seldom observed in the simulation or physical-test-ground experiments. This therefore leads to  rare-event estimation in a Monte Carlo or ``pseudo" Monte Carlo setting, with sophisticated system logic or rare-event boundaries lacking  analytical tractability, thus motivating our investigation.

 \subsection{Overview of our Framework}
Connecting to our introduction earlier, we aim to study an approach to obtain efficient rare-event estimators for black-box problems by endowing versatile samplers with theoretical guarantees. On a high level, our approach comprises a two-stage framework, where the first stage is a set-learning phase to gather information about the black-box rare event, while the second stage runs an efficient IS based on the collected knowledge. To endow statistical guarantees in this ``learning-to-reduce-variance" approach, our framework utilizes three key ingredients:
\begin{enumerate}[leftmargin=*]
\item\textbf{Relaxed efficiency certificate: }We shift the estimation of target rare-event probability to an upper (and lower) bound, in a way that supports the integration of learning errors into variance reduction without giving up estimation correctness.
 
\item\textbf{Set-learning with one-sided error: }We design learning algorithms based on deep neural network classifier to create outer (or inner) approximations of rare-event sets. This classifier has a special property that, under a geometric property called orthogonal monotonicity, it exhibits zero false negative rates. 

\item\textbf{Deep-learning-based IS: }With the deep-learning based rare-event set approximation, we search the so-called \emph{dominating points} in rare-event analysis to create IS that achieves the relaxed efficiency certificate.
\end{enumerate}

We call our framework consisting of the three ingredients above \emph{Deep Probabilistic Accelerated Evaluation (Deep-PrAE)}, where ``accelerated evaluation'' follows terminologies in recent approaches for the safety-testing of autonomous vehicles \citep{zhao2016accelerated, 8116682}. In the set-learning step in Deep-PrAE, the samples fed into our deep classifier can be generated by any general-purpose algorithms including the cross-entropy (CE) method \citep{de2005tutorial,rubinstein2013cross} and particle approaches such as adaptive multi-level splitting (AMS) \citep{au2001estimation,cerou2007adaptive, webb2018statistical}. Deep-PrAE turns these samples into an IS with the relaxed efficiency certificate against undetected under-estimation. Our approach is robust in the sense that it provides a tight bound for the target rare-event probability if the underlying classifier is expressive enough, while it still provides a correct, though conservative, bound if the classifier is weak. To our best knowledge, such type of guarantees and robustness features is the first of its kind in the rare-event simulation literature. We envision our work to lay the foundation for further improvements to design certified methods for evaluating more sophisticated intelligent designs via rare-event simulation. 

In the following, Section \ref{sec:existing} first describes the statistical risks in black-box rare-event estimation, starting from a review of the established rare-event estimation literature, and then leading the discussion into the under-estimation peril for black-box problems. Section \ref{sec:Deep-PrAE} presents in detail our Deep-PrAE framework to address the peril and its underpinning mechanism and guarantees. Then, Section \ref{sec:numerics} demonstrates and compares our framework with existing baselines in several experiments, including the safety testing of an intelligent driving model. Additional materials and all proofs are provided in the Appendix.

\section{Statistical Challenges in Black-Box Rare-Event Simulation}\label{sec:existing}
We first review some basics on Monte Carlo (Section \ref{sec:MC}) and the standard challenges in rare-event simulation (Section \ref{sec:challenges}). We then discuss variance reduction including IS (Section \ref{sec:var red}) and the related large deviations literature (Section \ref{sec:LD}). With these backgrounds in place, we describe, as the key message of this section, the main statistical risk when applying existing variance reduction schemes to tackle black-box systems (Sections \ref{sec:perils} and \ref{sec:past}).

Our evaluation goal is a rare-event probability $\mu=P(X\in\mathcal S_\gamma)$, where $X$ is a random vector in $\mathbb R^d$ and distributed according to $p$. $\mathcal S_\gamma$ denotes the rare-event set. In a safety testing task for instance, $X$ could denote a stochastic environment, and $\mathcal S_\gamma$ a safety-critical set on the interaction between the considered physical system and the environment. The ``rarity'' parameter $\gamma\in\mathbb R$ is considered a large number, with the property that as $\gamma\to\infty$, $\mu\to0$ (Think of, e.g., $\mathcal S_\gamma=\{x:g(x)\geq\gamma\}$ for some risk function $g$ and exceedance threshold $\gamma$).

\subsection{Monte Carlo Efficiency}\label{sec:MC}
Suppose we use a Monte Carlo estimator $\hat\mu_n$ to estimate $\mu$, by running $n$ simulation runs in total. Since $\mu$ is tiny, the error of a meaningful estimation must be measured in relative term, i.e., we would like 
\begin{equation}
P(|\hat\mu_n-\mu|>\epsilon \mu)\leq\delta\label{efficiency certificate}
\end{equation}where $\delta$ is some confidence level (e.g., $\delta=5\%$) and $0<\epsilon<1$. 

Suppose that $\hat\mu_n$ is unbiased and is an average of $n$ i.i.d. simulation runs, i.e., $\hat\mu_n=(1/n)\sum_{i=1}^n Z_i$ for some random unbiased output $Z_i$. We define the \emph{relative error} $RE=Var(Z_i)/\mu^2$ as the ratio of variance (per-run) and squared mean. The Markov inequality gives that
$$
P(|\hat\mu_n-\mu|>\epsilon\mu)\leq\frac{Var(\hat\mu_n)}{\epsilon^2\mu^2}=\frac{Var(Z_i)}{n\epsilon^2\mu^2}$$
so that
$$\frac{Var(Z_i)}{n\epsilon^2\mu^2}\leq\delta$$
ensures \eqref{efficiency certificate}. Equivalently, 
$$n\geq\frac{Var(Z_i)}{\delta\epsilon^2\mu^2}=\frac{RE}{\delta\epsilon^2}$$
is a sufficient condition to achieve \eqref{efficiency certificate}. So, when $RE$ is large, the required Monte Carlo size is also large. As a remark, note that replacing the second $\mu$ with $\hat\mu_n$ in the left hand side of \eqref{efficiency certificate} does not change the condition fundamentally, as either is equivalent to saying the ratio $\hat{\mu}_n/\mu$ should be close to 1. Also, note that we focus on the nontrivial case that the target probability $\mu$ is non-zero but tiny; if $\mu=0$, then no non-zero Monte Carlo estimator can achieve a good relative error. 

\subsection{Challenges in Naive Monte Carlo}\label{sec:challenges}
Consider naive Monte Carlo, which means we let $Z_i=\mathbb 1 (X_i\in\mathcal S_\gamma)$ where $\mathbb 1(\cdot)$ denotes the indicator function, and $X_i$ is an i.i.d. copy of $X$. Since $Z_i$ follows a Bernoulli distribution, $RE=(1-\mu)/\mu$. Thus, when $\mu$ is tiny, the sufficient condition for $n$ to attain \eqref{efficiency certificate} scales at least linearly in $1/\mu$. This demanding condition is a manifestation of the difficulty in hitting $\mathcal S_\gamma$. In the standard large deviations regime \citep{dembo2010large,dupuis2011weak} where $\mu$ is exponentially small in $\gamma$, the required Monte Carlo size $n$ would grow \emph{exponentially} in $\gamma$. 

To be clearer, we mention that the observation above is in fact tight (i.e., not because we have used a potentially loose Markov inequality). This can be seen by analyzing $n\hat\mu_n$ as a binomial variable. To be more specific, we know that $P(|\hat{\mu}_n-\mu|>\varepsilon\mu)=P(|n\hat{\mu}_n-n\mu|>\varepsilon n\mu)$ and that $n\hat{\mu}_n$ takes values in $\{0,1,\dots,n\}$. Therefore, if $n\mu\to0$, then $P(|\hat{\mu}_n-\mu|>\varepsilon\mu)\to1$, and hence \eqref{efficiency certificate} does not hold. Moreover, the following provides a concrete general statement that an $n$  that grows only polynomially in $\gamma$ would fail to estimate $\mu$ that decays exponentially in $\gamma$ with enough relative accuracy, of which \eqref{efficiency certificate} fails to hold is an implication. 
\begin{proposition}[Failure of Naive Monte Carlo]
Suppose that $\mu=P(X\in\mathcal{S}_{\gamma})$ is exponentially decaying in $\gamma$ and $n$ is polynomially growing in $\gamma$. Define $\hat{\mu}_n=(1/n)\sum_{i=1}^n \mathbb 1(X_i\in \mathcal{S}_{\gamma})$. Then for any $0<\varepsilon<1$, 
$$
\lim_{\gamma\rightarrow\infty}P(|\hat{\mu}_n-\mu|>\varepsilon\mu)=1.
$$
\label{naive Monte Carlo}
\end{proposition}

\subsection{Variance Reduction}\label{sec:var red}
The severe burden when using naive Monte Carlo motivates techniques to drive down $RE$. 

First we introduce the following notion:

\begin{definition}[Efficiency Certificate]
We say an estimator $\hat\mu_n$ satisfies an \emph{efficiency certificate} to estimate $\mu$ if it achieves \eqref{efficiency certificate} with $n=\tilde{O}(\log(1/\mu))$, for given $0<\epsilon,\delta<1$.\label{def:ec}
\end{definition}

In the above definition, $\tilde O(\cdot)$ denotes a polynomial growth in $\cdot$. If $\hat\mu_n$ is constructed from $n$ i.i.d.~samples, then the efficiency certificate can be attained with $RE=\tilde O(\log(1/\mu))$. In the large deviations regime, the sample size $n$ used in a certifiable estimator is reduced from exponential in $\gamma$ in naive Monte Carlo to \emph{polynomial} in $\gamma$.

We have used the term ``efficiency certificate" to denote an estimator that achieves \eqref{efficiency certificate} with $n=\tilde{O}(\log(1/\mu))$. In the rare-event literature, such an estimator is known as ``logarithmically efficient" or ``weakly efficient" \citep{juneja2006rare,blanchet2012state}. We use a new terminology of ``certificate" here because of the convenience when we introduce the ``relaxed" efficiency notion in Section \ref{sec:Deep-PrAE}, and also in the actual safety testing of a physical system a methodology or system indeed needs to be ``certified" to be safe. We comment that there are other related efficiency criteria (e.g., \citealt{juneja2006rare,l2010asymptotic}), including bounded relative error that is satisfied if the relative error is bounded from above as $\gamma\to\infty$ and is a stronger condition than the efficiency certificate that we use.

To achieve efficiency certificate like Definition \ref{def:ec}, Importance sampling (IS) stands as a prominent technique by sampling from an alternative distribution that puts more emphasis on the rare event region \citep{siegmund1976importance,glynn1989importance}. IS generates $X$ from another distribution $\tilde p$ (called IS distribution), and outputs $\hat\mu_n=(1/n)\sum_{i=1}^n L(X_i) \mathbb 1(X_i\in\mathcal S_\gamma)$ where $L=dp/d\tilde p$ is the likelihood ratio, or the Radon-Nikodym derivative, between $p$ and $\tilde p$. Via a change of measure, it is easy to see that $\hat\mu_n$ is unbiased for $\mu$. The key is to carefully choose $\tilde{p}$ to control the $RE$, which involves increasing the probability of hitting rare event and also ensuring that the likelihood ratio is properly bounded over the rare event region. The choice of $\tilde{p}$ is studied extensively in the IS literature; see the surveys \cite{Bucklew2004,juneja2006rare,blanchet2012state}. 
In particular, a vast literature on IS design is based on large deviations theory \citep{budhiraja2019analysis}, which leads to the notion of dominating points \citep{sadowsky1990large,dieker2005asymptotically} that we will utilize heavily in this paper, and also subsolution approaches and Lyapunov inequalities \citep{dupuis2009importance,blanchet2012lyapunov}. In settings involving heavy tails where the large deviations is of sub-exponential type, common schemes include mixture-based IS with mixing ``big jumps" \citep{blanchet2008efficient,blanchet2008state,chen2019efficient,BLANCHET20123361,murthy2014state,hult2012importance} and conditional Monte Carlo \citep{asmussen2006improved}.

\subsection{Large Deviations and Dominating Points}\label{sec:LD}
Unfortunately, in black-box settings where complete model knowledge and analytical tractability are unavailable, the classical IS methodology described above faces severe challenges. To explain this, we first present how efficiency certificate can be obtained based on the concept of large deviations and dominating points.

From now on, we consider input $X\in\mathbb{R}^d$ with density function $f$. We define $\lambda(s)=\log Ee^{s^T X},s\in\mathbb R^d$ as the cumulant generating function of $X$ and $I(y)=\sup_{s\in\mathbb R^d}\{y^T s-\lambda(s)\},y\in\mathbb R^d$ as the so-called rate function, given as the Legendre transform of $\lambda(s)$. First, we make some assumptions on $\lambda(s)$, which imply useful properties of $I(y)$.
\begin{assumption}
$\lambda(s)$ satisfies the following conditions:
\begin{enumerate}
    \item $\lambda(s)$ is a closed proper convex function;
    \item $\mathcal{D}(\lambda):=\{s\in\mathbb{R}^d:\lambda(s)<\infty\}$ has non-empty interior including 0;
    \item $\lambda(s)$ is strictly convex and differentiable on $\mathcal{D}(\lambda)^{\circ}$;
    \item $\lim_{n\to\infty}\|\nabla\lambda(s_n)\|_2=\infty$ for any sequence $\{s_n\}_{n=1}^{\infty}$ in $\mathcal{D}(\lambda)^{\circ}$ converging to a boundary point of $\mathcal{D}(\lambda)$.
\end{enumerate}
\label{asm:lambda}
\end{assumption}

Assumption \ref{asm:lambda} is standard in large deviations and applies to most light-tailed distributions \citep{dembo2010large}. To facilitate understanding, if we take multivariate Gaussian distribution $N(m,\Sigma)$ as an example, then by definition $\lambda(s)=m^Ts+\frac12s^T\Sigma s$ and $I(y)=\frac12(y-m)^T\Sigma^{-1}(y-m)$, and it is easy to verify that all the conditions in Assumption \ref{asm:lambda} hold. From Assumption \ref{asm:lambda} we have the following:

\begin{lemma}
Under Assumption \ref{asm:lambda}, $I(y)$ has the following properties:
\begin{enumerate}
    \item $\mathcal{D}(I):=\{y\in\mathbb{R}^d:I(y)<\infty\}$ has non-empty interior;
    \item $I(y)$ is strictly convex and differentiable on $\mathcal{D}(I)^{\circ}$;
    \item $I(y)\geq 0$ with $I(y)=0$ if and only if $y=\nabla\lambda(0)$;
    \item For any $y\in \mathcal{D}(I)^{\circ}$, there exists a unique $s=s_y\in\mathcal{D}(\lambda)^{\circ}$ such that $\nabla \lambda(s_y)=y$ and $I(y)=s_y^T y-\lambda(s_y)$.

\end{enumerate}
\label{thm:properties_I}
\end{lemma}

With these, we introduce the concept of dominating set:
\begin{definition}[Dominating Set]
Suppose that Assumption \ref{asm:lambda} holds and $0<\inf_{y\in \mathcal S_{\gamma}}I(y)<\infty$. We call $A_{\gamma}\subset  \mathcal S_{\gamma}$ a dominating set for $\mathcal S_{\gamma}$ associated with the distribution $p$ if
\begin{enumerate}
    \item For any $x\in \mathcal S_{\gamma}$, there exists at least one $a\in A_{\gamma}$ such that $s_a^T(x-a)\geq 0$ where $s_a$ is as defined in Lemma \ref{thm:properties_I};
    \item For any $a\in A_{\gamma}$, $A_{\gamma}\setminus\{a\}$ does not satisfy the above condition.
\end{enumerate}
Moreover, we call any point in $A_\gamma$ a dominating point.
\label{dominating point def}
\end{definition}

\begin{figure}[htbp]
  \begin{subfigure}{.4\textwidth}
  \centering
    \includegraphics[width=\textwidth]{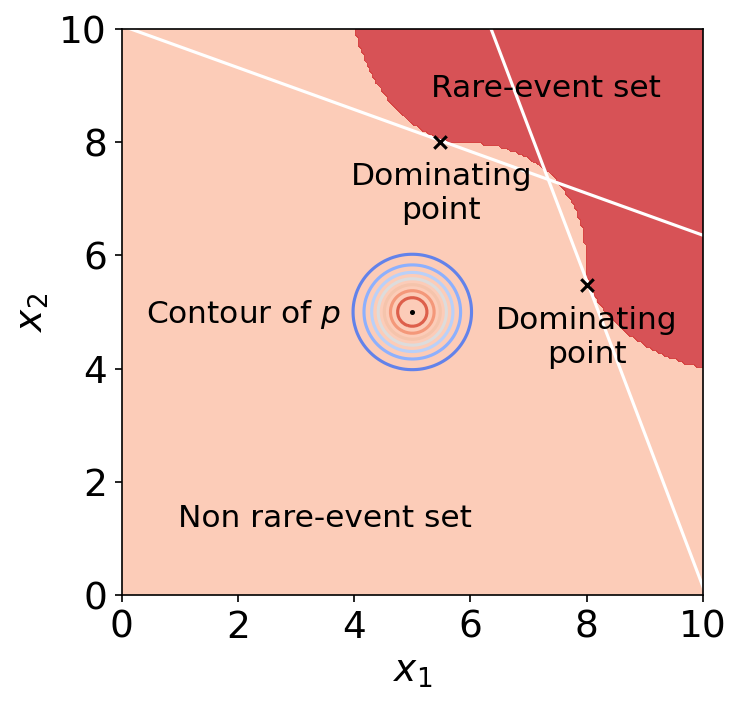}
    \caption{}
  \end{subfigure}%
  \begin{subfigure}{.4\textwidth}
  \centering
    \includegraphics[width=\textwidth]{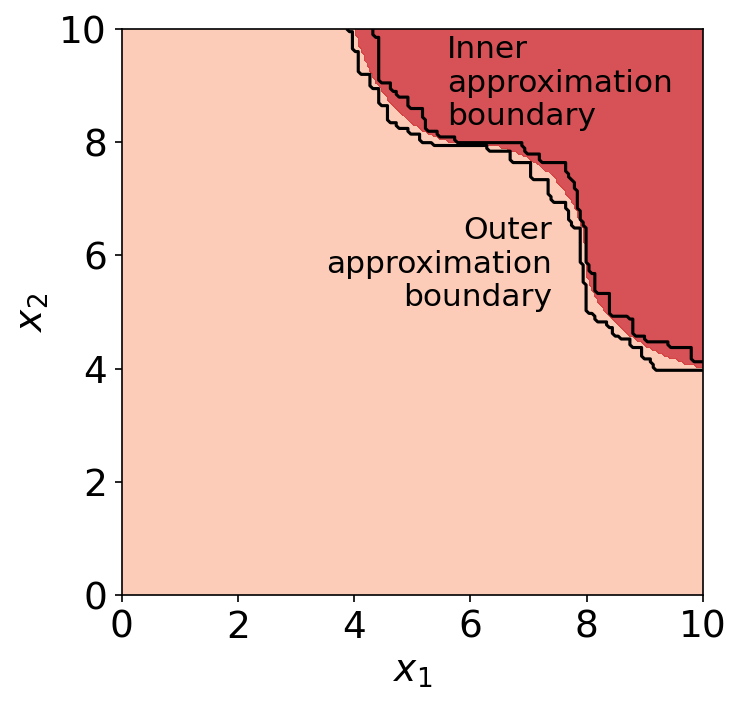}
    \caption{}
  \end{subfigure}
  
    \begin{subfigure}{.4\textwidth}
  \centering
    \includegraphics[width=\textwidth]{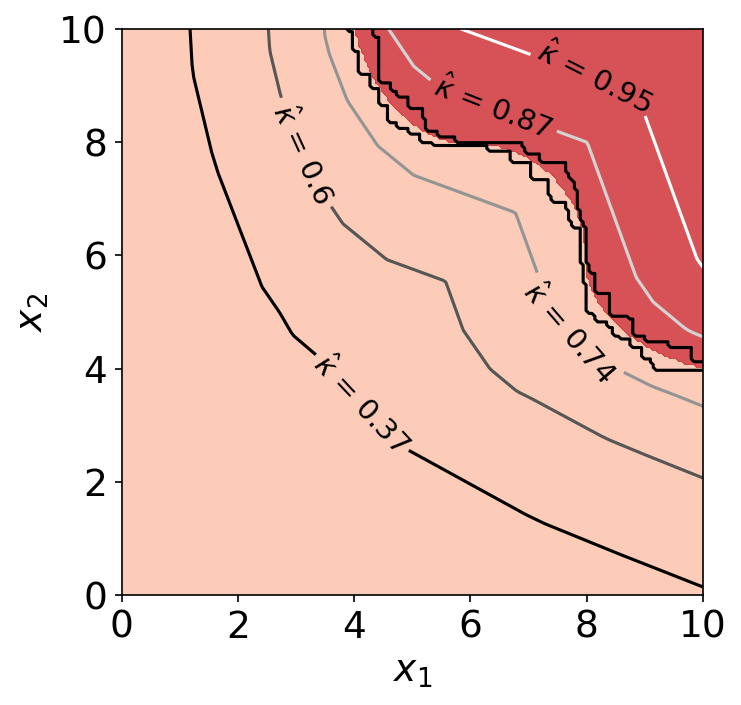}
    \caption{}
  \end{subfigure}
  \begin{subfigure}{.4\textwidth}
  \centering
    \includegraphics[width=\textwidth]{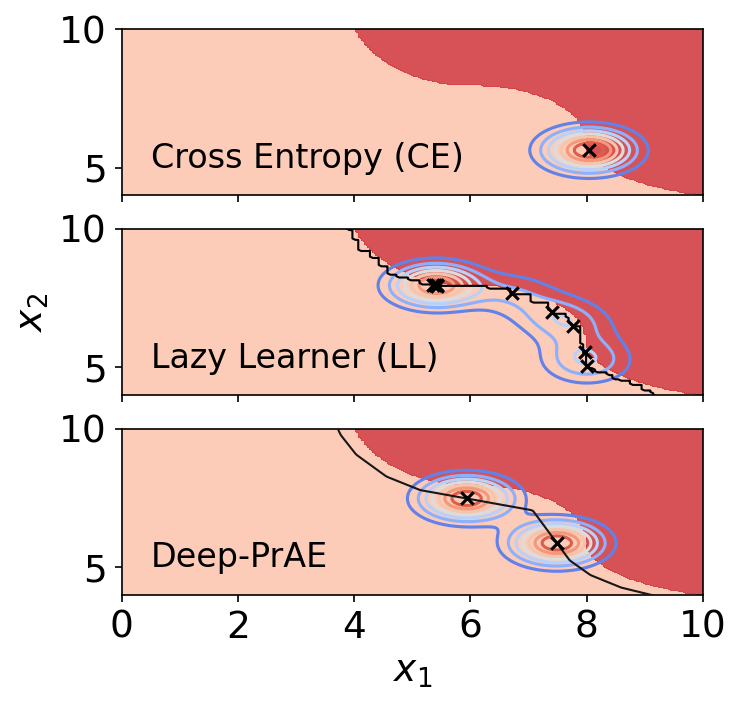}
    \caption{}
  \end{subfigure}
  
\caption{(a) An example of $\mathcal S_\gamma$ with two dominating points (b) Outer- and inner- approximations of $\mathcal S_\gamma$ (c) $\hat \kappa$ tuning for Stage 1 in Algorithm \ref{algo:stage1}  (d) IS proposals: too few dominating points for CE with too simple parametric class, too many for LL, and a balance for Deep-PrAE.}
\label{fig:illustration}
\end{figure}

The dominating set comprises the ``corner'' cases where the rare event occurs \citep{sadowsky1990large}. In other words, each dominating point $a$ encodes, in a local region, the most likely scenario should the rare event happen, and this typically corresponds to the highest-density point in this region. Locality here refers to the portion of the rare-event set that is on one side of the hyperplane cutting through $a$ (see Figure \ref{fig:illustration}(a)). For instance, in the example of multivariate Gaussian $N(m,\Sigma)$, we get $s_a=\Sigma^{-1}(a-m)$ and then $s_a^T(x-a)\geq 0$ implies that $(x-m)^T\Sigma^{-1}(x-m)\geq (a-m)^T\Sigma^{-1}(a-m)$. That is, $a$ minimizes the rate function $I$ or equivalently, maximizes the density function over $\mathcal{S}_{\gamma}\cup\{x:s_a^T(x-a)\geq 0\}$.

Intuitively, to increase the frequency of hitting the rare-event set (and subsequently to reduce variance), an IS would translate the distributional mean 
to the global highest-density point in the rare-event set. The delicacy, however, is that this is \emph{insufficient} to control the variance, due to the ``overshoots'' arising from sampling randomness. In order to properly control the overall variance, one needs to divide the rare-event set into local regions governed by dominating points, and using a mixture IS distribution that accounts for \emph{all} of them.

The above delicacy is well-known in the literature (e.g., \citealt{glasserman1997counterexamples}), and here we provide an intuitive explanation. Roughly speaking, for $X\sim p$ and a rare-event set $\mathcal S_\gamma$, we have $P(X\in\mathcal S_\gamma)\approx e^{-I(a^*)}$ where $a^*=\arg\min_{y\in \mathcal S_{\gamma}}I(y)$ and $\approx$ is up to some factor polynomial in $I(a^*)$ (see, e.g., Theorem \ref{general IS} in the sequel).
Thus, to obtain an efficiency certificate, IS estimator given by $Z=L(X)\mathbb 1 (X\in\mathcal S_\gamma)$, where $X\sim\tilde p$ and $L=dp/d\tilde p$, needs to have $\widetilde{Var}(Z)\leq\tilde E[Z^2]\approx e^{-2I(a^*)}$ where $\widetilde{Var}(\cdot)$ and $\tilde E[\cdot]$ denote the variance and expectation under $\tilde p$. Note that the last equality relation cannot be improved, as otherwise it would imply that $\widetilde{Var}(Z)=\tilde E[Z^2]-(\tilde E[Z])^2<0$. Now consider an IS with likelihood ratio $L(x)=e^{\lambda(s^*)-s^{*T}x}$ where $s^*:=s_{a^*}$, giving
\begin{equation}\label{second moment}
\tilde E[Z^2]=\tilde E[L(X)^2 \mathbb 1(X\in\mathcal S_\gamma)]=e^{-2(s^{*T}a^*-\lambda(s^*))}\tilde E[e^{-2s^{*T}(x-a^*)}\mathbb 1(X\in\mathcal S_\gamma)]
\end{equation}
If the ``overshoot''  $s^{*T}(x-a^*)$, i.e., the remaining term in the exponent of $L(x)$ after moving out $I(a^*)=s^{*T}a^*-\lambda(s^*)$, satisfies $s^{*T}(x-a^*)\geq0$ for all $x\in\mathcal S_\gamma$, then the expectation in the right hand side of (\ref{second moment}) is bounded by 1, and an efficiency certificate is achieved. This, however, is not true for all points in the set $\mathcal S_\gamma$ (which motivates the notion of dominating sets and points in Definition \ref{dominating point def}). For instance, if $\mathcal S_\gamma$  is convex, then, noting that $s_y=\nabla I(y)$, we get that $a^*$ gives a singleton dominant set since  $s^{*T}(x-a^*)\geq0$ for all $x\in\mathcal S_\gamma$ is precisely the first order optimality condition of the involved optimization. In general, if we can decompose 
$\mathcal S_\gamma=\bigcup_j\mathcal S_\gamma^j$ where $\mathcal S_\gamma^j\subset \{x:s_{a_j}^T(x-a_j)\geq0\}$ for a dominating point $a_j\in A_\gamma$, then each $\mathcal S_\gamma^j$ can be viewed as a ``local'' region where the dominating point $a_j$ is the highest-density, or the most likely point such that the rare event occurs.

Next, to present an efficient IS using dominating sets, we first introduce an additional geometric notion that we call \emph{orthogonal monotonicity}:
\begin{definition}[Orthogonal monotonicity]
We call a set $\mathcal S\subset\mathbb R_+^d$ \emph{orthogonally monotone} if for any two points $x,x'\in\mathbb R_+^d$, we have $x\leq x'$ (where the inequality is defined coordinate-wise) and $x\in\mathcal S$ implies $x'\in\mathcal S$ too. \label{OM def}
\end{definition}

Definition \ref{OM def} when applied to a rare-event set means that any point that is more ``extreme'' than a point in the set must also lie inside it. This is an intuitive assumption that holds for interesting safety-critical  settings (see Section \ref{sec:numerics}) and will be used heavily in Section \ref{sec:Deep-PrAE}. For now, its main use is to facilitate the design of a certifiable IS. To proceed, we make the following assumptions on the rare-event set $\mathcal{S}_{\gamma}$:

\begin{assumption}
Denote $a^*=\arg\min_{y\in\mathcal S_{\gamma}}I(y)$ and $s^*=s_{a^*}$. Suppose that 
\begin{enumerate}
    \item $S_{\gamma}$ is orthogonally monotone;
    \item $0<I(a^*)<\infty$;
    \item As $\gamma\to\infty$, $s_i^*$ at most grows polynomially in $\gamma$ for any $i=1,\dots,d$;
    \item There exists a constant $\varepsilon>0$ such that $f^*(x):=f(x)e^{s^{*T}x-\lambda(s^*)}$ has a uniform lower bound for $a^*\leq x\leq a^*+\varepsilon\mathbf{1}$ which at most decays polynomially in $\gamma$.
\end{enumerate}
\label{asm:S}
\end{assumption}

 Under the above assumptions, the dominating set approach gives a certifiable IS as follows:

\begin{theorem}[Certifiable IS]
Suppose Assumptions \ref{asm:lambda} and \ref{asm:S} hold. Also suppose that $A_{\gamma}$ is the dominating set for $\mathcal S_\gamma$ associated with the distribution $p$. Then we can decompose $\mathcal S_\gamma=\bigcup_j \mathcal S_\gamma^j$ where $\mathcal S_\gamma^j$'s are disjoint, $a_j\in \mathcal S_\gamma^j$ and $\mathcal S_\gamma^j\subset\{x:s_{a_j}^T(x-a_j)\geq 0\}$ for $a_j\in A_{\gamma}$. Moreover, for any constants $\alpha_j$'s such that $\alpha_j>0,\forall j$ and $\sum_j \alpha_j=1$, the IS distribution $f(x)\sum_j \alpha_j e^{s_{a_j}^Tx-\lambda(s_{a_j})}$ achieves an efficiency certificate in estimating $\mu=P(X\in S_\gamma)$, i.e., if we let $Z=\mathbb 1(X\in \mathcal S_\gamma)L(X)$ where $L$ is the corresponding likelihood ratio, then $\tilde{E}[Z^2]/\tilde{E}[Z]^2$ is at most polynomially growing in $\gamma$. 
\label{general IS}
\end{theorem}

The dominating point concept and mixture-type IS described in Theorem \ref{general IS} have been known in the literature (e.g.,  \citealt{sadowsky1990large,dieker2006fast}). Nonetheless, our setting bears some technical distinction in that we focus on the geometry of the rare-event set instead of using a scaling regime for the G\"artner-Ellis Theorem \citep{Gartner1977,Ellis1984}, the latter requiring the rare-event set to scale proportionately with the rarity parameter. To support the generality of Theorem \ref{general IS}, we provide several examples in Appendix \ref{app:examples} to show the validity of Assumption \ref{asm:S} for various common input probability distributions.

\subsection{Perils of Black-Box Variance Reduction Algorithms}\label{sec:perils}
Theorem \ref{general IS} has presented a mixture IS that possesses efficiency guarantees by including all the dominating points in the mixture. The main risk in black-box settings, however, lies in what happens when the mixture IS misses any of the dominating points. The latter can occur since in black-box settings we may not have full information or analytical capability to locate all these points. 

We stipulate that, in the case that the mixture IS could miss dominating points, the resulting estimate may be utterly unreliable for two reasons. First is that its RE can be arbitrarily large or the efficiency certificate can fail to hold, a consequence intuited naturally from our discussion in Section \ref{sec:LD}. Second, more importantly, this poor performance can be empirically hidden and leads to a systematic \emph{under-estimation} of the rare-event probability \emph{without being detected}. In other words, in a given experiment, we may observe a reasonable empirical relative error (i.e., sample variance over squared sample mean), yet the estimate is much lower than the correct value. Thus, a user could be trapped into making a big mistake in the rare-event probability estimation, but not knowing it.

We support our claim above with the following example. To connect to Theorem \ref{general IS}, the example is constructed under Assumption \ref{asm:S}, where the orthogonal monotonicity assumption is slightly generalized to the case where there is a ``modal point'' (in this case the origin) and the set is orthogonally monotone in each quadrant.

\begin{theorem}[Perils of under-estimation]
Suppose we estimate $\mu=P(X\geq\gamma\text{ or }X\leq -k\gamma)$ where $X\sim p=N(0,1)$ and $0<k<3$. We choose $\tilde p=N(\gamma,1)$ as the IS distribution to obtain $\hat\mu_n$. To be specific, we generate $X_1,\dots,X_n$ i.i.d. from $\tilde p$, compute $Z_i=\mathbb 1(X_i\geq\gamma\text{ or }X_i\leq -k\gamma)L(X_i)$ for each $i$ where $L(x)=e^{-\gamma x+\frac12\gamma^2}$, and set $\hat{\mu}_n=\frac1n \sum_{i=1}^n Z_i$. Then
\begin{enumerate}
    \item $\frac{\tilde E[Z^2]}{\tilde E[Z]^2}$ grows exponentially in $\gamma$.
    \item If $n$ is polynomial in $\gamma$, we have $\tilde P\left(\left|\hat\mu_n-\bar{\Phi}(\gamma)\right|>\varepsilon\bar{\Phi}(\gamma)\right)=O\left(\frac{\gamma}{n\varepsilon^2}\right)$ for any $\varepsilon>0$ where $\bar\Phi(\gamma)=P(X\geq\gamma)<\mu$, and 
    $$
    \tilde{P}\left(\frac{\frac1n \sum_{i=1}^n Z_{i}^2}{\left(\frac1n\sum_{i=1}^n  Z_i\right)^2}> \frac{1+\varepsilon}{(1-\varepsilon)^2}\frac{e^{\gamma^2}\bar{\Phi}(2\gamma)}{\bar{\Phi}^2(\gamma)}\right)=O\left(\frac{\gamma}{n\varepsilon^2}\right)
    $$
    for any $0<\varepsilon<1$ where $\frac{e^{\gamma^2}\bar{\Phi}(2\gamma)}{\bar{\Phi}^2(\gamma)}=\Theta(\gamma)$.
\end{enumerate}
\label{counterexample}
\end{theorem}

Theorem \ref{counterexample} gives several implications. First of all, in the considered setting there are two dominating points $\gamma$ and $-k\gamma$ but the IS design only considers the first one. As a result, efficiency certificate fails to hold. Mathematically (evident from the proof), this happens because there could exist ``unlucky'' scenario where the sample falls into the rare-event set, so that $\mathbb 1(X\in\mathcal S_\gamma)=1$, while the likelihood ratio $L(X)$ explodes, which leads to a tremendous estimation variance.

Though Part 1 is undesirable, the real risk, however, is revealed in Part 2 of Theorem \ref{counterexample}. The first half of Part 2, namely $\tilde P\left(\left|\hat\mu_n-\bar{\Phi}(\gamma)\right|>\varepsilon\bar{\Phi}(\gamma)\right)=O\left(\frac{\gamma}{n\varepsilon^2}\right)$, shows that the estimate concentrates (with deviation probability growing only polynomially in $\gamma$) at a value that can be severely under the correct one, especially when $k<1$. On the other hand, the empirical relative error, captured by $\frac{\frac1n \sum_{i=1}^n Z_{i}^2}{\left(\frac1n\sum_{i=1}^n  Z_i\right)^2}$, is also small with high probability (both the relative error magnitude and deviation probability are polynomial in $\gamma$). This means the empirical result would suggest a convergence supported confidently by a small relative error, when in reality the user could have committed a significant under-estimation mistake.

We mention that, mathematically, Part 2 of Theorem \ref{counterexample} arises because all samples most likely land on the neighborhood of the solely considered dominating point. If the missed dominating point is a significant contributor to the rare-event probability, then the empirical performance would look as if the rare-event set is \emph{smaller}, leading to a systematic under-estimation. Note that this phenomenon occurs even if the estimator is unbiased, which is guaranteed by IS by default.

\subsection{Existing Black-Box Algorithms}\label{sec:past}
With this, we now explain why using black-box variance reduction algorithms can be dangerous -- in the sense of not having an efficiency certificate and, more importantly, the risk of an unnoticed systematic under-estimation. 

In the literature, there are two lines of techniques that apply to black-box problems. The first line is the CE method \citep{de2005tutorial,rubinstein2013cross}, which uses a sequential optimization approach to iteratively solve for the optimal parameter in a parametric class of IS distributions. The objective in this optimization sequence is to minimize the Kullback–Leibler divergence between the IS distribution and the zero-variance IS distribution (the latter is theoretically known to be the conditional distribution given the occurrence of the rare event, but is unimplementable as it requires knowing the rare-event probability itself). Specifically, assume we are interested in estimating $P(g(X)>\gamma)$ and a parametric class $p_\theta$ is considered. The cross-entropy method adaptively chooses $\gamma_1<\gamma_2<...<\gamma$. At each intermediate level $k$, we use the updated IS distribution $p_{\theta^*_k}$, designed for simulating $P(g(X)>\gamma_k)$, as the sampling distribution to draw samples of $X$ that sets up an empirical optimization, from which the next $\theta^*_{k+1}$ is obtained. While flexible and easy to use, the efficiency of CE depends crucially on the expressiveness of the parametric class $p_\theta$ and the parameter convergence induced by the empirical optimization sequence. There are good approaches to determine the parametric classes (e.g., \citealt{botev2016semiparametric}), and also studies on the efficiency of IS distributions parametrized by empirical optimization \citep{tuffin2012probabilistic}. However, it is challenging to obtain an efficiency certificate for CE that requires iterative empirical optimization in the common form depicted above. Insufficiency on either the choice of the parametric class or the parameter convergence may lead to the undetectable under-estimation issue (e.g., as in Theorem~\ref{counterexample}). 

under-estimation (see Appendix E

The second line of methods is the multi-level splitting or subset simulation \citep{au2001estimation,cerou2007adaptive}, a particle method in lieu of IS, which 
decomposes the rare-event estimation problem into estimating a sequence of conditional probabilities. We adaptively choose a threshold sequence $\gamma_1<\gamma_2<...<\gamma_{K}=\gamma$. Then $P(g(x)>\gamma)$ can be rewritten as  $P(g(x)>\gamma)=P(g(x)>\gamma_1) \prod_{k=2}^{K} P(g(x)>\gamma_{k}|g(x)>\gamma_{k-1})$. AMS then aims to estimate $P(g(x)>\gamma_1)$ and $P(g(x)>\gamma_{k}|g(x)>\gamma_{k-1})$ for each intermediate level $k=2,...,K$. In standard implementation, these conditional probabilities are estimated using samples from $p(g(x)>\gamma_{k}|g(x)>\gamma_{k-1})$ through variants of the Metropolis-Hasting (MH) algorithms.

Theoretical studies have shown some nice properties of AMS, including unbiasedness and asymptotic normality (e.g., see \citealt{cerou2019adaptive}). However, the variance of the estimator depends on the mixing property of the proposal distribution in the MH steps \citep{cerou2016fluctuation}. Under ideal settings when direct sampling from $P(g(x)>\gamma_{k}|g(x)>\gamma_{k-1})$ is possible, it is shown that AMS is ``almost'' asymptotically optimal \citep{guyader2011simulation}. However, to our best knowledge, there is yet any study on provable efficiency of rare-event estimators with consideration of both AMS and MH sampling. In practice, to achieve a good performance, AMS requires a proposal distribution in the MH algorithm that can efficiently generate samples with low correlations.

Finally, we also mention there are other variants of CE and AMS. The former include enhanced CE such as Markov chain IS \citep{botev2013markov,botev2016semiparametric,grace_kroese_sandmann_2014}, neural network IS \citep{muller2019neural} and nonparametric CE \citep{rubinstein2005stochastic}.

The latter include RESTART which works similarly as subset simulation and splitting but performs a number of simulation retrials after entering regions with a higher importance function value \citep{VILLENALTAMIRANO2010156}. Despite their versatility, these methods encounter similar challenges as standard CE and AMS in satisfying an efficiency certificate in black-box settings described above.

\section{The Deep Probabilistic Accelerated Evaluation Framework}\label{sec:Deep-PrAE}
 We propose the Deep-PrAE framework to overcome the challenges faced by black-box variance reduction algorithms presented in Sections \ref{sec:perils} and \ref{sec:past}. We first give an overview of the approach and our target guarantees (Section \ref{sec:overview}). Then we present our resulting IS and accompanying properties (Section \ref{sec:learning}).
 
\subsection{Overview and Guarantees}\label{sec:overview}
Our framework comprises two stages: First is to learn the rare-event set from a first-stage sample batch, by viewing set learning as a classification task. These first-stage samples can be drawn from any rare-event sampling methods including CE and AMS. Second is to apply an efficiency-certified IS on the rare-event probability over the learned set. Algorithm \ref{algo:stage1} shows our main procedure. The key to achieving an ultimate efficiency certificate lies in how we learn the rare-event set in Stage 1, which requires two properties: 

\begin{enumerate}[leftmargin=*]
\item
 \textbf{Small one-sided generalization error: }``One-sided'' generalization error here means the learned set is either an outer or an inner approximation of the unknown true rare-event set, with probability 1. Converting this into a classification, this means the false negative (or positive) rate is exactly 0. ``Small'' here then refers to the other type of error being controlled.
\item
\textbf{Decomposability: }The learned set is decomposable according to dominating points in the form of Theorem \ref{general IS}, so that an efficient mixture IS can apply.
\end{enumerate}

\begin{algorithm}[htbp]
\KwIn{Black-box evaluator $I(\cdot\in\mathcal S_\gamma)$, initial Stage 1 samples $\{(\tilde X_i, Y_i) \}_{i=1,\ldots,n_1}$, Stage 2 sampling budget $n_2$, input density function $f(x)$.}
\KwOut{IS estimate $\hat\mu_n$.}

\nl \textbf{Stage 1 (Set Learning):}\\

\nl Train classifier with positive decision region $\overline{\mathcal  S}_\gamma^{\kappa}=\{x:\hat{g}(x) \geq\kappa\}$ using $\{(\tilde X_i, Y_i) \}_{i=1,\ldots,n_1}$;\\
\nl Replace $\kappa$ by $\hat\kappa=\max\{\kappa\in\mathbb R: (\overline{\mathcal  S}_\gamma^{\kappa})^c\subset\mathcal H(T_0)\}$;\\

\nl \textbf{Stage 2 (Mixture IS based on Searched dominating points):}\\
\nl Start with $\hat A_\gamma = \emptyset$;\\

\nl {\bf While } $\{x: \hat g(x) \geq \hat \kappa , s_{x_j^*}^T(x-x^*_j) <0, \mbox{ $\forall x^*_j \in \hat A_\gamma$} \} \neq  \emptyset$ {\bf do }\\ \label{algo:while_line}
\nl \ \ \ \ \ \ Find a dominating point $x^*$ by solving the optimization problem \begin{align*} \label{eq:opt_ite}
    x^* =\arg \min_{x} &\ \  I(x) \ \ \ \\
\text{s.t.}\ \ \  &\hat g(x) \geq \hat \kappa,\ \  \\ &s_{x_j^*}^T(x-x^*_j) <0, \ \mbox{$\forall x^*_j \in\hat A_\gamma$}
\end{align*}

\ \ \ \ \ \ and update $\hat A_\gamma \leftarrow\hat A_\gamma \cup \{x^*\}$;\\
\nl {\bf End}\\

\nl Sample $X_1,...,X_{n_2}$ from the mixture distribution $ f(x)\sum_{a\in\hat A_\gamma} (1/|\hat A_\gamma|) e^{s_a^Tx-\lambda(s_a)} $.\\

\nl Compute the IS estimator $\hat\mu_n=( 1/n_2) \sum_{i =1}^ {n_2 } L(X_i) \mathbb 1(X_i \in \bar{\mathcal{S}}_{\gamma}^{\hat{\kappa}})$, where the likelihood ratio $L(X_i)=1/( \sum_{a\in \hat A_\gamma} (1/|\hat A_\gamma|)e^{s_a^TX_i-\lambda(s_a)} )$.

    \caption{{\bf Deep-PrAE to estimate $\mu=P(X\in\mathcal S_\gamma)$.} 
    \label{algo:stage1}}
\end{algorithm}

The first property ensures that, even though the learned set can contain errors, the learned rare-event probability is either an upper or lower bound of the truth. This requirement is important as it is difficult to translate the impact of generalization errors into rare-event estimation errors. By Theorem \ref{counterexample}, we know that any non-zero error implies the risk of missing out on important regions of the rare-event set, undetectably. The one-sided generalization error allows a shift of our target to valid upper and lower bounds that can be correctly estimated, which is the core novelty of Deep-PrAE. 

To this end, we introduce a new efficiency notion:

\begin{definition}[Relaxed efficiency certificate]
We say an estimator $\hat\mu_n$ satisfies an upper-bound relaxed efficiency certificate to estimate $\mu$ if
$P(\hat\mu_n-\mu<-\epsilon\mu)\leq\delta$
with $n\geq\tilde O(\log(1/\mu))$, for given $0<\epsilon,\delta<1$. \label{relaxed certificate}
\end{definition}

Compared with the efficiency certificate in \eqref{efficiency certificate}, Definition \ref{relaxed certificate} is relaxed to only requiring $\hat\mu_n$ to be an upper bound of $\mu$, up to an error of $\epsilon\mu$. An analogous lower-bound relaxed efficiency certificate can be seen in Appendix \ref{app:lowerbound}. From a risk quantification viewpoint, the upper bound for $\mu$ is more crucial, and the lower bound serves to assess an estimation gap. The following provides a handy certification:

\begin{proposition}[Achieving relaxed efficiency certificate]
Suppose $\hat\mu_n$ is upward biased, i.e., $\overline\mu:=E[\hat\mu_n]\geq\mu$. Moreover, suppose $\hat\mu_n$ takes the form of an average of $n$ i.i.d. simulation runs $Z_i$, with $RE=Var(Z_i)/\overline\mu^2=\tilde O(\log(1/\overline\mu))$. Then $\hat\mu_n$ possesses the upper-bound relaxed efficiency certificate. \label{certificate prop simple}
\end{proposition}

Proposition \ref{certificate prop simple} stipulates that a relaxed efficiency certificate can be attained by an upward biased estimator that has a logarithmic relative error with respect to the biased mean. 

Proposition \ref{prop:extend}
shows an extension of Proposition \ref{certificate prop simple} to two-stage procedures, where the first stage determines the upward biased mean. 

\begin{proposition}[Extended relaxed efficiency certificate] \label{prop:extend}
Suppose constructing $\hat\mu_n=\hat\mu_{n_2}(D_{n_1})$ consists of two stages, with $n=n_1+n_2$: First we sample $D_{n_1}=\{\tilde X_1,\ldots,\tilde X_{n_1}\}$, where $\tilde X_i$ are i.i.d. (following some sampling distribution), and given $D_{n_1}$, we construct $\hat\mu_{n_2}(D_{n_1})=(1/n_2)\sum_{i=1}^{n_2}Z_i$ where $Z_i$ are i.i.d. conditional on $D_{n_1}$ (following some distribution). Suppose $\hat\mu_n$ is conditionally upward biased almost surely, i.e., $\overline\mu(D_{n_1}):=E[\hat\mu_n|D_{n_1}]\geq\mu$, and the conditional relative error given $D_{n_1}$ in the second stage satisfies $RE(D_{n_1}):=Var(Z_i|D_{n_1})/\overline\mu(D_{n_1})^2=\tilde O(\log(1/\overline\mu(D_{n_1})))$. If $n_1=\tilde O(\log(1/\mu))$ (such as a constant number), then $\hat\mu_n$ possesses the upper-bound relaxed efficiency certificate.
\end{proposition}

This upward biased mean, in turn, can be obtained by learning an outer approximation for the rare-event set, giving:

\begin{corollary}[Set-learning + IS]
Consider estimating $\mu=P(X\in\mathcal S_\gamma)$. Suppose we can learn a set $\overline{\mathcal S}_\gamma$ with any number $n_1$ of i.i.d. samples $D_{n_1}$ (drawn from some distribution) such that $\overline{\mathcal S}_\gamma\supset\mathcal S_\gamma$ with probability 1. Also suppose that there is an efficiency certificate for an IS estimator for $\overline\mu(D_{n_1}):=P(X\in\overline{\mathcal S}_\gamma)$. Then a two-stage estimator where a constant $n_1$ number of samples $D_{n_1}$ are first used to construct $\overline{\mathcal S}_\gamma$, and $n_2=\tilde O(\log(1/\overline\mu(D_{n_1}))$ samples are used for the IS in the second stage, achieves the upper-bound relaxed efficiency certificate.\label{relaxed prob}
\end{corollary}

To execute the procedure in Corollary \ref{relaxed prob}, we need to learn an outer approximation of the rare-event set. To this end, consider set learning as a classification problem. Suppose we have collected $n_1$ Stage 1 samples $\{(\tilde X_i, Y_i) \}_{i=1,\ldots,n_1}$, where $Y_i=1$ if $\tilde X_i$ is in the rare-event set $\mathcal S_\gamma$, and 0 otherwise. Here, it is beneficial to use Stage 1 samples that have sufficient presence in $\mathcal S_\gamma$, which can be achieved via any black-box variance reduction methods.
We then consider the pairs $\{(\tilde X_i, Y_i)\}$ where $\tilde X_i$ is regarded as the feature and $Y_i$ as the binary label, and construct a classifier, say $\hat g(x):\mathbb R^d\to[0,1]$, from some hypothesis class $\mathcal G$ that (nominally) signifies $P(Y=1|X=x)$. The learned rare-event set $\overline{\mathcal S}_\gamma$ is taken to be $\{x:\hat g(x)\geq\kappa\}$ for some threshold $\kappa\in\mathbb R$.

The outer approximation requirement $\overline{\mathcal S}_\gamma\supset\mathcal S_\gamma$ means that all true positive (i.e., 1) labels must be correctly classified, or in other words, the false negative (i.e., 0) rate is zero, i.e.,
\begin{equation}
P(X\in\overline{\mathcal S}_\gamma^c,Y=1)=0
\label{zero mis}
\end{equation}
Typically, achieving such a zero ``Type I'' misclassification rate is impossible for any finite sample except in degenerate cases. 
However, this is achievable under orthogonal monotonicity in Definition \ref{OM def}. To facilitate discussion, suppose from now on that the rare-event set is known to lie entirely in the positive quadrant $\mathbb R_+^d$, so in learning the set, we only consider sampling points in $\mathbb R_+^d$ (analogous development can be extended to the entire space). Note that, even with such a monotonicity property, the boundary of the rare-event set can still be very complex. The key is that, with orthogonal monotonicity, we can now produce a classification procedure that satisfies \eqref{zero mis}. In fact, the simplest approach is to use what we call an orthogonally monotone hull:

\begin{definition}[Orthogonally monotone hull]
For a set of points $D=\{x_1,\ldots,x_n\}\subset\mathbb R_+^d$, we define the \emph{orthogonally monotone hull} of $D$ (with respect to the origin) as $\mathcal H(D)=\cup_i\mathcal R(x_i)$, where $\mathcal R(x_i)$ is the rectangle that contains both $x_i$ and the origin as two of its corners.
\label{OM hull def}
\end{definition}

In other words, the orthogonally monotone hull consists of the union of all the rectangles each wrapping each point $x_i$ and the origin $0$. Now, denote $T_0=\{\tilde X_i:Y_i=0\}$ as the non-rare-event sampled points. Evidently, if $\mathcal S_\gamma$ is orthogonally monotone, then $\mathcal H(T_0)\subset\mathcal S_\gamma^c$ (where complement is with respect to $\mathbb R_+^d$), or equivalently, $\mathcal H(T_0)^c\supset\mathcal S_\gamma$, i.e., $\mathcal H(T_0)^c$ is an outer approximation of the rare-event set $\mathcal S_\gamma$. Figure \ref{fig:illustration}(b) shows this outer approximation (and also the inner counterpart). Moreover, $\mathcal H(T_0)^c$ is the smallest region (in terms of set volume) such that \eqref{zero mis} holds, because any smaller region could exclude a point that has label 1 with positive probability.

\subsection{Learning-Based Importance Sampling}\label{sec:learning}
With the above developments, we now consider two approaches to construct IS based on the learned rare-event set, one via a direct creation of an orthogonally monotone hull, and another via neural networks to construct rare-event approximating boundaries.

\subsubsection{Lazy-Learner IS.}Consider an estimator for $\mu$ where, given the $n_1$ samples in Stage 1, we build the mixture IS depicted in Theorem 1 to estimate $P(X\in\mathcal H(T_0)^c)$ in Stage 2. Since $\mathcal H(T_0)^c$ takes the form $(\cup_{i:Y_i=0}\mathcal R(\tilde X_i))^c$, it has a finite number of dominating points, which can be found by a sequential algorithm (similar to the one that we will discuss momentarily). We call this the ``lazy-learner'' approach. Its problem, however, is that $\mathcal H(T_0)^c$ tends to have a very rough boundary. This generates a large number of dominating points, many of which are unnecessary in that they do not correspond to any ``true'' dominating points in the original rare-event set $\mathcal S_\gamma$ (see the middle of Figure \ref{fig:illustration}(d)). This in turn leads to a large number of mixture components that degrades the IS efficiency, as the RE bound in Theorem 1 scales linearly with the number of mixture components.

\subsubsection{Deep-Learning-Based IS.}Our main approach is a deep-learning alternative that resolves the statistical degradation of the lazy learner. We train a neural network classifier, say $\hat g$, using all the Stage 1 samples $\{(\tilde X_i,Y_i)\}$, and obtain an approximate non-rare-event region $(\overline{\mathcal S}_\gamma^\kappa)^c=\{x:\hat g(x)<\kappa\}$, where $\kappa$ is say $1/2$. Then we adjust $\kappa$ minimally away from $1/2$, say to $\hat\kappa$, so that $(\overline{\mathcal  S}_\gamma^{\hat\kappa})^c\subset\mathcal H(T_0)$, i.e., $\hat\kappa=\max\{\kappa\in\mathbb R: (\overline{\mathcal  S}_\gamma^{\kappa})^c\subset\mathcal H(T_0)\}$. Then $\overline{\mathcal  S}_\gamma^{\hat\kappa}\supset\mathcal H(T_ 0)^c\supset\mathcal S_\gamma$, so that $\overline{\mathcal  S}_\gamma^{\hat\kappa}$ is an outer approximation for $\mathcal S_\gamma$ (see Figure \ref{fig:illustration}(c), where $\hat\kappa=0.74$). Stage 1 in Algorithm \ref{algo:stage1} shows this procedure. With this, we can run mixture IS to estimate $P(X\in\overline{\mathcal  S}_\gamma^{\hat\kappa})$ in Stage 2. 

The execution of this algorithm requires the set  $\overline{\mathcal S}_\gamma^{\hat\kappa}=\{x:\hat g(x)\geq\hat\kappa\}$ to be in a form susceptible to Theorem \ref{general IS} and the search of all its dominating points. When $\hat g$ is a ReLU-activated neural net, the boundary of $\hat g(x)\geq\hat\kappa$ is piecewise linear and $\overline{\mathcal S}_\gamma^{\hat\kappa}$ is a union of polytopes, and Theorem \ref{general IS} applies. Finding all dominating points is done by a sequential ``cutting-plane'' method that iteratively locates the next dominating point by minimizing $I(x)$ over the remaining portion of $\overline{\mathcal S}_\gamma^{\hat\kappa}$ not covered by the local region of any previously found points $x_j^*$. These optimization sequences can be solved via mixed integer program (MIP) formulations for ReLU networks \citep{tjeng2017evaluating,huang2018designing}. 

\subsubsection{Implementation via Mixed Integer Programming.}
We provide further details on implementing Algorithm \ref{algo:stage1}. In particular, we present how to solve the optimization problem
\begin{align} 
    x^*=\arg \min_{x} &\ \  I(x) \ \ \ 
\text{s.t.}\ \ \  \hat g(x) \geq \hat \kappa,\ \   s_{x_j^*}^T(x-x^*_j) <0\ \mbox{$\forall x^*_j \in\hat A_\gamma$}\label{MIP problem}
\end{align}
to obtain the next dominating point in the sequential cutting-plane approach in Stage 2. Moreover, we also present how to tune \begin{equation}
\hat\kappa=\max\{\kappa\in\mathbb R:(\overline{\mathcal S}_\gamma^\kappa)^c\subset\mathcal H(T_0)\}\label{tune kappa}
\end{equation}
in Stage 1.

Problem \eqref{MIP problem} can be reformulated into a mixed integer program (MIP), in the case where $\hat g(x)$ is trained via a ReLU-activated neural net classifier, which is used in our deep-learning-based IS. 
Note that the objective is convex and second set of constraints is linear in \eqref{MIP problem}. If we can convert the first constraint $\hat g(x) \geq \gamma$ into linear mixed integer constraints, then the resulting formulation is a convex MIP with linear constraints, which can be solved much more efficiently than general nonlinear MIP with specialized algorithms (e.g., in the Gaussian case, this becomes a quadratic MIP) \citep{bonami2012algorithms}.
Focusing on the first constraint, the neural net structure $\hat g (x)$ in our approach (say with $n_g$ layers) can be represented as $\hat g(x)=(\hat g_{n_g}\circ ...\circ \hat g_1) (x)$, where each $\hat g_i (\cdot)$ denotes a ReLU-activated layer with linear transformation, i.e. $\hat g_i (\cdot)=\max\{ LT(\cdot),0 \}$, with $LT(\cdot)$ denoting a certain linear transformation in the input. In order to convert $\hat g(\cdot)$ into an MIP constraint, we introduce $M$ as a practical upper bound for $x_1,...,x_n$ such that $|x_i| < M$. The key step is to reformulate the ReLU function
$y=\max\{x,0 \}$ into \begin{align*}
     & y\leq x + M (1-z) \\
     & y\geq x\\
     & y\leq M z\\
     & y\geq 0\\
     & z \in \{0,1\}.
\end{align*} 
For simple ReLU networks, the size of the resulting MIP formulation depends linearly on the number of neurons in the neural network. In particular, the number of binary decision variables is linearly dependent on the number of ReLU neurons, and the number of constraints is linearly dependent the total number of all neurons (here we consider the linear transformations as independent neurons). 

The MIP reformulation we discussed can be generalized to many other popular piecewise linear structures in deep learning. For instance, linear operation layers, such as normalization and convolutional layers, can be directly used as constraints; some non-linear layers, such as ReLU and max-pooling layers, introduce  non-linearity by the ``max'' functions. A general reformulation for the max functions can be used to convert these non-linear layers to mixed integer constraints. More concrete, the equality defined by a max operation $y=\max\{x_1,x_2,...,x_n\}$ can be converted into \begin{align*}
     & y\leq x_i + 2M (1-z_i), i=1,...,n  \\
     & y\geq x_i, i=1,...,n\\
     & \sum_{i=1,...,n} z_i = 1\\
     & z_i \in \{0,1\}.
\end{align*}

Next, we illustrate how to tune $\hat\kappa$ to achieve \eqref{tune kappa}. This requires checking, for a given $\kappa$, whether $(\overline{\mathcal  S}_\gamma^{\kappa})^c\subset\mathcal H(T_0)$. Then, by discretizing the range of $\kappa$ or using a bisection algorithm, we can leverage this check to obtain \eqref{tune kappa}.

We use an MIP to check  $(\overline{\mathcal  S}_\gamma^{\kappa})^c\subset\mathcal H(T_0)$. Recall that $\mathcal H(T_0)=\bigcup_{i:Y_i=0}\{x\in\mathbb R_+^d:x\leq\tilde X_i\}$. We want to check if $\{x\in\mathbb R_+^d:\hat g(x)\leq \kappa \}$ for a given $\kappa$ lies completely inside the hull, where $\hat g(x)$ is trained with a ReLU-activated neural net. This can be done by solving an optimization problem as follows. First, we rewrite $\mathcal H(T_0)$ as $\{x\in\mathbb R_+^d:\min_{i=1,\ldots,n}\max_{j=1,\ldots,d}\{x^j-\tilde X_i^j\}\leq0\}$, where $x^j$ and $x_i^j$ refer to the $j$-th components of $x$ and $\tilde X_i$ respectively. Then we solve
\begin{equation}
\begin{array}{ll}
\max_{x\in\mathbb R^d}&\min_{i=1,\ldots,n}\max_{j=1,\ldots,d}\{x^j-\tilde X_i^j\}\\
\text{subject to}&\hat g(x)\leq \kappa\\
&x\geq0
\end{array}\label{opt}
\end{equation}
If the optimal value is greater than 0, this means $\{x\in\mathbb R_+^d:\hat g(x)\leq \kappa \}$ is not completely inside $\mathcal H(T_0)$, and vice versa. Now, we rewrite \eqref{opt} as

\begin{equation}
\begin{array}{ll}
\max_{x\in\mathbb R^d,\beta\in\mathbb R}&\beta\\
\text{subject to}&\max_{j=1,\ldots,d}\{x^j-\tilde X_i^j\}\geq\beta\ \forall i=1,\ldots,n\\
&\hat g(x)\leq \kappa\\
&x\geq0
\end{array}\label{opt1}
\end{equation}
We then rewrite \eqref{opt1} as an MIP by introducing a large real number $M$ as a practical upper bound for all coordinates of $x$:

\begin{equation}
\begin{array}{ll}
\max_{x\in\mathbb R^d,\beta\in\mathbb R}&\beta\\
\text{subject to}& x^j-\tilde X_i^j + 4M (1-z_{ij})\geq\beta\ \ \  \forall i=1,\ldots,n, j=1,\ldots,d\\
&\sum_{j=1,...,d} z_{ij} \geq 1\ \ \  \forall i=1,\ldots,n\\
&z_{ij}\in \{0,1\}\ \ \  \forall i=1,\ldots,n, j=1,\ldots,d\\
&\hat g(x)\leq \kappa\\
&x\geq0
\end{array}\label{opt2}
\end{equation}

Note that the set of points $T_0$ to be considered in constructing $\mathcal H(T_0)$ can be reduced to its ``extreme points''. More concretely, we call a point $x\in T_0$ an extreme point if there does not exist any other point $x'\in T_0$ such that $x\leq x'$.  We can eliminate all points $x\in T_0$ such that $x\leq x'$ for another $x'\in T_0$, and the resulting orthogonal monotone hull would remain the same. If we carry out this elimination, then in \eqref{opt1} we need only consider $\tilde X_i$ that are extreme points in $\mathcal H(T_0)$, which can reduce the number of integer variables needed to add. In practice, we can also randomly remove points in $T_0$ to further reduce the number of integer variables. This would not affect the correctness of our approach, but would increase the conservativeness of the final estimate. 

\subsubsection{Efficiency Guarantees.}
Regardless of the expressiveness of the ReLU networks presented above, Algorithm \ref{algo:stage1} enjoys the following guarantee:
\begin{theorem}[Relaxed efficiency certificate for deep-learning-based mixture IS]
Suppose $\mathcal S_\gamma$ is orthogonally monotone, and $\overline{\mathcal S}_\gamma^{\hat\kappa}$ satisfies the same conditions for $\mathcal S_{\gamma}$ in Theorem \ref{general IS}. Then Algorithm \ref{algo:stage1} attains the upper-bound relaxed efficiency certificate by using a constant number of Stage 1 samples.\label{NN main}
\end{theorem}

Figure \ref{fig:illustration}(d) shows how our deep-learning-based IS achieves superior efficiency compared to other alternatives. CE can miss a dominating point (1) and result in systematic under-estimation. The lazy-learner IS, on the other hand, generates too many dominating points (64) and degrades efficiency. Algorithm \ref{algo:stage1} finds the right number (2) and approximate locations of the dominating points.

Moreover, whereas the upper-bound certificate is guaranteed in our design, in practice, the deep-learning-based IS also appears to work well in controlling the conservativeness of the bound, as dictated by the false positive rate  $P(X\in \overline{\mathcal S}_\gamma^{\hat\kappa},Y=0)$.
Here we will provide a finite-sample bound on the false positive rate to support this phenomenon. We assume the use of a sampling distribution $q$ in generating independent Stage 1 samples.
Suppose that $f_{\theta}$ is the output of a neural network with 2 neurons in the output layer, and denote them as $f_{\theta,0},f_{\theta,1}$. Let $\mathcal{F}:=\{f_{\theta}\}$ denote the corresponding function class.
Let the loss function evaluated at the $i$-th sample be $\ell(f_{\theta}(\tilde X_{i}),Y_{i})$. 
For example, the cross-entropy loss is given by $-\left[\mathbb 1(Y_{i}=0)\log\frac{e^{f_{\theta,0}(\tilde X_{i})}}{e^{f_{\theta,0}(\tilde X_{i})}+e^{f_{\theta,1}(\tilde X_{i})}}+\mathbb 1(Y_{i}=1)\log\frac{e^{f_{\theta,1}(\tilde X_{i})}}{e^{f_{\theta,0}(\tilde X_{i})}+e^{f_{\theta,1}(\tilde X_{i})}}\right]$.
We compute the empirical minimizer of risk, which is
$\hat f:=\text{argmin}_{f\in\mathcal{F}}\{ {R}_{n_1}(f):=\frac{1}{n_1}\sum_{i=1}^{n_1}\ell(f(\tilde X_{i}),Y_{i})\}$.
For each function $f_{\theta}\in\mathcal{F}$, define function $g_{\theta}$ as $g_{\theta}:=f_{\theta,1}-f_{\theta,0}$. In
this approach, the learned rare-event set would be given
by $\tilde{\mathcal{S}}_{\gamma}^{\kappa}:=\{x:g_{\hat{\theta}}(x)\geq\kappa\}$,
and to make sure that $\mathcal{S}_{\gamma}\subset\tilde{\mathcal{S}}_{\gamma}^{\kappa}$,
we would replace $\kappa$ by $\hat{\kappa}=\min\{g_{\hat{\theta}}(x):x\notin\mathcal{H}(T_{0})\}$ as in Step 1 of Algorithm \ref{algo:stage1}.

We introduce several notations. Let $R(f_{\theta}):=E_{X\sim q}\ell(f_{\theta}(X),\mathbb 1(X\in\mathcal{S}_{\gamma}))$ denote the
true risk function. Let $f^{*}=\arg\min_{f\in\mathcal{F}}R(f)$ denote
the true risk minimizer within function class $\mathcal{F}$. Define
$g^{*}=f_{1}^{*}-f_{0}^{*}$ accordingly and let $\kappa^{*}:=\min_{x\in \mathcal{S}_{\gamma}}g^{*}(x)$
denote the true threshold associated with $f^*$ in obtaining the smallest outer rare-event approximation. Then we have the following result.

\begin{theorem}[Conservativeness of neural-network-generated set approximation] \label{thm: set_ERM_modified}  Suppose that the density $q$ has bounded support $K\subset[0,M]^d$ and $0<q_l\leq q(x)\leq q_u$ for any $x\in K$. Also suppose that there exists a function $h$
such that for any $f_{\theta}\in\mathcal{F}$, if $g_{\theta}(x)\geq\kappa$,
we have $\ell(f_{\theta}(x),0)\geq h(\kappa)>0$ (for the cross entropy loss, this happens if we know that $f_{\theta}$ has a bounded range). Then, for the set $\tilde{\mathcal{S}}_{\gamma}^{\hat{\kappa}}$, with probability at
least $1-\delta$,
\begin{align*}
 & P_{X\sim q}\left(X\in\tilde{\mathcal{S}}_{\gamma}^{\hat{\kappa}},X\in\mathcal{S}_{\gamma}^{c}\right)\\
\leq & \left(h(\kappa^{*}-t(\delta,n_1)\sqrt{d}\text{Lip}(g^{*})-\left\Vert \hat{g}-g^{*}\right\Vert _{\infty})\right)^{-1}\left(R(f^{*})+2\sup_{f_{\theta}\in\mathcal{F}}\left|R_{n_1}(f_{\theta})-R(f_{\theta})\right|\right).
\end{align*}
Here $\text{Lip}(g^*)$ is the Lipschitz parameter of $g^*$, and $t(\delta,n_1)=3\left(\frac{\log(n_1q_{l})+d\log M+\log\frac{1}{\delta}}{n_1q_{l}}\right)^{\frac{1}{d}}$.
\end{theorem}

 Theorem \ref{thm: set_ERM_modified} reveals a tradeoff between overfitting (measured by $\sup_{f_{\theta}\in\mathcal{F}}\left|R_{n_1}(f_{\theta})-R(f_{\theta})\right|$ and $\left\Vert \hat{g}-g^{*}\right\Vert _{\infty}$) and underfitting (measured by $R(f^*)=\inf_{f_{\theta}\in\mathcal{F}}R(f_{\theta})$). If the function class $\mathcal{F}$ is not rich (which happens when the neural network does not have enough neurons/layers), then the underfitting error may be big because of the lack of expressive power. On the other hand, if $\mathcal{F}$ contains too many functions, then the overfitting error will be large. In Appendix \ref{app: conservativeness}, we give related results on the sharp estimates of the quantities in Theorem \ref{thm: set_ERM_modified} for deep neural networks, the corresponding bounds for standard empirical risk minimization (which is viewed as an intermediate step towards showing Theorem \ref{thm: set_ERM_modified}) and the lazy learner, as well as results to interpret Theorem \ref{thm: set_ERM_modified} under the original distribution $p$.

To close this section, we point out the connections and distinctions of some works in the literature that relate to our orthogonal monotonicity notion and IS construction approach. First, \cite{Wu2018EfficientlyAT,legriel2010approximating} approximate the Pareto frontier of a monotone function. While the  boundary of an orthogonally monotone set looks similar to the Pareto frontier, our recipe (outer/inner approximation using piecewise-linear-activation NN) is designed to minimize the number of dominating points while simultaneously achieve the relaxed efficiency certificate for rare-event estimation. Such a guarantee is novelly beyond these previous works. 

Second, a recent work \cite{deo2021achieving} proposes an elegant IS method based on the self-similarity property of the optimal IS distribution to achieve logarithmic efficiency. They require the imposition of a risk function that is so-called asymptotically homogeneous and has a known order of growth. Compared to \cite{deo2021achieving}, we do not introduce risk functions or impose any accompanying analytical assumptions, but we require light-tailedness and the geometric premise that the rare-event set is orthogonally monotone. Our paper is connected to \cite{deo2021achieving} in that under their asymptotic homogeneity condition, we have, in our notation, $x\in\mathcal S_\gamma$ implies $tx\in\mathcal S_\gamma$ for any $t>1$ in $\mathbb R$ when $x$ is large enough in a suitable sense. Orthogonal monotonicity implies more points than $tx$ to be included in $\mathcal S_\gamma$. At the same time, our algorithm is parameter-free in that it does not utilize a risk function nor its growth properties.

\section{Numerical Experiments}\label{sec:numerics}
In this section, we present four experiments: a 2-dimensional rare event (shown in Figure \ref{fig:illustration}), complement of a 5-dimensional ball, random walk excursion, and safety testing of intelligent driving model. 
These experiments are chosen to demonstrate how our proposed framework works and gains insights under various settings. The first experiment is low-dimensional (visualizable) yet with \textit{extremely} rare target set, while the second one is very challenging for classical approaches due to its infinitely many dominating points. The third example is a classical setting in rare-event simulation which allows an analytic approach to design IS. Finally, the last experiment exemplifies a potential use-case of our method for safety evaluations of complex models. 
The codes and environment settings for the experiments are available at \href{https://github.com/safeai-lab/Deep-PrAE/}{https://github.com/safeai-lab/Deep-PrAE/}.

\subsection{Example 1: A 2D Example} 
We first consider a linear combination of sigmoid functions
$g(x) = \|\theta_1\psi(x-c_1-\gamma) +\theta_2\psi(x-c_2-\gamma)+\theta_3\psi(x-c_3-\gamma)+\theta_4\psi(x-c_4-\gamma)\|$ where $\theta,c$ are some constant vectors and $\psi(x) = \frac{ \exp(x)}{1+\exp(x)}$. A point $x$ is a rare-event if $g(x) > \gamma$, where we take $\gamma=1.8$ in Figure \ref{fig:illustration}.  We estimate $\mu=P(X \in \mathcal S_\gamma)$ where $X \sim N([5,5]^T, 0.25I_{2\times2})$, and $\gamma$ ranging from 1.0 to 2.0.  The target rare-event probability is microscopically small (e.g., $\gamma = 1.8$  gives $\mu=4.1\times 10^{-24}$) and serves to investigate our performance in ultra-extreme situations. We use a total number of samples $n=30,000$ ($n_1 = 10,000$ for Stage 1 and $n_2=20,000$ for Stage 2) and use the CE samples as our Stage 1 samples. 

We implement and compare the estimated probabilities and the REs of deep-learning-based IS for the upper bound (Deep-PrAE UB) and lazy-learner IS (LL UB). We also show the corresponding lower-bound estimator (Deep-PrAE LB and LL LB) and benchmark with the cross entropy method (CE), adaptive multilevel splitting (AMS), and naive Monte Carlo (NMC). For CE, we run a few variations testing different parametric classes and report two in the following figures: CE that uses a single Gaussian distribution (CE Naive), representing an overly-simplified CE implementation, and CE with Gaussian Mixture Model with $k$ components (CE GMM-$k$), representing a more sophisticated CE implementation. For Deep-PrAE, we use the samples from CE Naive as the Stage 1 samples. We also run a modification of Deep-PrAE (Deep-PrAE Mod) that replaces $\overline{\mathcal S}_\gamma^{\hat\kappa}$ by $\mathcal S_\gamma$ in the last step of Algorithm \ref{algo:stage1} as an additional comparison.

\begin{figure}[htbp]
    \begin{subfigure}{\textwidth}
      \begin{subfigure}{.33\textwidth}
      \centering
        \includegraphics[width=\textwidth]{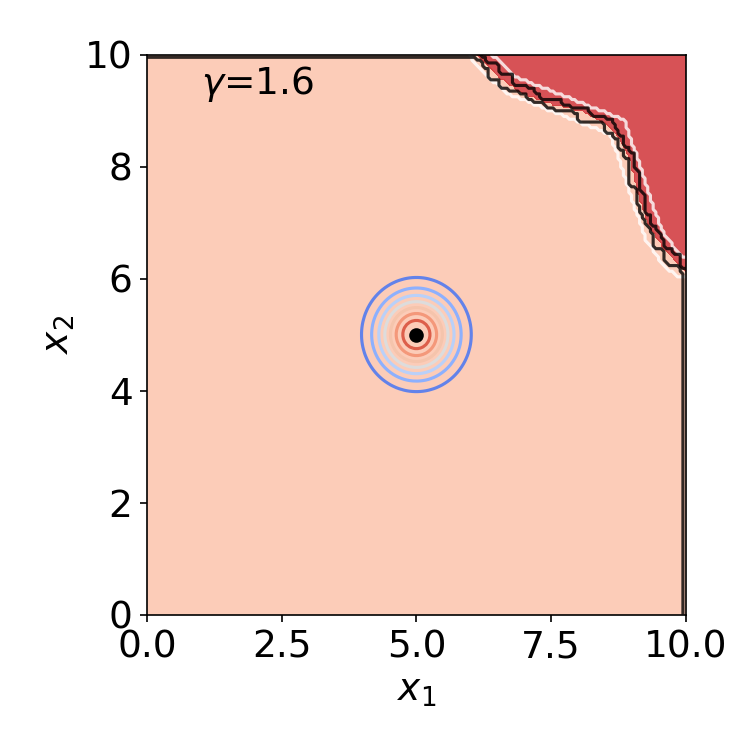}
        \caption{}
      \end{subfigure}
       \begin{subfigure}{.33\textwidth}
      \centering
        \includegraphics[width=\textwidth]{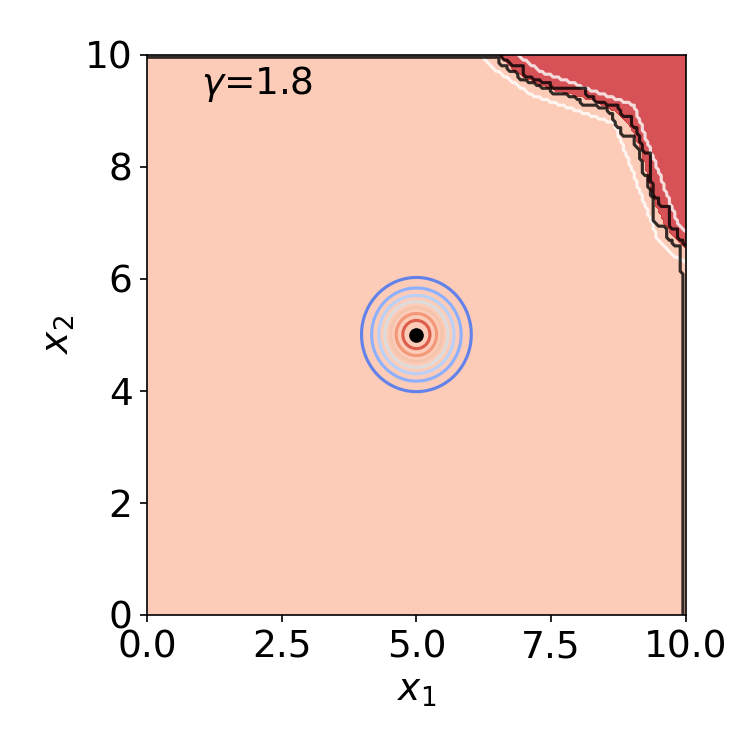}
        \caption{}
     \end{subfigure}
      \begin{subfigure}{.33\textwidth}
      \centering
        \includegraphics[width=\textwidth]{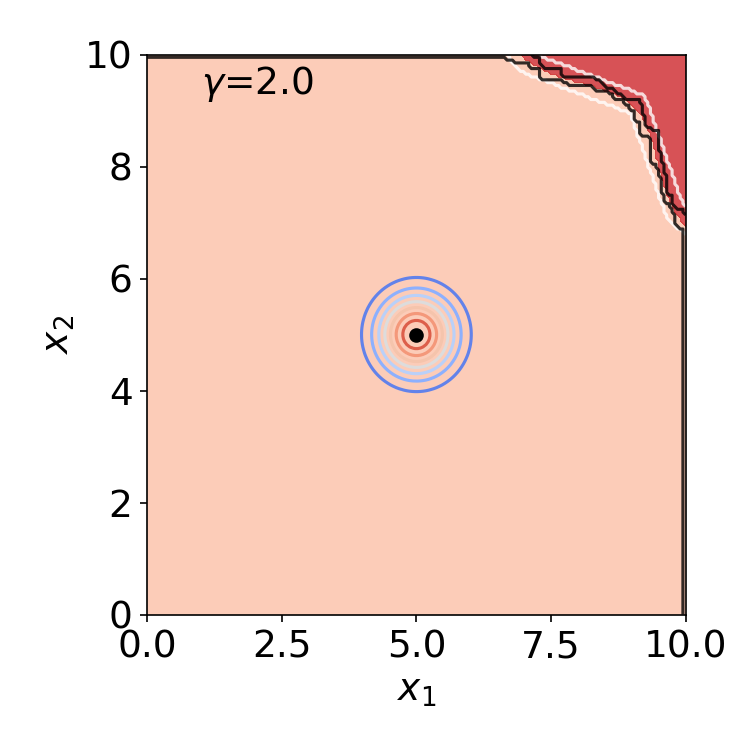}
        \caption{}
     \end{subfigure}
    \end{subfigure}
     
    \begin{subfigure}{\textwidth}
      \begin{subfigure}{.33\textwidth}
      \centering
        \includegraphics[width=\textwidth]{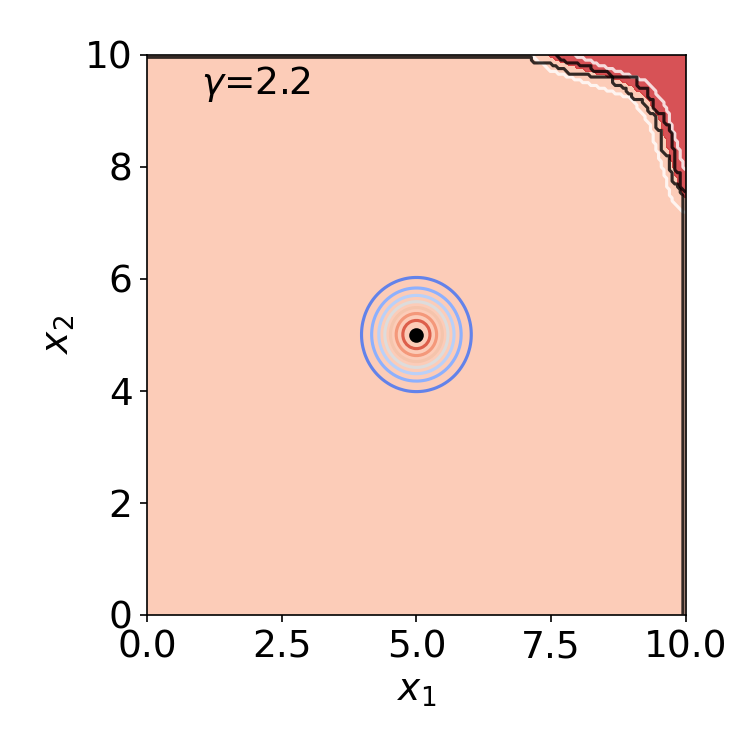}
        \caption{}
      \end{subfigure}
       \begin{subfigure}{.33\textwidth}
      \centering
        \includegraphics[width=\textwidth]{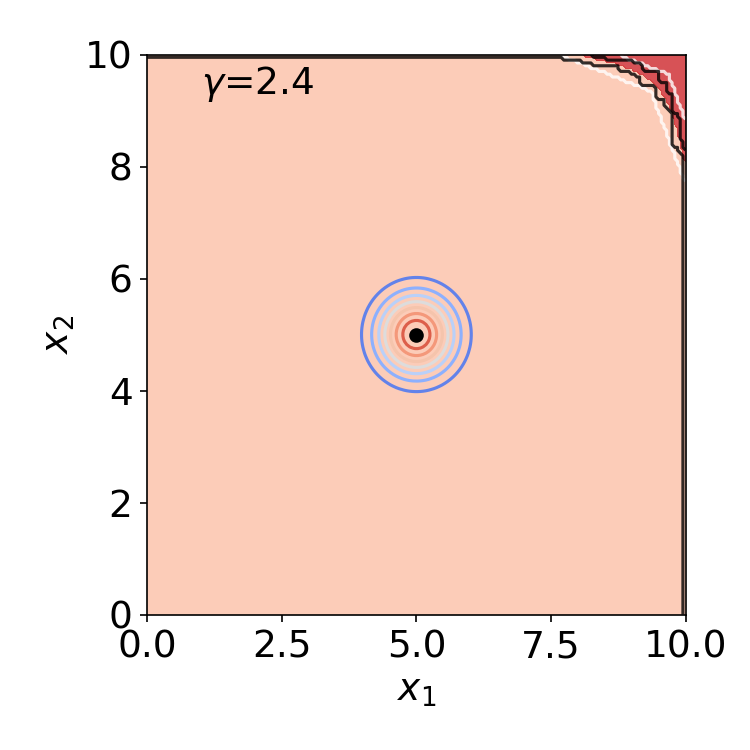}
        \caption{}
     \end{subfigure}
      \begin{subfigure}{.33\textwidth}
      \centering
        \includegraphics[width=\textwidth]{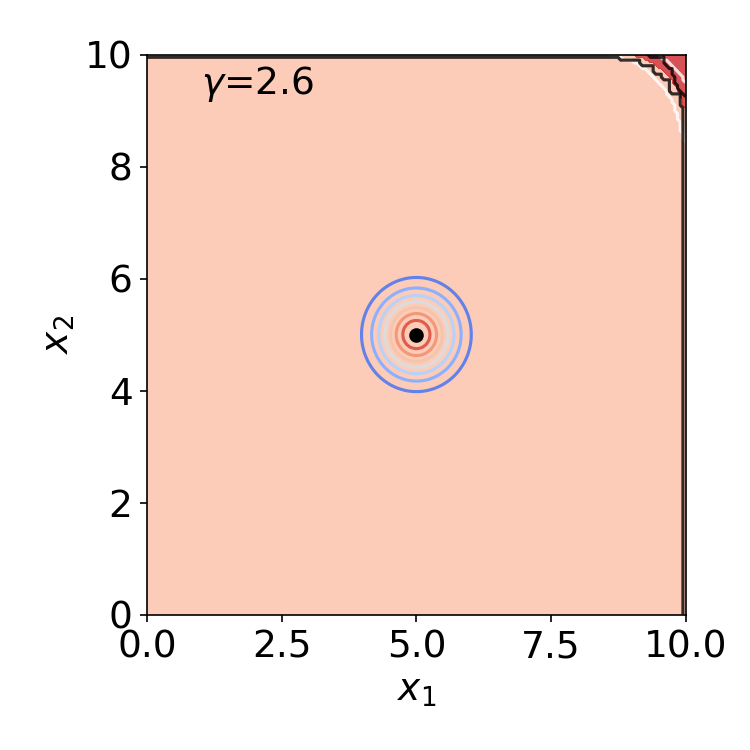}
        \caption{}
     \end{subfigure}
     
   \end{subfigure}
  
  \caption{The contour of $p$, rare-event set $\mathcal S_\gamma$ (dark reddish region), outer- and inner- approximation boundaries (black lines) and Deep-PrAE UB and LB decision boundaries (white lines) for some $\gamma$ values in Example 1. }
  \label{fig:exp_2d_rare_set}
\end{figure}

\begin{figure}[htbp]
  \begin{subfigure}{\textwidth}
  \centering
    \includegraphics[width=0.6\textwidth]{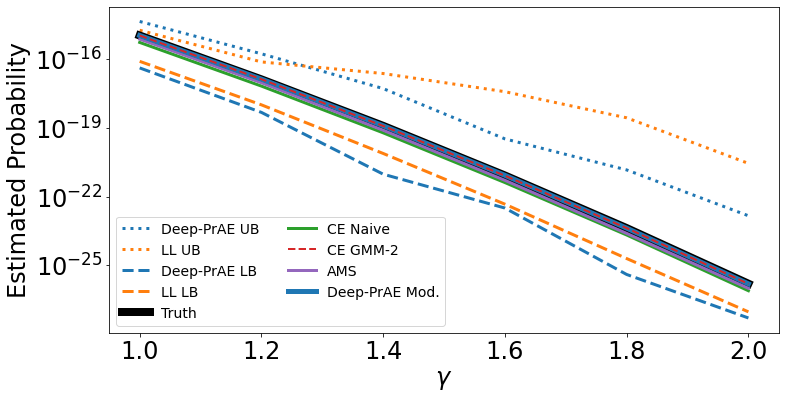}
    \caption{Estimated rare-event probability (Example 1)}
  \end{subfigure}
  
  \begin{subfigure}{\textwidth}
  \centering
    \includegraphics[width=0.6\textwidth]{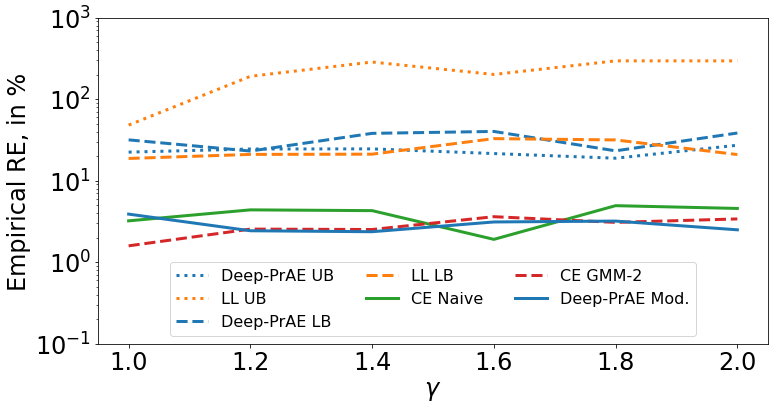}
    \caption{Estimator's empirical relative error (Example 1)}
  \end{subfigure}
  
  \caption{Results in Example 1. Naive Monte Carlo failed in all cases and hence not shown.}
  \label{fig:nonstd_result}
\end{figure}

\begin{figure}[htbp]
  \centering
    \includegraphics[width=0.6\textwidth]{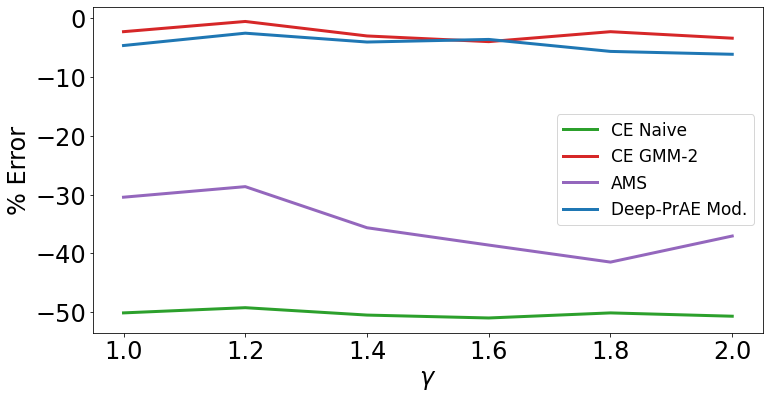}
  \caption{Percentage error of CE, AMS, and modified Deep-PrAE for Example 1 (minus \% error means under-estimation)}
  \label{fig:pe_result}
\end{figure}

Figure \ref{fig:exp_2d_rare_set} shows the rare-event set and its approximations for various $\gamma$'s. The Deep-PrAE boundaries seem tight in most cases, attributed to both the sufficiently trained NN classifier and the bisection algorithm implemented for tuning $\hat \kappa$ after the NN training. Figure \ref{fig:nonstd_result} compares all approaches to the true value, which we compute via a proper mixture IS with 50,000 samples assuming the full knowledge of $\mathcal S_\gamma$. It shows several observations. First, Deep-PrAE and LL (both UB and LB) always provide valid bounds that contain the truth. Second, the UB for LL is more conservative than Deep-PrAE in up to two orders of magnitudes, which is attributed to the overly many (redundant) dominating points. Correspondingly, the RE of LL UB is tremendously high, reaching over $500\%$ when $\gamma = 2.0$, compared to around 40\% for Deep-PrAE UB, Deep-PrAE LB, and LL LB. Third, CE Naive, which finds only one dominating point, consistently under-estimates the truth by about $50\%$, yet it gives an over-confident RE, e.g., $<5\%$ when $\gamma < 2$. This shows a systematic undetected under-estimation issue when CE is implemented overly-naively. AMS also underestimates the true value by 30\%-40\%, while CE GMM-2 and Deep-PrAE Mod perform empirically well.  Figure \ref{fig:pe_result} summarizes the zoomed-in performances of CE Naive, CE GMM-2, AMS, and Deep-PrAE Mod in terms of percentage error, which is the difference between the estimated and true probability as a percentage of the true value. It shows that while CE performs well when the IS parametric class is well-chosen (CE GMM-2), a poor CE parametric class (CE Naive) as well as AMS could under-estimate. Yet our Deep-PrAE, despite using samples from a poor CE class in Stage 1, can recover valid results: Deep-PrAE provides a valid UB, and Deep-PrAE Mod gives an estimate as good as the good CE class.

\subsection{Example 2: Complement of a Ball}

\begin{figure}
    \centering
    \includegraphics[width=0.9\textwidth]{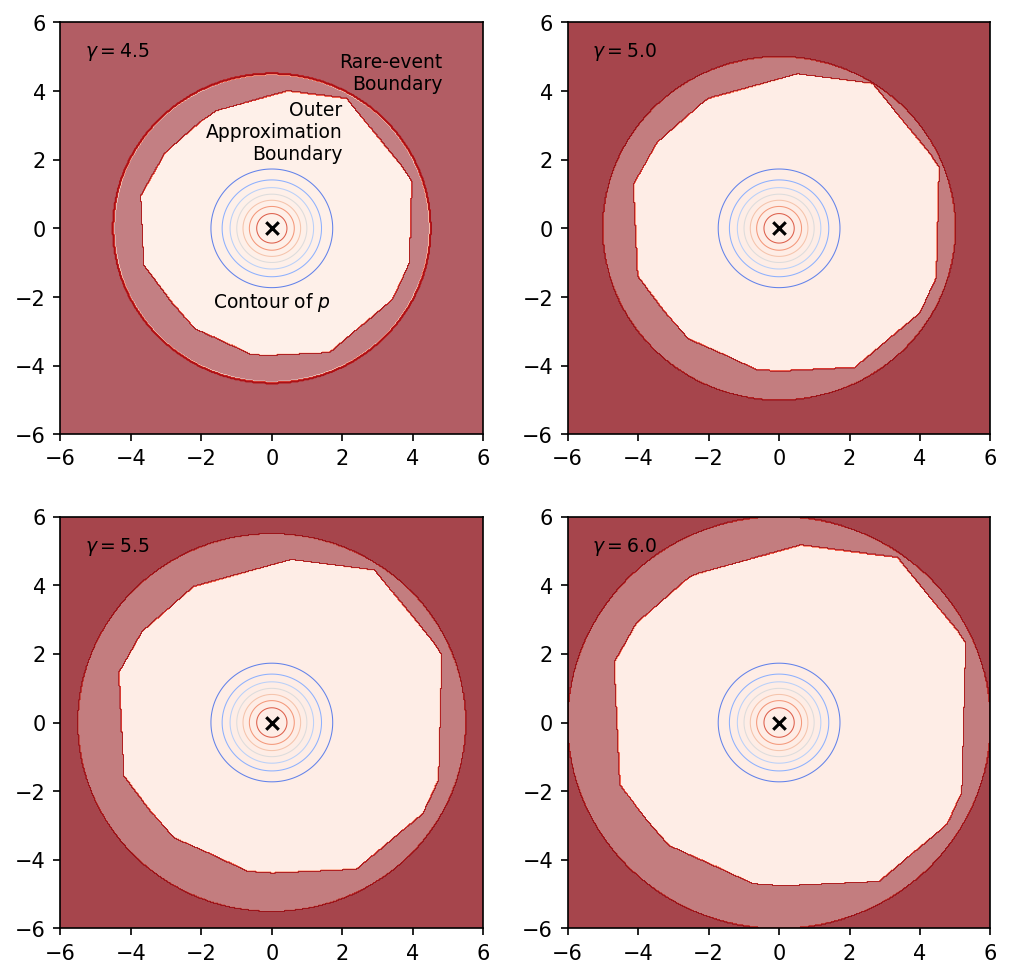}
    \caption{2-dimensional slices depicting the rare-event set $\mathcal S_\gamma$ and Deep-PrAE outer-approximation 
    for various rarity levels $\gamma$ in Example 2}
    \label{fig:circle_settings}
\end{figure}

\begin{figure}
  \centering
    \includegraphics[width=\linewidth]{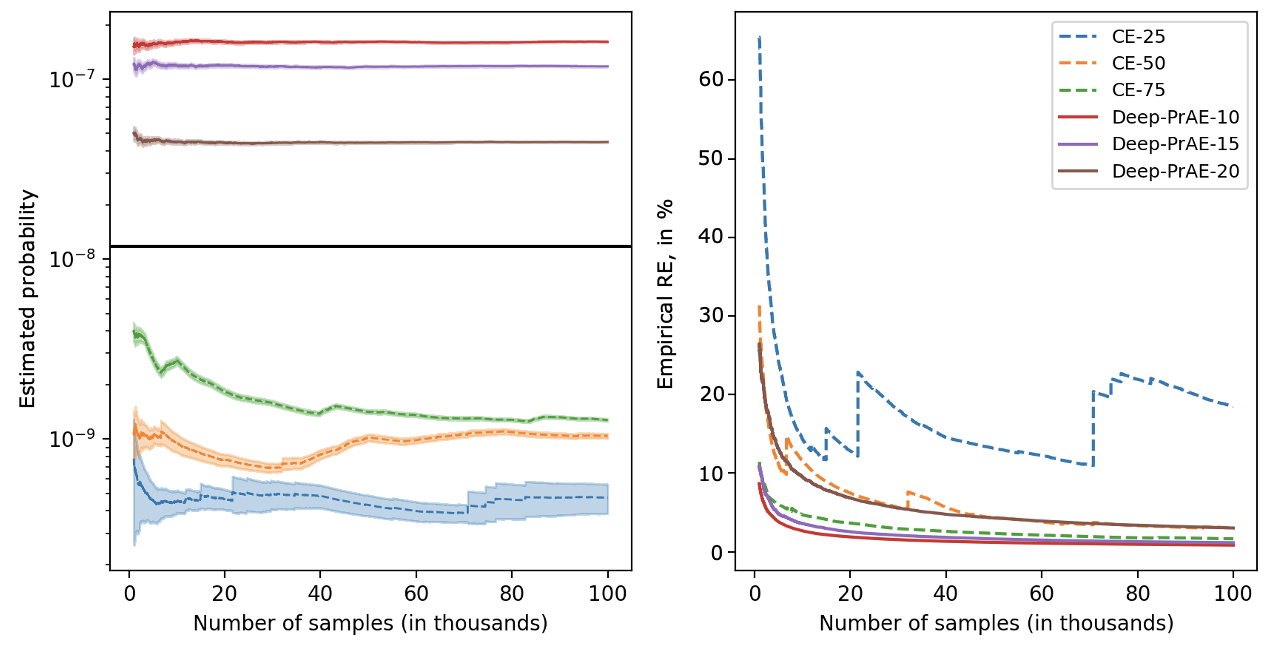}
    \caption{Estimated rare-event probability and empirical relative error for $\gamma=4.75 ~(\mu \approx 1.18 \times 10^{-8})$ for Example 2}
    \label{fig:circle_res_1}    
\end{figure}

\begin{figure}
  \centering
    \includegraphics[width=\textwidth]{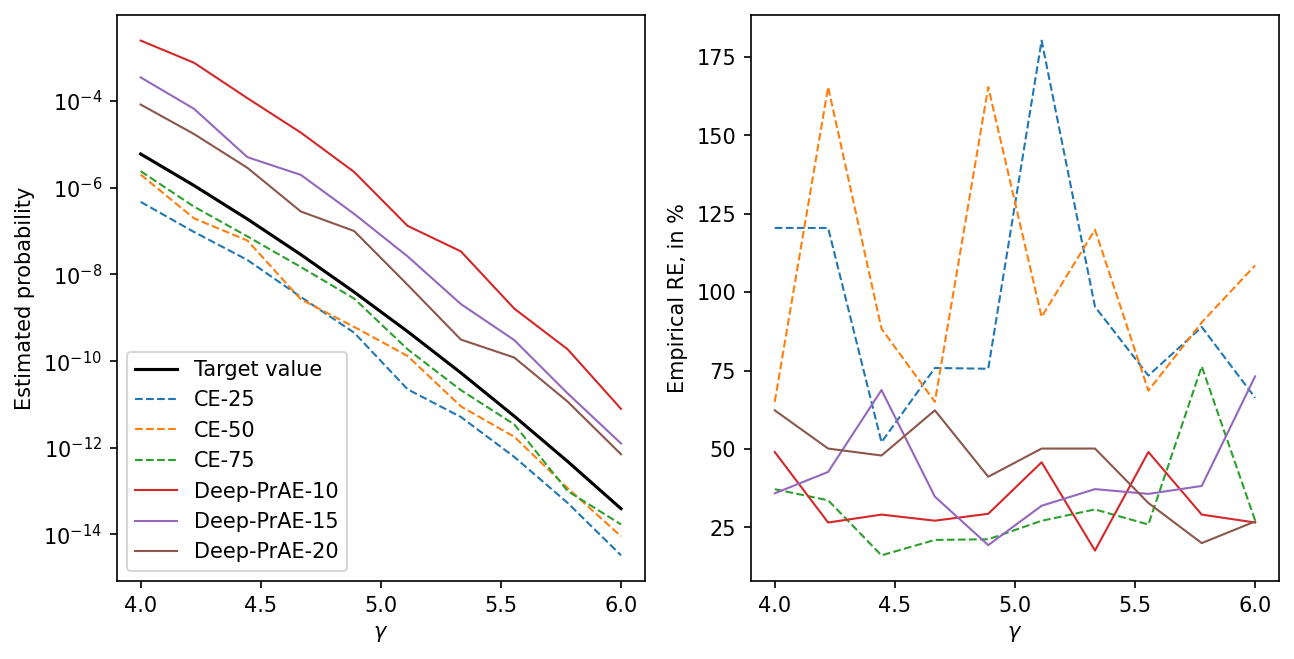}  
    \caption{Estimated rare-event probability and empirical relative error for various rarity levels $\gamma$ in Example 2}
    \label{fig:circle_res_2}      
\end{figure}

In this example, our target rare-event set is $\mathcal S_\gamma = \{x \in \mathbb R^5: \|x\|_2 \geq \gamma\}$, the complement of a $5$-dimensional ball of radius $\gamma$. The interesting aspect of this example is that the target rare-event set has infinitely many dominating points (any points in its circumference), making it extremely hard to design efficient IS using tools from the existing literature. We study how estimators from CE constructed using GMM with $k=25$, $50$ and $75$ components and Deep-PrAE using $h=10$, $15$ and $20$ hidden nodes perform in terms of RE and correctness. The original distribution is $p = N(0, 0.5I_{5 \times 5})$. Figure \ref{fig:circle_settings} shows the 2-dimensional slices of the space centered at the origin with the rare-event set and the outer approximations obtained by Deep-PrAE Stage 1. We show the estimated probabilities and REs for $\gamma=4.75$ across different samples sizes in Figure \ref{fig:circle_res_1} and across rarity levels in Figure \ref{fig:circle_res_2}. 

We see from Figure \ref{fig:circle_res_1} (left) that CE estimators, supposed to be unbiased, underestimate the target value. Meanwhile, Deep-PrAE estimators designed to overestimate the target indeed provide valid upper bounds. Again, CE underestimation issues are likely to be undetected had we not been able to compare the estimated probabilities with the target value, especially since all the REs are already relatively small signaling a confident estimation with number of samples $>$ 20,000  (from Figure \ref{fig:circle_res_1} (right)). In Figure \ref{fig:circle_res_2}, for rarity level $\gamma$ ranging from 4.0 to 6.0, we see that both the CE under-estimation and misleadingly high confidence issues persist. Across CE estimators, we see that increasing the number of components helps reduce the under-estimation gap and RE to some extent. Meanwhile, across Deep-PrAE, increasing the number of hidden nodes helps reduce the conservativeness of the estimators.

We also observe that the number of dominating points found using Deep-PrAE increases significantly in the number of hidden nodes $h$. We have 49 dominating points for $h=10$, 80 for $h=15$, and 153 for $h=20$, thus implying that Deep-PrAE tends to construct more complex IS proposal with more complex architecture. Comparing Deep-PrAE-10 and CE-50 that use IS proposals with roughly the same complexity (49-component vs 50-component GMM), we see that CE-50 suffers from under-estimation while Deep-PrAE-10 obtains an upper-bound with a much lower RE, highlighting the superiority of the proposed method.

\subsection{Example 3: Random Walk}
In this example, our goal is to estimate the excursion probability of a $T$-step random walk
$\mu = P \left(  \max_{t=1, \cdots, T} S_t > \gamma \right),$
where $S_t = \sum_{i=1}^t X_i$, $X_i$ for $i=1,\ldots,T$ are i.i.d. following distribution $p$, and the rarity parameter is the cross level $\gamma$. We use $T=10$, $p=N(0, \sigma^2I_{T \times T})$, and $\gamma=11$. 

\begin{figure}[htbp]
  \begin{subfigure}{\textwidth}
  \centering
    \includegraphics[width=0.8\textwidth]{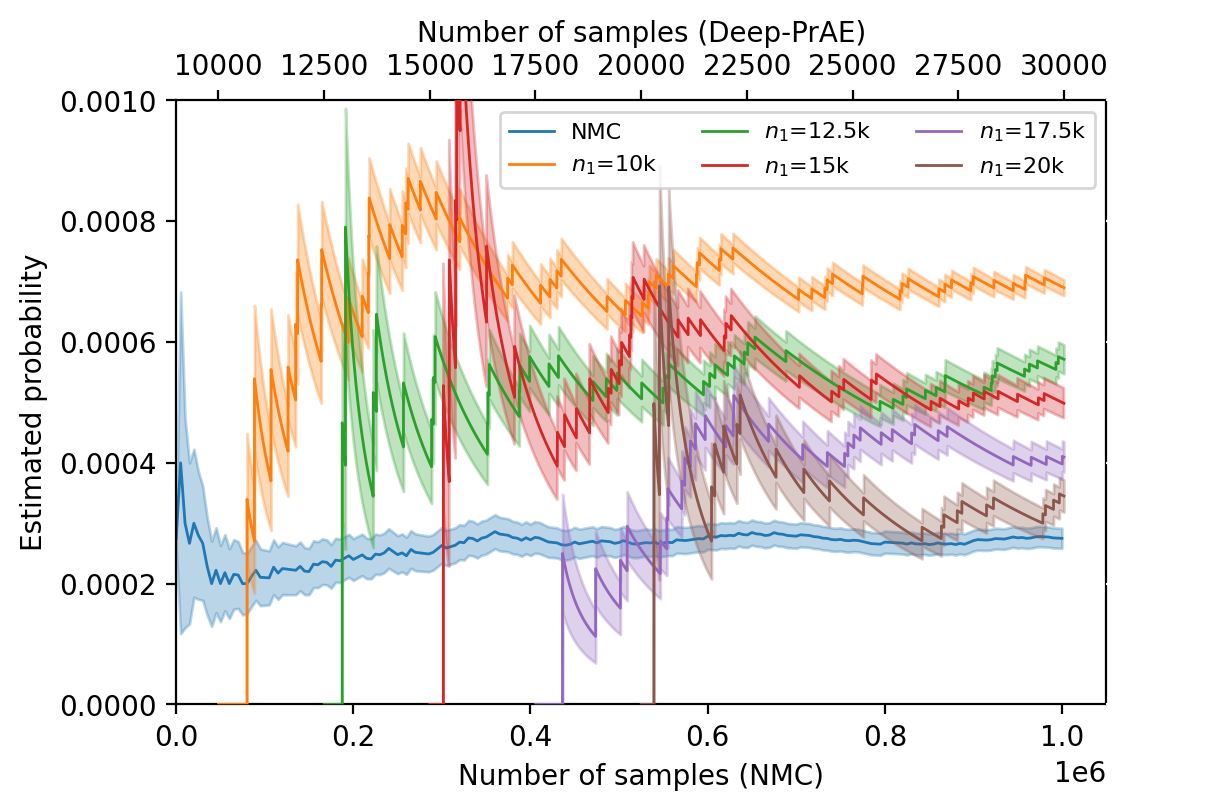}
    \caption{Estimated rare-event probability (Example 3)}
  \end{subfigure}%
  
  \begin{subfigure}{\textwidth}
  \centering
    \includegraphics[width=0.8\textwidth]{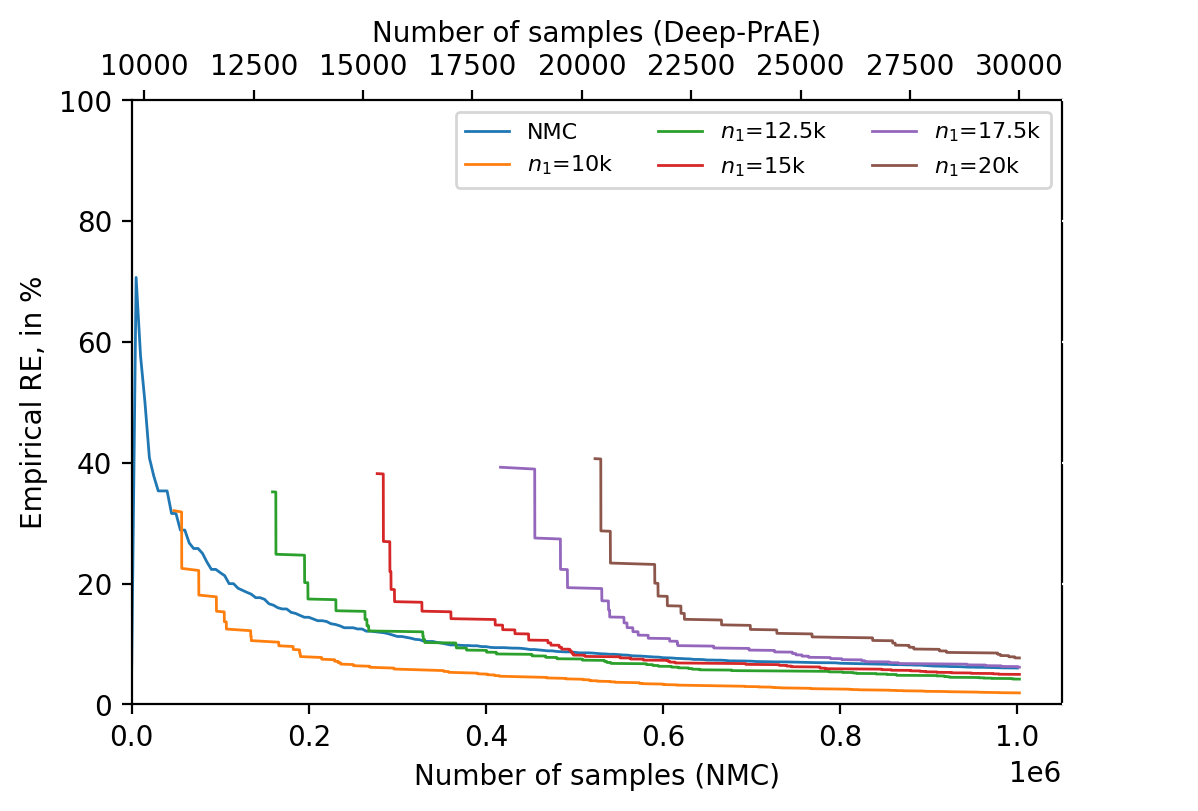}
    \caption{Estimator's empirical relative error (Example 3)}
  \end{subfigure}
  \caption{Estimated rare-event probability and empirical relative error for Example 3}
  
  \label{fig:rw-main}
\end{figure}

Figure \ref{fig:rw-main} compares the estimators and REs obtained using NMC and Deep-PrAE with various Stage 1 sample sizes ($n_1$). We set the maximum total sample size for any of the Deep-PrAE estimators to be $n=30,000$, and the Stage 2 sample size is $n_2 = n - n_1$. In the figure, we show the  sample sizes for NMC and Deep-PrAE separately due to their significant difference in magnitude and usage: NMC sample size in the bottom $x$-axis (all used to drive down estimator variance) and Deep-PrAE sample size in the top $x$-axis (used for both $n_1$ and $n_2$ with different proportions). The starting $x$-value of Deep-PrAE estimator curve marks the portion of sample budget used for $n_1$ while the rest used for $n_2$. Therefore, high variances occur at the beginning of each curve, which then saturate as larger $n_2$ is allocated to drive down the variance.

We observe that all Deep-PrAE estimators output valid upper bounds for the target probability, up to 2.5 times larger than the target value, with much better relative errors (all converge to less than 5\% with total sample size $n=3\times10^4$) compared to NMC which has a relative error of 15\% when sample size $n=10^6$). We also observe that allocating more samples to Stage 1 (using $n_1=17.5\times10^3$ and $n_2=12.5\times10^3$) leads to less conservative estimate and more confident estimation in this example. However, it is expected that when $n_1$ increases too much further, then there could be a deterioration in performances because of a strained $n_2$. 

\subsection{Example 4: Intelligent Driving Example} 
\begin{figure}[htbp]
  \centering
    \includegraphics[width=0.8\textwidth]{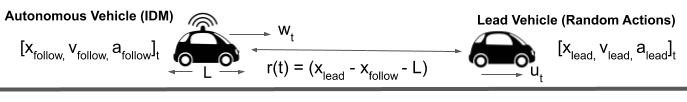}
    \caption{The states $s_t$ and input $u_t$ of the self-driving safety-testing simulation. $w_t$ denotes the throttle input of the AV from the IDM.
    }
    \label{fig:sim_state}
\end{figure}

In this example, we evaluate the crash probability of car-following scenarios involving a human-driven lead vehicle (LV) followed by an autonomous vehicle (AV). 
The AV is controlled by the Intelligent Driver Model (IDM), widely used for autonomy evaluation and microscopic transportation simulation \citep{Treiber_2000, dlforcf, rssidm}, that maintains a safety distance while ensuring smooth ride and maximum efficiency. The states of the AV are $s_t = [x_{\text{follow}}, x_{\text{lead}}, v_{\text{follow}}, v_{\text{lead}}, a_{\text{follow}}, a_{\text{lead}}]_t$ which are the position, velocity and acceleration of the AV and LV respectively. The throttle input to the AV is defined as $w_t$  which has an affine relationship with the acceleration of the vehicle. Similarly, the randomized throttle of the LV is represented by $u_t$. With a car length of $L$, the distance between the LV and AV at time $t$ is given by $r_t=x_{\text{lead}, t}-x_{\text{follow}, t}-L$, which has to remain below the crash threshold for safety. We consider an evaluation horizon $T=60$ seconds and draw a sequence of 15 Gaussian random actions at a 4-second epoch, leading to a 15-dimensional LV action space. A (rare-event) crash occurs at time $t \leq T$ if the longitudinal distance $r_t$ between the two vehicles is negative, with $\gamma$ parameterizing the AV maximum throttle and brake pedals. This rare-event set is analytically challenging (see \citealt{zhao2017accelerated} for a similar setting). We describe the dynamics in more detail below. Figure \ref{fig:sim_state} gives a pictorial overview of the interaction.

\subsubsection{LV Actions.}
The LV action contains human-driving uncertainty in decision-making modeled as Gaussian increments. For every $\Delta t$ time-steps, a Gaussian random variable is generated with the mean centered at the previous action $u_{t-\Delta t}$. We initialize $u_0 = 10$ (unitless) and $\Delta t = 4$ sec, which corresponds to zero initial acceleration and an acceleration change in the LV once every 4 seconds.

\subsubsection{Intelligent Driver Model (IDM) for AV.} The IDM is governed by the following equations (the subscripts ``follow'' and ``lead'' defined in Figure \ref{fig:sim_state} is abbreviated to ``f'' and ``l'' for conciseness):
\begin{align*}
    \dot x_{f} &= v_{f}\\
    \dot x_{l} &= v_{l}\\
     \dot v_{f} & = \max \left(a(1-\left(\frac{v_{f}}{v_0}\right)^{\delta}-\left(\frac{s^{*}(v_{f},\Delta v_{f})}{s_{f}}\right)^{2}), -d\right)\\
    s^{*}(v_{f},\Delta v_{f}) &= s_0 + v_{f}\bar T + \frac{v_{f}\Delta v_{f}}{2\sqrt{ab}}\\
    s_{f} &= x_{l} - x_{f} - L\\
    \Delta v_{f} &= v_{f} - v_{l},
\end{align*}
The parameters are presented in Table \ref{idm-params}. The randomness of LV actions $u_t$'s propagates into the system and affects all the simulation states $s_t$.
The IDM is governed by simple first-order kinematic equations for the position and velocity of the vehicles. The acceleration of the AV is the decision variable where it is defined by a sum of non-linear terms which dictate the ``free-road'' and ``interaction'' behaviors of the AV and LV. The acceleration of the AV is constructed in such a way that certain terms of the equations dominate when the LV is far away from the AV to influence its actions and other terms dominate when the LV is in close proximity to the AV.

\begin{table}
\caption{Parameters of the Intelligent Drivers Model (IDM)}
\label{idm-params}
\centering
\begin{tabular}{ll}
\toprule
Parameters                         & Value              \\
\midrule
Safety distance, $s_0$                                    & 2 m                         \\ 
Speed of AV in free traffic, $v_0$                        & 30 m/s                      \\ 
Maximum acceleration of AV, $a$                          & $2\gamma$ m/s$^2$    \\ 
Comfortable deceleration of AV, $b$                      & 1.67 m/s$^2$ \\ 
Maximum deceleration of AV, $d$                          & $2\gamma$ m/s$^2$    \\ 
Safe time headway, $\bar T$                                   & 1.5 s                       \\ 
Acceleration exponent parameter, $\delta$ & 4                           \\
Car length, $L$                                          & 4 m                         \\
\bottomrule
\end{tabular}
\end{table}

\subsubsection{Rarity Parameter $\gamma$.} 
\begin{figure}[htbp]
    \begin{subfigure}{\textwidth}
       \begin{subfigure}{.49\textwidth}
      \centering
        \includegraphics[width=\textwidth]{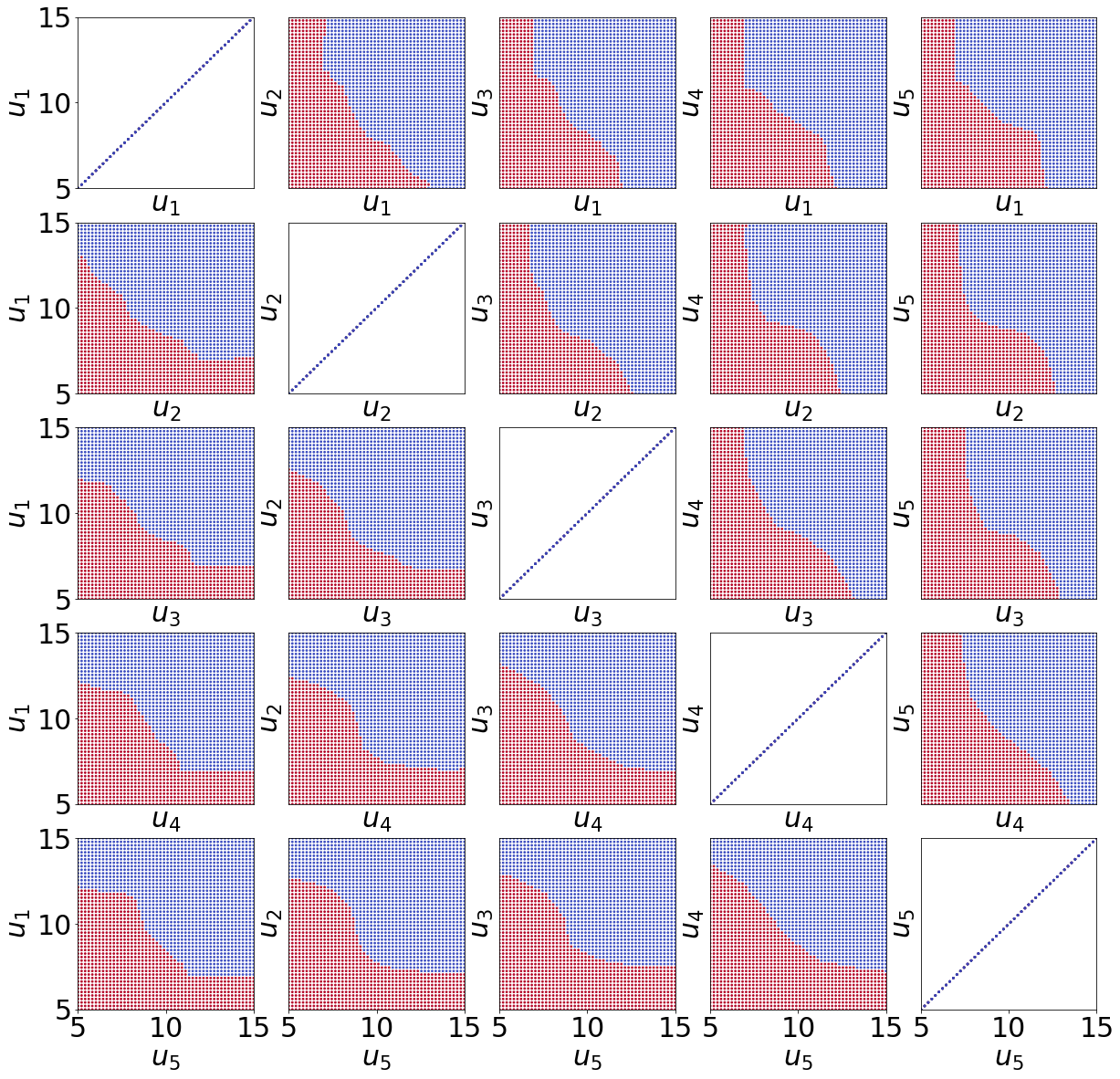}
        \caption{Case $\gamma=1.0$}
        \label{fig:lv_slices_gamma10}
     \end{subfigure}
      \begin{subfigure}{.49\textwidth}
      \centering
        \includegraphics[width=\textwidth]{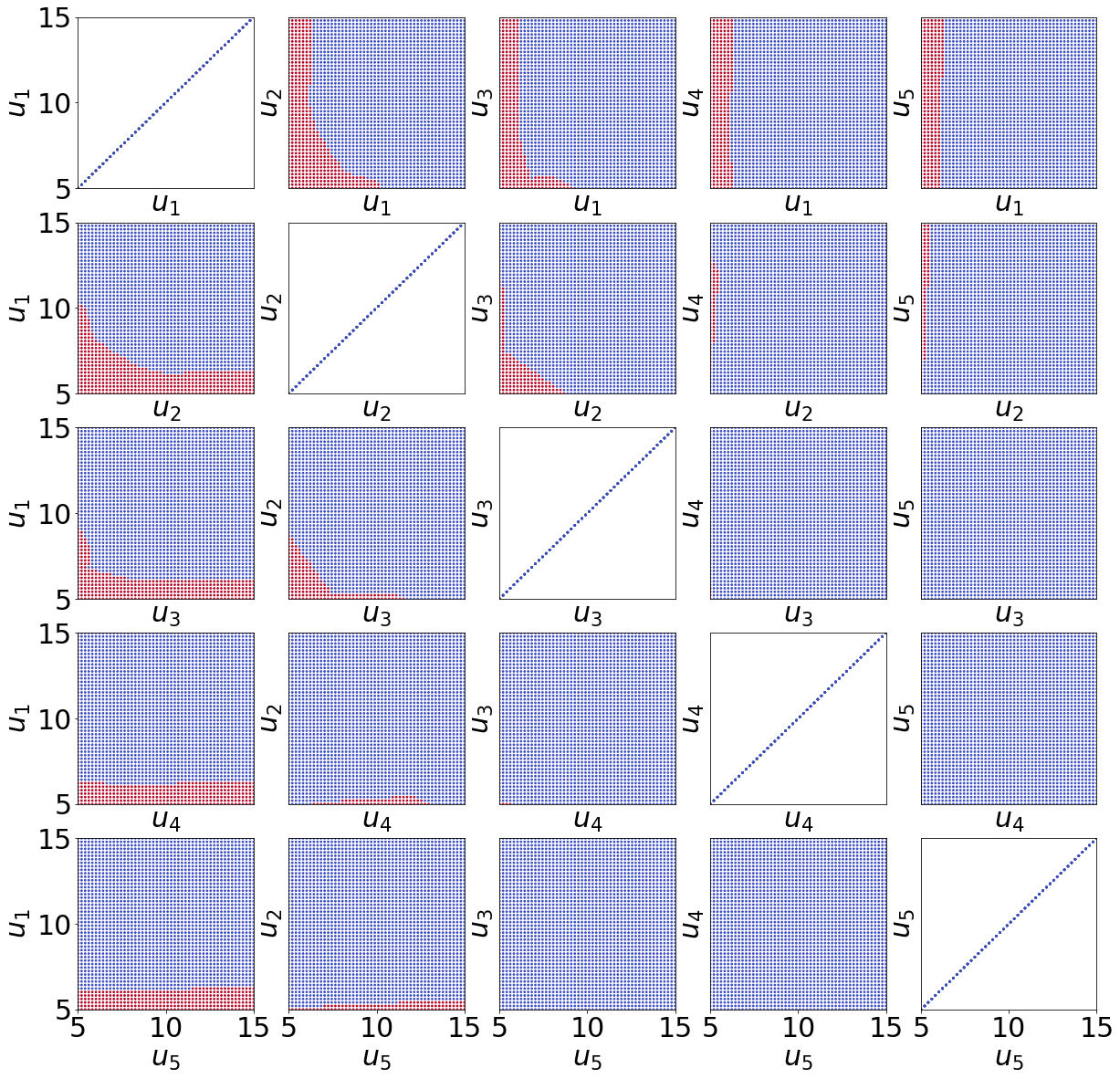}
        \caption{Case $\gamma=2.0$}
        \label{fig:lv_slices_gamma20}
      \end{subfigure}%
   \end{subfigure}
  \centering
    \caption{Slice of pairs of the first 5 dimensions of LV action space. For any $(u_i, u_{i^\prime})$ shown, $u_j, j \not \in \{i, i^\prime\}$ is fixed at a constant value. Blue dots = non-crash cases, red dots= crash cases.}
  
  \label{fig:lv_slices_gamma}
\end{figure}

Parameter $\gamma$ signifies the range invoked by the AV acceleration and deceleration pedals. Increasing $\gamma$ implies that the AV can have sudden high deceleration and hence avoid crash scenarios better and making crashes rarer. In contrast, decreasing $\gamma$ reduces the braking capability of the AV and more easily leads to crashes. For instance, $\gamma=1.0$ corresponds to AV actions in the range $[5, 15]$ or correspondingly $a_{\text{follow}, t} \in [-2, 2]$, and  $\gamma=2.0$ corresponds to $a_{\text{follow}, t} \in [-4, 4]$. Figure \ref{fig:lv_slices_gamma} shows the approximate rare-event set by randomly sampling points and evaluating the inclusion in the set, for the two cases of $\gamma=1.0$ and $\gamma=2.0$. In particular, we slice the 15-dimensional space onto pairs from five of the dimensions. In all plots, we see that the crash set (red) are monotone, thus supporting the use of our Deep-PrAE framework. Although the crash set is not located in the ``upper-right corner'', we can implement Deep-PrAE framework for such problems by simple re-orientation.

\subsubsection{Sample Trajectories. }
\begin{figure}[htbp]
  \centering
      \begin{subfigure}{.22\textwidth}
      \centering
        \includegraphics[width=\textwidth]{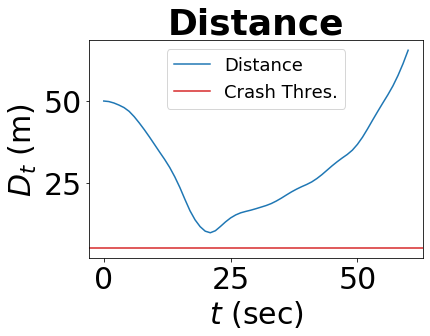}
        \caption{}
      \end{subfigure}
      \begin{subfigure}{.22\textwidth}
      \centering
        \includegraphics[width=\textwidth]{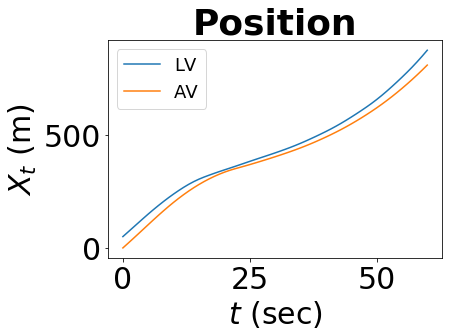}
        \caption{}
      \end{subfigure}
      \begin{subfigure}{.22\textwidth}
      \centering
        \includegraphics[width=\textwidth]{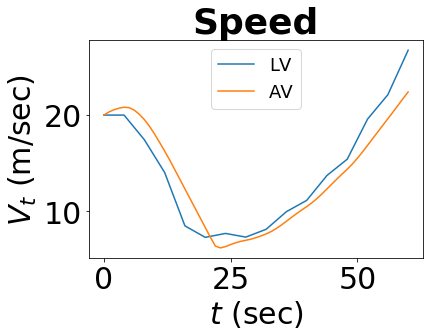}
        \caption{}
      \end{subfigure}
       \begin{subfigure}{.22\textwidth}
      \centering
        \includegraphics[width=\textwidth]{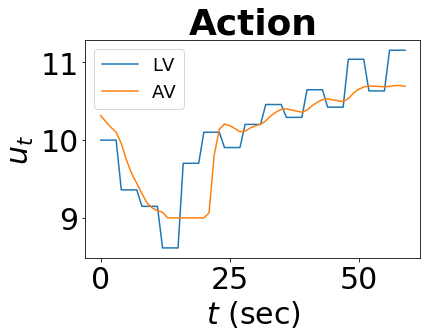}
        \caption{}
      \end{subfigure}
        \begin{subfigure}{.22\textwidth}
      \centering
        \includegraphics[width=\textwidth]{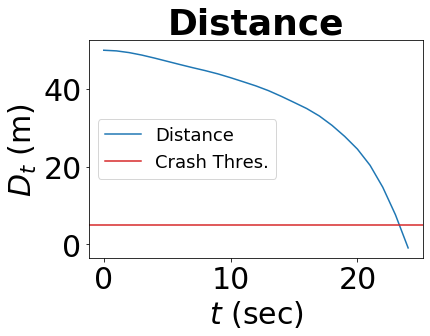}
        \caption{}
      \end{subfigure}
      \begin{subfigure}{.22\textwidth}
      \centering
        \includegraphics[width=\textwidth]{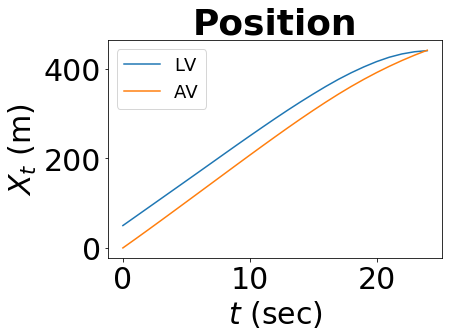}
        \caption{}
      \end{subfigure}
      \begin{subfigure}{.22\textwidth}
      \centering
        \includegraphics[width=\textwidth]{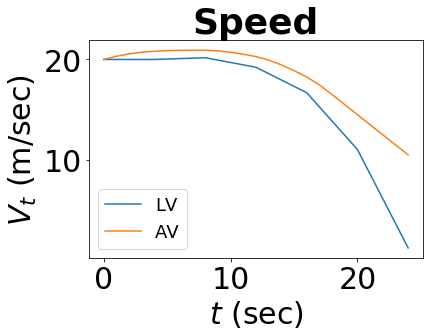}
        \caption{}
      \end{subfigure}
       \begin{subfigure}{.22\textwidth}
      \centering
        \includegraphics[width=\textwidth]{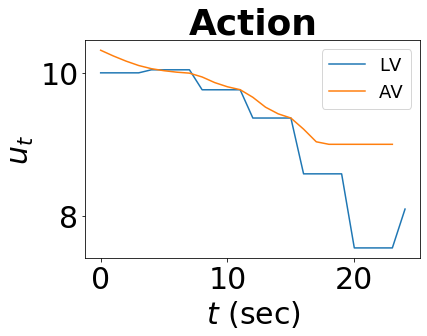}
        \caption{}
      \end{subfigure}
    \caption{Autonomous Car Following Experiment Trajectories. Figures (a) - (d) represent a simulation episode without a crash occurring where the AV follows the LV successfully at a safe distance. Figures (e) - (h) represents a simulation episode where crash occurs at $t=23$ seconds due to the repeated deceleration of the LV.}
  \label{fig:carfollowtraj}
\end{figure}

Figure \ref{fig:carfollowtraj} shows two examples of sample trajectories, one successfully maintaining a safe distance, and the other leading to a crash. In Figure \ref{fig:carfollowtraj}(e)-(h) where we show the crash case, the AV maintains a safe distance behind the LV until the latter starts rapidly decelerating (Figure \ref{fig:carfollowtraj}(h)). Here the action corresponds to the throttle input that has an affine relationship with the acceleration of the vehicle. The LV ultimately decelerates at a rate that the AV cannot attain and its deceleration saturates after a point which leads to the crash.

\begin{figure}[htbp]
  \begin{subfigure}{\textwidth}
  \centering
    \includegraphics[width=0.6\textwidth]{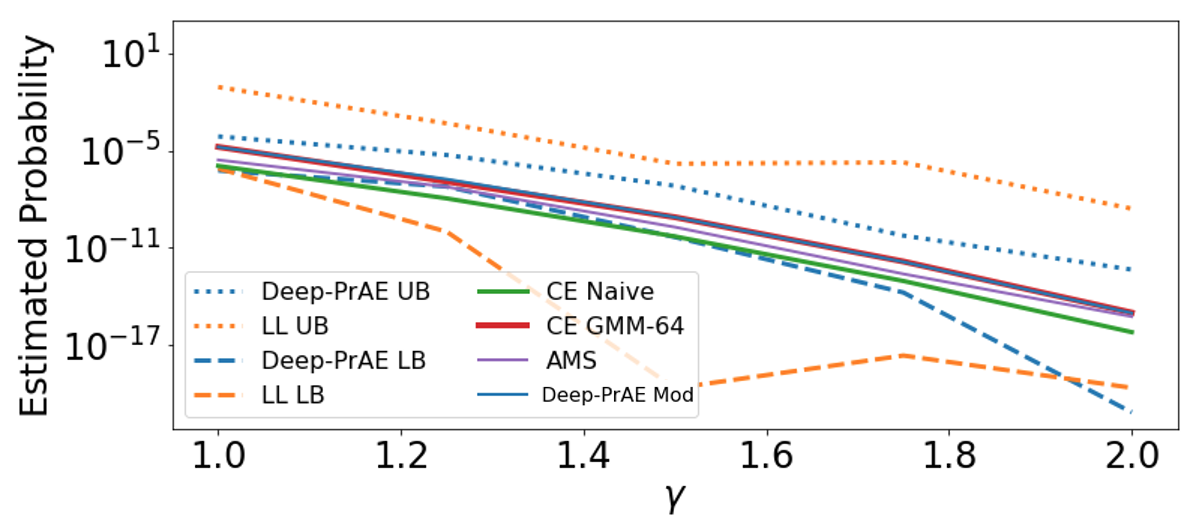}
    \caption{Estimated rare-event probability (Example 4)}
  \end{subfigure}%
  
  \begin{subfigure}{\textwidth}
  \centering
    \includegraphics[width=0.6\textwidth]{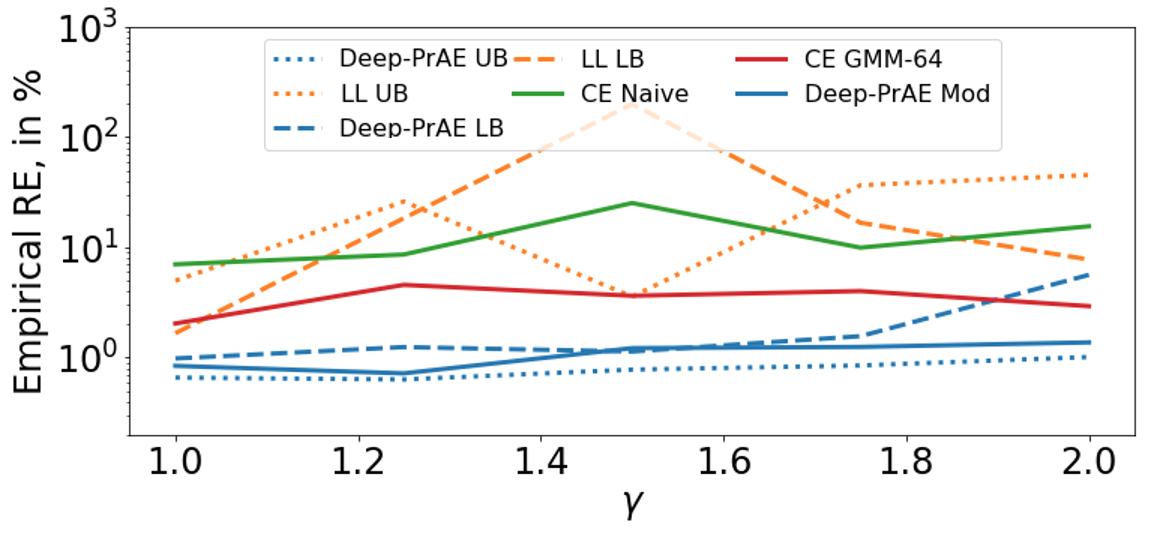}
    \caption{Estimator's empirical relative error (Example 4)}
  \end{subfigure}
  \caption{Results in Example 4: intelligent driving example. Naive Monte Carlo failed in all cases and hence not shown.}
  \label{fig:nonstd_result_highdim}
\end{figure}

\subsubsection{Results and Discussion.}Figure  \ref{fig:nonstd_result_highdim} shows the performances of competing approaches, using $n = 10,000$. For CE, we use a single Gaussian (CE Naive) and a large number of mixtures (CE GMM-64). Deep-PrAE and LL (UB and LB) appear consistent in giving upper and lower bounds for the target probability, and Deep-PrAE produces tighter bounds than LL ($10^{-2}$ vs $10^{-6}$ in general). LL UB has $5,644$ dominating points when $\gamma=1$ vs 42 in Deep-PrAE, and needs 4 times more time to search for them than Deep-PrAE. Moreover, the RE of Deep-PrAE is 3 times lower than LL across the range (in both UB and LB). Thus, Deep-PrAE outperforms LL in both tightness, computation speed, and RE. CE Naive and AMS seem to give a stable estimation, but evidence shows they are under-estimated: Deep-PrAE Mod and CE GMM-64 lack efficiency certificates and thus could under-estimate, and the fact that their estimates are higher than CE Naive and AMS suggests both CE Naive and AMS are under-estimated. Lastly, NMC fails to give a single hit in all cases (thus not shown on the graphs). Thus, among all approaches, Deep-PrAE stands out as giving reliable and tight bounds with low RE.

To conclude, our investigation shows evidence of the practical benefits of using Deep-PrAE for rare-event estimation. It generates valid bounds for the target probability with low RE and improved efficiency. The use of classifier prediction helps reduce the computational effort from running more simulations. For example, to assess whether the AV crash rate is below $10^{-8}$ for $\gamma = 1.0$, only 1000 simulation runs would be needed by Deep-PrAE UB or LB to get around $1\%$ RE, which takes about $400$ seconds in total. This is in contrast to 3.7 months for naive Monte Carlo.

\section{Discussion and Future Work}

In this paper, we proposed a robust certifiable approach to estimate rare-event probabilities in black-box settings that arise in safety-critical applications. The proposed approach designs efficient IS distribution by combining the dominating point machinery with deep-learning-based rare-event set learning. We study the theoretical guarantees and present numerical examples. The key property that distinguishes our approach with existing black-box rare-event simulation methods is our correctness guarantee. Leveraging on a new notion of relaxed efficiency certificate and the orthogonal monotonicity assumption, our approach avoids the perils of undetected under-estimation as potentially encountered by other methods.

We discuss some key assumptions in our approach and related prospective follow-up works. First, the orthogonal monotonicity assumption appears an important first step to give new theories on black-box rare-event estimation beyond the existing literature. Indeed, we show that even with this assumption, black-box approaches such as CE and splitting can suffer from the dangerous pitfall of undiagnosed under-estimation, and our approach corrects for it. The real-world values of our approach are: (1) We rigorously show why our method has better performances in the orthogonally monotone cases; (2) For tasks close to being orthogonal monotonic (e.g., the IDM example), our method is empirically more robust; (3) For non-orthogonally-monotone tasks, though directly using our approach does not provide guarantees, we could potentially train mappings to latent spaces that are orthogonally monotone. Further work also includes relaxing notions from orthogonal monotonicity to attain efficiency guarantees. We believe such type of geometric assumptions comprises a key ingredient towards a rigorous theory for black-box rare-event estimation that warrants much further developments.

Second, the dominating point search algorithm in our approach requires information on the cumulant generating function as well as its Legendre transform. For many multivariate distributions, such information is unavailable or hard to analyze. To this end, we can consider, for instance, fitting a GMM with a sufficiently large number of components. These extensions will again be left for future work.
Finally, the tightness of the upper bound depends on the sample quality. An ideal method in Stage 1 would generate samples close to the rare-event boundary to produce good approximations. Cutting the Stage 1 effort by, e.g., designing iterative schemes between Stages 1 and 2, will also be a topic for future investigation.

\section*{Acknowledgments}
We gratefully acknowledge support from the National Science Foundation under grants CAREER CMMI-1834710,  IIS-1849280 and IIS-1849304. A preliminary conference version of this work has appeared in the International Conference on Artificial Intelligence and Statistics 2021 \citep{arief2021deep}.

\bibliographystyle{informs2014} 
\bibliography{references}

\begin{thebibliography}{97}
\providecommand{\natexlab}[1]{#1}
\providecommand{\url}[1]{\texttt{#1}}
\providecommand{\urlprefix}{URL }

\bibitem[{Anil et~al.(2019)Anil, Lucas, \protect\BIBand{}
  Grosse}]{pmlr-v97-anil19a}
Anil C, Lucas J, Grosse R (2019) Sorting out {L}ipschitz function
  approximation. Chaudhuri K, Salakhutdinov R, eds., \emph{Proceedings of the
  36th International Conference on Machine Learning}, volume~97 of
  \emph{Proceedings of Machine Learning Research}, 291--301 (Long Beach,
  California, USA: PMLR).

\bibitem[{Arief et~al.(2018)Arief, Glynn, \protect\BIBand{}
  Zhao}]{arief2018accelerated}
Arief M, Glynn P, Zhao D (2018) An accelerated approach to safely and
  efficiently test pre-production autonomous vehicles on public streets.
  \emph{2018 21st International Conference on Intelligent Transportation
  Systems (ITSC)}, 2006--2011 (IEEE).

\bibitem[{Arief et~al.(2021)Arief, Huang, Kumar, Bai, He, Ding, Lam,
  \protect\BIBand{} Zhao}]{arief2021deep}
Arief M, Huang Z, Kumar GKS, Bai Y, He S, Ding W, Lam H, Zhao D (2021) Deep
  probabilistic accelerated evaluation: A robust certifiable rare-event
  simulation methodology for black-box safety-critical systems.
  \emph{International Conference on Artificial Intelligence and Statistics},
  595--603 (PMLR).

\bibitem[{Asmussen(1985)}]{pASM85a}
Asmussen S (1985) Conjugate processes and the simulation of ruin problems.
  \emph{Stochastic Processes and their Applications} 20:213--229.

\bibitem[{Asmussen \protect\BIBand{} Albrecher(2010)}]{ASM00Ruin}
Asmussen S, Albrecher H (2010) \emph{Ruin Probabilities}, volume~14 (World
  scientific).

\bibitem[{Asmussen \protect\BIBand{} Kroese(2006)}]{asmussen2006improved}
Asmussen S, Kroese DP (2006) Improved algorithms for rare event simulation with
  heavy tails. \emph{Advances in Applied Probability} 38(2):545–558.

\bibitem[{Au \protect\BIBand{} Beck(2001)}]{au2001estimation}
Au SK, Beck JL (2001) Estimation of small failure probabilities in high
  dimensions by subset simulation. \emph{Probabilistic Engineering Mechanics}
  16(4):263--277.

\bibitem[{Blanchet \protect\BIBand{} Glynn(2008)}]{blanchet2008efficient}
Blanchet J, Glynn P (2008) {Efficient rare-event simulation for the maximum of
  heavy-tailed random walks}. \emph{The Annals of Applied Probability}
  18(4):1351 -- 1378.

\bibitem[{Blanchet et~al.(2009)Blanchet, Glynn, \protect\BIBand{} Lam}]{BGL10}
Blanchet J, Glynn P, Lam H (2009) Rare event simulation for a slotted time
  {$M/G/s$} model. \emph{Queueing Systems} 63:33--57.

\bibitem[{Blanchet et~al.(2012{\natexlab{a}})Blanchet, Glynn, \protect\BIBand{}
  Leder}]{blanchet2012lyapunov}
Blanchet J, Glynn P, Leder K (2012{\natexlab{a}}) On lyapunov inequalities and
  subsolutions for efficient importance sampling. \emph{ACM Trans. Model.
  Comput. Simul.} 22(3), ISSN 1049-3301.

\bibitem[{Blanchet \protect\BIBand{} Lam(2012)}]{blanchet2012state}
Blanchet J, Lam H (2012) State-dependent importance sampling for rare-event
  simulation: An overview and recent advances. \emph{Surveys in Operations
  Research and Management Science} 17(1):38--59.

\bibitem[{Blanchet \protect\BIBand{} Lam(2014)}]{blanchet2014rare}
Blanchet J, Lam H (2014) Rare-event simulation for many-server queues.
  \emph{Mathematics of Operations Research} 39(4):1142--1178.

\bibitem[{Blanchet et~al.(2012{\natexlab{b}})Blanchet, Lam, \protect\BIBand{}
  Zwart}]{BLANCHET20123361}
Blanchet J, Lam H, Zwart B (2012{\natexlab{b}}) Efficient rare-event simulation
  for perpetuities. \emph{Stochastic Processes and their Applications}
  122(10):3361--3392, ISSN 0304-4149.

\bibitem[{Blanchet \protect\BIBand{} Mandjes(2009)}]{BMRT09}
Blanchet J, Mandjes M (2009) Rare event simulation for queues. \emph{Rare Event
  Simulation Using Monte Carlo Methods}, 87--124, chapter 5.

\bibitem[{Blanchet \protect\BIBand{} Liu(2008)}]{blanchet2008state}
Blanchet JH, Liu J (2008) State-dependent importance sampling for regularly
  varying random walks. \emph{Advances in Applied Probability}
  40(4):1104--1128.

\bibitem[{Bonami et~al.(2012)Bonami, Kilin{\c{c}}, \protect\BIBand{}
  Linderoth}]{bonami2012algorithms}
Bonami P, Kilin{\c{c}} M, Linderoth J (2012) Algorithms and software for convex
  mixed integer nonlinear programs. \emph{Mixed integer nonlinear programming},
  1--39 (Springer).

\bibitem[{Botev et~al.(2013)Botev, L'Ecuyer, \protect\BIBand{}
  Tuffin}]{botev2013markov}
Botev ZI, L'Ecuyer P, Tuffin B (2013) Markov chain importance sampling with
  applications to rare event probability estimation. \emph{Statistics and
  Computing} 23(2):271--285.

\bibitem[{Botev et~al.(2016)Botev, Ridder, \protect\BIBand{}
  Rojas-Nandayapa}]{botev2016semiparametric}
Botev ZI, Ridder A, Rojas-Nandayapa L (2016) Semiparametric cross entropy for
  rare-event simulation. \emph{Journal of Applied Probability} 53(3):633--649.

\bibitem[{Bucklew(2013)}]{bucklew2013introduction}
Bucklew J (2013) \emph{Introduction to Rare Event Simulation} (Springer Science
  \& Business Media).

\bibitem[{Bucklew(2004)}]{Bucklew2004}
Bucklew JA (2004) \emph{Rare Event Simulation for Level Crossing and Queueing
  Models}, 195--206 (New York, NY: Springer New York), ISBN 978-1-4757-4078-3.

\bibitem[{Budhiraja \protect\BIBand{} Dupuis(2019)}]{budhiraja2019analysis}
Budhiraja A, Dupuis P (2019) \emph{Analysis and Approximation of Rare Events:
  Representations and Weak Convergence Methods} (Springer).

\bibitem[{Cao \protect\BIBand{} Gu(2019)}]{NIPS2019_9246}
Cao Y, Gu Q (2019) Tight sample complexity of learning one-hidden-layer
  convolutional neural networks. \emph{Advances in Neural Information
  Processing Systems 32}, 10612--10622 (Curran Associates, Inc.).

\bibitem[{C{\'e}rou \protect\BIBand{} Guyader(2007)}]{cerou2007adaptive}
C{\'e}rou F, Guyader A (2007) Adaptive multilevel splitting for rare event
  analysis. \emph{Stochastic Analysis and Applications} 25(2):417--443.

\bibitem[{C{\'e}rou et~al.(2019)C{\'e}rou, Guyader, \protect\BIBand{}
  Rousset}]{cerou2019adaptive}
C{\'e}rou F, Guyader A, Rousset M (2019) Adaptive multilevel splitting:
  Historical perspective and recent results. \emph{Chaos: An Interdisciplinary
  Journal of Nonlinear Science} 29(4):043108.

\bibitem[{Chen et~al.(2019)Chen, Blanchet, Rhee, \protect\BIBand{}
  Zwart}]{chen2019efficient}
Chen B, Blanchet J, Rhee CH, Zwart B (2019) Efficient rare-event simulation for
  multiple jump events in regularly varying random walks and compound poisson
  processes. \emph{Mathematics of Operations Research} 44(3):919--942.

\bibitem[{Clarke et~al.(2018)Clarke, Henzinger, Veith, \protect\BIBand{}
  Bloem}]{clarke2018handbook}
Clarke EM, Henzinger TA, Veith H, Bloem R (2018) \emph{Handbook of Model
  Checking}, volume~10 (Springer).

\bibitem[{Claybrook \protect\BIBand{} Kildare(2018)}]{claybrook2018autonomous}
Claybrook J, Kildare S (2018) Autonomous vehicles: No driver… no regulation?
  \emph{Science} 361(6397):36--37.

\bibitem[{Collamore(2002)}]{collamore2002importance}
Collamore JF (2002) Importance sampling techniques for the multidimensional
  ruin problem for general markov additive sequences of random vectors.
  \emph{The Annals of Applied Probability} 12(1):382--421.

\bibitem[{Corso et~al.(2020)Corso, Moss, Koren, Lee, \protect\BIBand{}
  Kochenderfer}]{corso2020survey}
Corso A, Moss RJ, Koren M, Lee R, Kochenderfer MJ (2020) A survey of algorithms
  for black-box safety validation. \emph{arXiv preprint arXiv:2005.02979} .

\bibitem[{Cérou \protect\BIBand{} Guyader(2016)}]{cerou2016fluctuation}
Cérou F, Guyader A (2016) {Fluctuation analysis of adaptive multilevel
  splitting}. \emph{The Annals of Applied Probability} 26(6):3319 -- 3380.

\bibitem[{De~Boer et~al.(2005)De~Boer, Kroese, Mannor, \protect\BIBand{}
  Rubinstein}]{de2005tutorial}
De~Boer PT, Kroese DP, Mannor S, Rubinstein RY (2005) A tutorial on the
  cross-entropy method. \emph{Annals of Operations Research} 134(1):19--67.

\bibitem[{Dean \protect\BIBand{} Dupuis(2009)}]{dean2009splitting}
Dean T, Dupuis P (2009) Splitting for rare event simulation: A large deviation
  approach to design and analysis. \emph{Stochastic Processes and Their
  Applications} 119(2):562--587.

\bibitem[{Dembo \protect\BIBand{} Zeitouni(2010)}]{dembo2010large}
Dembo A, Zeitouni O (2010) \emph{Large Deviations Techniques and Applications}
  (Springer-Verlag).

\bibitem[{Deo \protect\BIBand{} Murthy(2021)}]{deo2021achieving}
Deo A, Murthy K (2021) Achieving efficiency in black box simulation of
  distribution tails with self-structuring importance samplers. \emph{arXiv
  preprint arXiv:2102.07060} .

\bibitem[{Dieker \protect\BIBand{} Mandjes(2005)}]{dieker2005asymptotically}
Dieker A, Mandjes M (2005) On asymptotically efficient simulation of large
  deviation probabilities. \emph{Advances in applied probability}
  37(2):539--552.

\bibitem[{Dieker \protect\BIBand{} Mandjes(2006)}]{dieker2006fast}
Dieker AB, Mandjes M (2006) Fast simulation of overflow probabilities in a
  queue with gaussian input. \emph{ACM Transactions on Modeling and Computer
  Simulation (TOMACS)} 16(2):119--151.

\bibitem[{Dupuis \protect\BIBand{} Ellis(2011)}]{dupuis2011weak}
Dupuis P, Ellis RS (2011) \emph{A Weak Convergence Approach to the Theory of
  Large Deviations}, volume 902 (John Wiley \& Sons).

\bibitem[{Dupuis et~al.(2009)Dupuis, Leder, \protect\BIBand{}
  Wang}]{dupuis2009importance}
Dupuis P, Leder K, Wang H (2009) Importance sampling for
  weighted-serve-the-longest-queue. \emph{Mathematics of Operations Research}
  34(3):642--660.

\bibitem[{Ellis(1984)}]{Ellis1984}
Ellis RS (1984) Large deviations for a general class of random vectors.
  \emph{Ann. Probab.} 12(1):1--12.

\bibitem[{Evan(2016)}]{Evan2016FatalPerfect}
Evan A (2016) {Fatal Tesla self-driving car crash reminds us that robots aren't
  perfect}. \emph{IEEE Spectrum} .

\bibitem[{Feng et~al.(2018)Feng, Yu, Xu, Liu, \protect\BIBand{} Peng}]{8500545}
Feng Y, Yu C, Xu S, Liu HX, Peng H (2018) An augmented reality environment for
  connected and automated vehicle testing and evaluation*. \emph{2018 IEEE
  Intelligent Vehicles Symposium (IV)}, 1549--1554,
  \urlprefix\url{http://dx.doi.org/10.1109/IVS.2018.8500545}.

\bibitem[{Glasserman(2004)}]{Glasserman04}
Glasserman P (2004) \emph{Monte Carlo Methods in Financial Engineering}
  (Springer).

\bibitem[{Glasserman et~al.(1999)Glasserman, Heidelberger, Shahabuddin,
  \protect\BIBand{} Zajic}]{glasserman1999multilevel}
Glasserman P, Heidelberger P, Shahabuddin P, Zajic T (1999) Multilevel
  splitting for estimating rare event probabilities. \emph{Operations Research}
  47(4):585--600.

\bibitem[{Glasserman et~al.(2008)Glasserman, Kang, \protect\BIBand{}
  Shahabuddin}]{glasserman2008fast}
Glasserman P, Kang W, Shahabuddin P (2008) Fast simulation of multifactor
  portfolio credit risk. \emph{Operations Research} 56(5):1200--1217.

\bibitem[{Glasserman \protect\BIBand{} Li(2005)}]{glasserman2005importance}
Glasserman P, Li J (2005) Importance sampling for portfolio credit risk.
  \emph{Management Science} 51(11):1643--1656.

\bibitem[{Glasserman \protect\BIBand{}
  Wang(1997)}]{glasserman1997counterexamples}
Glasserman P, Wang Y (1997) {Counterexamples in importance sampling for large
  deviations probabilities}. \emph{Annals of Applied Probability}
  7(3):731--746, ISSN 10505164.

\bibitem[{Glynn \protect\BIBand{} Iglehart(1989)}]{glynn1989importance}
Glynn PW, Iglehart DL (1989) Importance sampling for stochastic simulations.
  \emph{Management Science} 35(11):1367--1392.

\bibitem[{Grace et~al.(2014)Grace, Kroese, \protect\BIBand{}
  Sandmann}]{grace_kroese_sandmann_2014}
Grace AW, Kroese DP, Sandmann W (2014) Automated state-dependent importance
  sampling for markov jump processes via sampling from the zero-variance
  distribution. \emph{Journal of Applied Probability} 51(3):741–755.

\bibitem[{Guyader et~al.(2011)Guyader, Hengartner, \protect\BIBand{}
  Matzner-L{\o}ber}]{guyader2011simulation}
Guyader A, Hengartner N, Matzner-L{\o}ber E (2011) Simulation and estimation of
  extreme quantiles and extreme probabilities. \emph{Applied Mathematics \&
  Optimization} 64(2):171--196.

\bibitem[{Gärtner(1977)}]{Gartner1977}
Gärtner J (1977) On large deviations from the invariant measure. \emph{Theory
  of Probability \& Its Applications} 22(1):24--39.

\bibitem[{Harvey et~al.(2017)Harvey, Liaw, \protect\BIBand{}
  Mehrabian}]{pmlr-v65-harvey17a}
Harvey N, Liaw C, Mehrabian A (2017) Nearly-tight {VC}-dimension bounds for
  piecewise linear neural networks. Kale S, Shamir O, eds., \emph{Proceedings
  of the 2017 Conference on Learning Theory}, volume~65 of \emph{Proceedings of
  Machine Learning Research}, 1064--1068 (Amsterdam, Netherlands: PMLR).

\bibitem[{Heidelberger(1995)}]{Heidelberger95}
Heidelberger P (1995) Fast simulation of rare events in queueing and
  reliability models. \emph{ACM Transactions on Modeling and Computer
  Simulation (TOMACS)} 5:43--85.

\bibitem[{{Huang} et~al.(2018){Huang}, {Lam}, {LeBlanc}, \protect\BIBand{}
  {Zhao}}]{8116682}
{Huang} Z, {Lam} H, {LeBlanc} DJ, {Zhao} D (2018) Accelerated evaluation of
  automated vehicles using piecewise mixture models. \emph{IEEE Transactions on
  Intelligent Transportation Systems} 19(9):2845--2855, ISSN 1524-9050.

\bibitem[{Huang et~al.(2018)Huang, Lam, \protect\BIBand{}
  Zhao}]{huang2018designing}
Huang Z, Lam H, Zhao D (2018) Designing importance samplers to simulate machine
  learning predictors via optimization. \emph{2018 Winter Simulation Conference
  (WSC)}, 1730--1741 (IEEE).

\bibitem[{Hult \protect\BIBand{} Svensson(2012)}]{hult2012importance}
Hult H, Svensson J (2012) On importance sampling with mixtures for random walks
  with heavy tails. \emph{ACM Trans. Model. Comput. Simul.} 22(2), ISSN
  1049-3301.

\bibitem[{Juneja \protect\BIBand{} Shahabuddin(2006)}]{juneja2006rare}
Juneja S, Shahabuddin P (2006) Rare-event simulation techniques: An
  introduction and recent advances. \emph{Handbooks in Operations Research and
  Management Science} 13:291--350.

\bibitem[{Kalra \protect\BIBand{} Paddock(2016)}]{kalra2016driving}
Kalra N, Paddock SM (2016) Driving to safety: How many miles of driving would
  it take to demonstrate autonomous vehicle reliability? \emph{Transportation
  Research Part A: Policy and Practice} 94:182--193.

\bibitem[{Koopman \protect\BIBand{} Wagner(2017)}]{koopman2017autonomous}
Koopman P, Wagner M (2017) Autonomous vehicle safety: An interdisciplinary
  challenge. \emph{IEEE Intelligent Transportation Systems Magazine}
  9(1):90--96.

\bibitem[{Koopman \protect\BIBand{} Wagner(2018)}]{koopman2018toward}
Koopman P, Wagner M (2018) Toward a framework for highly automated vehicle
  safety validation. Technical report, SAE Technical Paper.

\bibitem[{Koren et~al.(2018)Koren, Alsaif, Lee, \protect\BIBand{}
  Kochenderfer}]{koren2018adaptive}
Koren M, Alsaif S, Lee R, Kochenderfer MJ (2018) Adaptive stress testing for
  autonomous vehicles. \emph{2018 IEEE Intelligent Vehicles Symposium (IV)},
  1--7 (IEEE).

\bibitem[{Kroese \protect\BIBand{} Nicola(1999)}]{KN99}
Kroese DP, Nicola VF (1999) Efficient estimation of overflow probabilities in
  queues with breakdowns. \emph{Performance Evaluation} 36-37:471--484.

\bibitem[{L'ecuyer et~al.(2010)L'ecuyer, Blanchet, Tuffin, \protect\BIBand{}
  Glynn}]{l2010asymptotic}
L'ecuyer P, Blanchet JH, Tuffin B, Glynn PW (2010) Asymptotic robustness of
  estimators in rare-event simulation. \emph{ACM Transactions on Modeling and
  Computer Simulation (TOMACS)} 20(1):1--41.

\bibitem[{Legriel et~al.(2010)Legriel, Le~Guernic, Cotton, \protect\BIBand{}
  Maler}]{legriel2010approximating}
Legriel J, Le~Guernic C, Cotton S, Maler O (2010) Approximating the pareto
  front of multi-criteria optimization problems. Esparza J, Majumdar R, eds.,
  \emph{Tools and Algorithms for the Construction and Analysis of Systems},
  69--83 (Berlin, Heidelberg: Springer Berlin Heidelberg).

\bibitem[{Lu et~al.(2017)Lu, Pu, Wang, Hu, \protect\BIBand{}
  Wang}]{NIPS2017_7203}
Lu Z, Pu H, Wang F, Hu Z, Wang L (2017) The expressive power of neural
  networks: A view from the width. Guyon I, Luxburg UV, Bengio S, Wallach H,
  Fergus R, Vishwanathan S, Garnett R, eds., \emph{Advances in Neural
  Information Processing Systems 30}, 6231--6239 (Curran Associates, Inc.).

\bibitem[{M\"{u}ller et~al.(2019)M\"{u}ller, Mcwilliams, Rousselle, Gross,
  \protect\BIBand{} Nov\'{a}k}]{muller2019neural}
M\"{u}ller T, Mcwilliams B, Rousselle F, Gross M, Nov\'{a}k J (2019) Neural
  importance sampling. \emph{ACM Trans. Graph.} 38(5), ISSN 0730-0301.

\bibitem[{Murthy et~al.(2014)Murthy, Juneja, \protect\BIBand{}
  Blanchet}]{murthy2014state}
Murthy KRA, Juneja S, Blanchet J (2014) {State-independent importance sampling
  for random walks with regularly varying increments}. \emph{Stochastic
  Systems} 4(2):321 -- 374.

\bibitem[{NHTSA(2007)}]{national2007new}
NHTSA (2007) The new car assessment program suggested approaches for future
  program enhancements. \emph{DOT HS} 810:698.

\bibitem[{Nicola et~al.(1993)Nicola, Nakayama, Heidelberger, \protect\BIBand{}
  Goyal}]{nicola1993fast}
Nicola VF, Nakayama MK, Heidelberger P, Goyal A (1993) Fast simulation of
  highly dependable systems with general failure and repair processes.
  \emph{IEEE Transactions on Computers} 42(12):1440--1452.

\bibitem[{Nicola et~al.(2001)Nicola, Shahabuddin, \protect\BIBand{}
  Nakayama}]{nicola2001techniques}
Nicola VF, Shahabuddin P, Nakayama MK (2001) Techniques for fast simulation of
  models of highly dependable systems. \emph{IEEE Transactions on Reliability}
  50(3):246--264.

\bibitem[{NTSB(2016)}]{PreliminaryHWY16FH018}
NTSB (2016) {Preliminary Report, Highway HWY16FH018}.
  \urlprefix\url{http://www.ntsb.gov/investigations/AccidentReports/Pages/HWY16FH018-preliminary.aspx}.

\bibitem[{Nyquist(2017)}]{nyquist2017moderate}
Nyquist P (2017) Moderate deviation principles for importance sampling
  estimators of risk measures. \emph{Journal of Applied Probability}
  54(2):490--506.

\bibitem[{O'Kelly et~al.(2018)O'Kelly, Sinha, Namkoong, Tedrake,
  \protect\BIBand{} Duchi}]{o2018scalable}
O'Kelly M, Sinha A, Namkoong H, Tedrake R, Duchi JC (2018) Scalable end-to-end
  autonomous vehicle testing via rare-event simulation. \emph{Advances in
  Neural Information Processing Systems}, 9827--9838.

\bibitem[{{Orzechowski} et~al.(2019){Orzechowski}, {Li}, \protect\BIBand{}
  {Lauer}}]{rssidm}
{Orzechowski} PF, {Li} K, {Lauer} M (2019) Towards responsibility-sensitive
  safety of automated vehicles with reachable set analysis. \emph{2019 IEEE
  International Conference on Connected Vehicles and Expo (ICCVE)}, 1--6.

\bibitem[{Ridder(2009)}]{Ridder09}
Ridder A (2009) Importance sampling algorithms for first passage time
  probabilities in the infinite server queue. \emph{European Journal of
  Operational Research} 199:176--186.

\bibitem[{Rockafellar(1970)}]{rockafellar1970convex}
Rockafellar RT (1970) \emph{{Convex Analysis}} (Princeton University Press).

\bibitem[{Rubino \protect\BIBand{} Tuffin(2009)}]{RT09}
Rubino G, Tuffin B (2009) Markovian models for dependability analysis.
  \emph{Rare Event Simulation Using Monte Carlo Methods}, 125--144, chapter 6.

\bibitem[{Rubinstein(2005)}]{rubinstein2005stochastic}
Rubinstein RY (2005) A stochastic minimum cross-entropy method for
  combinatorial optimization and rare-event estimation. \emph{Methodology and
  Computing in Applied Probability} 7(1):5--50.

\bibitem[{Rubinstein \protect\BIBand{} Kroese(2013)}]{rubinstein2013cross}
Rubinstein RY, Kroese DP (2013) \emph{The Cross-entropy Method: A Unified
  Approach to Combinatorial Optimization, Monte-Carlo Simulation and Machine
  Learning} (Springer Science \& Business Media).

\bibitem[{Sadowsky(1991)}]{Sadowsky91}
Sadowsky JS (1991) Large deviations theory and efficient simulation of
  excessive backlogs in a {$GI/GI/m$} queue. \emph{IEEE Transactions on
  Automatic Control} 36:1383--1394.

\bibitem[{Sadowsky \protect\BIBand{} Bucklew(1990)}]{sadowsky1990large}
Sadowsky JS, Bucklew JA (1990) On large deviations theory and asymptotically
  efficient monte carlo estimation. \emph{IEEE Transactions on Information
  Theory} 36(3):579--588.

\bibitem[{Siegmund(1976)}]{siegmund1976importance}
Siegmund D (1976) {Importance sampling in the Monte Carlo study of sequential
  tests}. \emph{The Annals of Statistics} 4(4):673 -- 684.

\bibitem[{Szechtman \protect\BIBand{} Glynn(2002)}]{SP02}
Szechtman R, Glynn P (2002) Rare event simulation for infinite server queues.
  \emph{Proceedings of the 2002 Winter Simulation Conference}, 416--423.

\bibitem[{Tjeng et~al.(2017)Tjeng, Xiao, \protect\BIBand{}
  Tedrake}]{tjeng2017evaluating}
Tjeng V, Xiao K, Tedrake R (2017) Evaluating robustness of neural networks with
  mixed integer programming. \emph{arXiv preprint arXiv:1711.07356} .

\bibitem[{Treiber et~al.(2000)Treiber, Hennecke, \protect\BIBand{}
  Helbing}]{Treiber_2000}
Treiber M, Hennecke A, Helbing D (2000) Congested traffic states in empirical
  observations and microscopic simulations. \emph{Physical Review E}
  62(2):1805–1824, ISSN 1095-3787.

\bibitem[{Tuffin(2004)}]{Tuffin04QEST}
Tuffin B (2004) On numerical problems in simulation of highly reliable
  {M}arkovian systems. \emph{Proceedings of the 1st International Conference on
  Quantitative Evaluation of SysTems (QEST)}, 156--164 (IEEE Computer Society
  Press).

\bibitem[{Tuffin \protect\BIBand{} Ridder(2012)}]{tuffin2012probabilistic}
Tuffin B, Ridder A (2012) Probabilistic bounded relative error for rare event
  simulation learning techniques. \emph{Proceedings of the 2012 Winter
  Simulation Conference (WSC)}, 1--12 (IEEE).

\bibitem[{Uesato et~al.(2018)Uesato, Kumar, Szepesvari, Erez, Ruderman,
  Anderson, Heess, \protect\BIBand{} Kohli}]{uesato2018rigorous}
Uesato J, Kumar A, Szepesvari C, Erez T, Ruderman A, Anderson K, Heess N, Kohli
  P (2018) Rigorous agent evaluation: An adversarial approach to uncover
  catastrophic failures. \emph{arXiv preprint arXiv:1812.01647} .

\bibitem[{van~der Vaart \protect\BIBand{} Wellner(1996)}]{WeakConvergence_VdV}
van~der Vaart AW, Wellner JA (1996) Weak convergence and empirical processes.
  \emph{Springer Series in Statistics} .

\bibitem[{Vill{\'e}n-Altamirano \protect\BIBand{}
  Vill{\'e}n-Altamirano(1994)}]{villen1994restart}
Vill{\'e}n-Altamirano M, Vill{\'e}n-Altamirano J (1994) Restart: a
  straightforward method for fast simulation of rare events. \emph{Proceedings
  of Winter Simulation Conference}, 282--289 (IEEE).

\bibitem[{Villén-Altamirano(2010)}]{VILLENALTAMIRANO2010156}
Villén-Altamirano J (2010) Importance functions for restart simulation of
  general jackson networks. \emph{European Journal of Operational Research}
  203(1):156--165, ISSN 0377-2217.

\bibitem[{{Wang} et~al.(2018){Wang}, {Jiang}, {Li}, {Lin}, {Zheng},
  \protect\BIBand{} {Wang}}]{dlforcf}
{Wang} X, {Jiang} R, {Li} L, {Lin} Y, {Zheng} X, {Wang} F (2018) Capturing
  car-following behaviors by deep learning. \emph{IEEE Transactions on
  Intelligent Transportation Systems} 19(3):910--920.

\bibitem[{Webb et~al.(2018)Webb, Rainforth, Teh, \protect\BIBand{}
  Kumar}]{webb2018statistical}
Webb S, Rainforth T, Teh YW, Kumar MP (2018) A statistical approach to
  assessing neural network robustness. \emph{arXiv preprint arXiv:1811.07209} .

\bibitem[{Wegener \protect\BIBand{} B{\"u}hler(2004)}]{wegener2004evaluation}
Wegener J, B{\"u}hler O (2004) Evaluation of different fitness functions for
  the evolutionary testing of an autonomous parking system. \emph{Genetic and
  Evolutionary Computation Conference}, 1400--1412 (Springer).

\bibitem[{Wu et~al.(2018)Wu, Gomes-Selman, Shi, Xue, Garc{\'i}a-Villacorta,
  Anderson, Sethi, Steinschneider, Flecker, \protect\BIBand{}
  Gomes}]{Wu2018EfficientlyAT}
Wu X, Gomes-Selman J, Shi Q, Xue Y, Garc{\'i}a-Villacorta R, Anderson E, Sethi
  S, Steinschneider S, Flecker A, Gomes CP (2018) Efficiently approximating the
  pareto frontier: Hydropower dam placement in the amazon basin. \emph{AAAI}.

\bibitem[{Zhao et~al.(2017)Zhao, Huang, Peng, Lam, \protect\BIBand{}
  LeBlanc}]{zhao2017accelerated}
Zhao D, Huang X, Peng H, Lam H, LeBlanc DJ (2017) Accelerated evaluation of
  automated vehicles in car-following maneuvers. \emph{IEEE Transactions on
  Intelligent Transportation Systems} 19(3):733--744.

\bibitem[{Zhao et~al.(2016)Zhao, Lam, Peng, Bao, LeBlanc, Nobukawa,
  \protect\BIBand{} Pan}]{zhao2016accelerated}
Zhao D, Lam H, Peng H, Bao S, LeBlanc DJ, Nobukawa K, Pan CS (2016) Accelerated
  evaluation of automated vehicles safety in lane-change scenarios based on
  importance sampling techniques. \emph{IEEE Transactions on Intelligent
  Transportation Systems} 18(3):595--607.

\bibitem[{Zhao \protect\BIBand{} Peng(2018)}]{zhao2018lab}
Zhao D, Peng H (2018) From the lab to the street: Solving the challenge of
  accelerating automated vehicle testing. \emph{Hitachi Review} 67(1):014--015.

\end{thebibliography}

\ECSwitch


\ECHead{Appendix}

\section{Examples of Applying Theorem \ref{general IS}}
\label{app:examples}

\begin{example}[One-dimensional Gamma distribution]
Suppose $X\sim Gamma(\alpha,\beta)$ where $\alpha,\beta>0$. The density function is
$$
f(x)=\frac{\beta^{\alpha}}{\Gamma(\alpha)}x^{\alpha-1}e^{-\beta x},x>0.
$$
We have that 
$$
\lambda(s)=
\begin{cases}
-\alpha\log(1-s/\beta)&s<\beta;\\
\infty &s\geq\beta.
\end{cases}
$$ 
In the one dimensional case, we consider $\mathcal{S}_{\gamma}=[a^*,\infty)$ where $a^*>\alpha/\beta$ such that $\mathcal S_{\gamma}$ is orthogonally monotone and $0<I(a^*)<\infty$. We have that $s^*=\beta-\alpha/a^*$. The dominating set is $\{a^*\}$ by definition and the IS distribution is defined by density function $f^*(x)=\frac{1}{\Gamma(\alpha)}(\alpha/a^*)^{\alpha}x^{\alpha-1}e^{-\frac{\alpha}{a^*}x}$. That is, the IS distribution is exactly $Gamma(\alpha,\alpha/a^*)$, where we move the mean value from $\alpha/\beta$ to $a^*$. If $a^*\leq x\leq a^*+\varepsilon$, then $f^*(x)\geq f^*(a^*+\varepsilon)=\frac{1}{\Gamma(\alpha)}(\alpha/a^*)^{\alpha}(a^*+\varepsilon)^{\alpha-1}e^{-\alpha\frac{a^*+\varepsilon}{a^*}}$. If $a^*\to\infty$ at most polynomially in $\gamma$, then this lower bound is also polynomial in $\gamma$. We could easily verify that Assumption \ref{asm:lambda} and $\ref{asm:S}$ hold and hence the efficiency certificate is achieved.
\end{example}

\begin{example}[Multivariate Gaussian distribution]
Suppose that $X\sim N(m,\Sigma)$ where $m\in\mathbb{R}^d$ and $\Sigma\in\mathbb{R}^{d\times d}$ is positive definite. We have that $\lambda(s)=m^T s+\frac{1}{2}s^T\Sigma s$ and $I(y)=\frac{1}{2}(y-m)^T\Sigma^{-1}(y-m)$. Then $a^*=\arg\min_{y\in\mathcal{S}_{\gamma}}(y-m)^T\Sigma^{-1}(y-m)$. We assume that $\mathcal S_{\gamma}$ is orthogonally monotone and also $m\notin \mathcal S_{\gamma}$ such that $I(a^*)>0$. We have that $s^*=\Sigma^{-1}(a^*-m)$. Then $f^*(x)=(2\pi)^{-\frac{d}{2}}|\Sigma|^{-\frac{1}{2}}\exp\{-\frac{1}{2}(x-a^*)^T\Sigma^{-1}(x-a^*)\}$, which is the density function of $N(a^*,\Sigma)$. Moreover, if $a^*\leq x\leq a^*+\varepsilon\mathbf{1}$, then we can find a constant $\varepsilon'>0$ such that $f^*(x)>\varepsilon'$. Thus, as long as the components of $s^*$ at most grow polynomially in $\gamma$, the efficiency certificate is guaranteed. 
\label{eg:gaussian}
\end{example}

\begin{example}[Gaussian mixture distribution]
Suppose that $X\sim \sum_{i=1}^k \pi_iN(m_i,\Sigma_i)$ where $m_i\in\mathbb R^d$, $\Sigma_i\in\mathbb R^{d\times d}$ is positive definite, $\pi_i>0$ and $\sum_{i=1}^k\pi_i=1$. Similarly, we assume that $\mathcal S_{\gamma}$ is orthogonally monotone and $m_i\notin\mathcal S_{\gamma},\forall i$. Suppose that $a_{ij}$'s are the dominating points for $\mathcal S_{\gamma}$ associated with the distribution $N(m_i,\Sigma_i)$. Denote $a^{(i)}:=\arg\min_{a_{ij}}(a_{ij}-m_i)^T\Sigma_i^{-1}(a_{ij}-m_i)$ and $s^{(i)}:=s_{a^{(i)}}$. We also assume that the components of $s^{(i)}$ at most grow polynomially in $\gamma$. From Example \ref{eg:gaussian}, we know that if the input distribution is $N(m_i,\Sigma_i)$, then the IS distribution $\sum_j\alpha_{ij}N(a_{ij},\Sigma_i)$ achieves the efficiency certificate for any $\alpha_{ij}$'s such that $\alpha_{ij}>0,\forall j$ and $\sum_j\alpha_{ij}=1$. Now we consider the input distribution $p=\sum_{i=1}^k \pi_iN(m_i,\Sigma_i)$. We will show that the IS distribution $\tilde{p}=\sum_{i=1}^k\pi_i\sum_j\alpha_{ij}N(a_{ij},\Sigma_i)$ achieves the efficiency certificate. First, we know that the likelihood ratio function is 
$$
\begin{aligned}
L(x)&=\frac{\sum_{i=1}^k\pi_i(2\pi)^{-d/2}|\Sigma_i|^{-1/2}\exp\{-(x-m_i)^T\Sigma_i^{-1}(x-m_i)/2\}}{\sum_{i=1}^k\pi_i\sum_j\alpha_{ij}(2\pi)^{-d/2}|\Sigma_i|^{-1/2}\exp\{-(x-a_{ij})^T\Sigma_i^{-1}(x-a_{ij})/2\}}\\
&\leq \sum_{i=1}^k\frac{\exp\{-(x-m_i)^T\Sigma_i^{-1}(x-m_i)/2\}}{\sum_j\alpha_{ij}\exp\{-(x-a_{ij})^T\Sigma_i^{-1}(x-a_{ij})/2\}}
\end{aligned}
$$
and hence 
$$
L^2(x)\leq k\sum_{i=1}^k\left(\frac{\exp\{-(x-m_i)^T\Sigma_i^{-1}(x-m_i)/2\}}{\sum_j\alpha_{ij}\exp\{-(x-a_{ij})^T\Sigma_i^{-1}(x-a_{ij})/2\}}\right)^2.
$$
Note that $\frac{\exp\{-(x-m_i)^T\Sigma_i^{-1}(x-m_i)/2\}}{\sum_j\alpha_{ij}\exp\{-(x-a_{ij})^T\Sigma_i^{-1}(x-a_{ij})/2\}}$ is the likelihood ratio function between $N(m_i,\Sigma_i)$ and $\sum_j\alpha_{ij}N(a_{ij},\Sigma_i)$. Following the proof of Theorem \ref{general IS}, we get that 
$$
\begin{aligned}
&\tilde{E}\left[I(X\in\mathcal S_{\gamma})\left(\frac{\exp\{-(X-m_i)^T\Sigma_i^{-1}(X-m_i)/2\}}{\sum_j\alpha_{ij}\exp\{-(X-a_{ij})^T\Sigma_i^{-1}(X-a_{ij})/2\}}\right)^2\right]\\
\leq &\left(\sum_j\frac{1}{\alpha_{ij}^2}\right)\exp\{-(a^{(i)}-m_i)^T\Sigma_i^{-1}(a^{(i)}-m_i)\}.
\end{aligned}
$$
Thus we have that 
$$
\tilde{E}[I(X\in\mathcal S_{\gamma})L^2(X)]\leq k\left(\sum_i\sum_j\frac{1}{\alpha_{ij}^2}\right)\exp\{-\min_i(a^{(i)}-m_i)^T\Sigma_i^{-1}(a^{(i)}-m_i)\}.
$$
Moreover, we know that (also see the proof for details)
$$
\begin{aligned}
\tilde{E}[I(X\in\mathcal S_{\gamma})L(X)]&=P(X\in\mathcal S_{\gamma})\\
&=\sum_{i=1}^k\pi_iP(N(m_i,\Sigma_i)\in\mathcal S_{\gamma})\\
&\sim \exp\{-\min_i(a^{(i)}-m_i)^T\Sigma_i^{-1}(a^{(i)}-m_i)/2\}
\end{aligned}
$$
where $\sim$ is up to polynomial factor in $\gamma$. Combining the results, we get that this IS distribution achieves the efficiency certificate.
\end{example}

\begin{example}[Multivariate Laplace distribution]
Suppose that $X=\sqrt{W}Y$ where $Y\sim N(0,I_d)$ and $W\sim Exp(1)$ is independent of $Y$. It is known that the density function is 
$$
f(x)=\frac{2}{(2\pi)^{d/2}}\left(\frac{x^Tx}{2}\right)^{\nu/2}K_{\nu}(\sqrt{2x^Tx})
$$
where $\nu=(2-d)/2$ and $K_{\nu}$ is the modified Bessel function of the second kind. We have that $$
\lambda(s)=
\begin{cases}
-\log(1-\frac12 s^Ts) & s^Ts<2;\\
\infty & s^Ts\geq 2.
\end{cases}
$$
We assume that $\mathcal{S}_{\gamma}$ is orthogonally monotone and $0\notin \mathcal S_{\gamma}$. By solving $\nabla\lambda(s^*)=a^*$, we get that 
$$
s^*=\frac{\sqrt{2a^{*T}a^*+1}-1}{a^{*T}a^*}a^*.
$$
Then 
$$
\begin{aligned}
f^*(x)&=f(x)\exp\{s^{*T}x-\lambda(s^*)\}\\
&=\frac{2}{(2\pi)^{d/2}}\left(\frac{x^Tx}{2}\right)^{\nu/2}K_{\nu}(\sqrt{2x^Tx})\exp\left\{\frac{\sqrt{2a^{*T}a^*+1}-1}{a^{*T}a^*}a^{*T}x\right\}\frac{\sqrt{2a^{*T}a^*+1}-1}{a^{*T}a^*}.
\end{aligned}
$$
We assume that $a_i^*\to\infty$ polynomially in $\gamma$ for any $i$. For $a^*\leq x\leq a^*+\varepsilon\mathbf{1}$, we have that $a^{*T}a^*\leq x^Tx\leq (a^*+\varepsilon\mathbf{1})^T(a^*+\varepsilon\mathbf{1})$ and $a^{*T}x\geq a^{*T}a^*$. Thus, we have that 
$$
\left(\frac{x^Tx}{2}\right)^{\nu/2}\geq 
\begin{cases}
\left(\frac{a^{*T}a^*}{2}\right)^{\nu/2} & d=1\\
\left(\frac{(a^*+\varepsilon\mathbf{1})^T(a^*+\varepsilon\mathbf{1})}{2}\right)^{\nu/2} &d\geq 2
\end{cases}
$$
and
$$
\begin{aligned}
&K_{\nu}(\sqrt{2x^Tx})\exp\left\{\frac{\sqrt{2a^{*T}a^*+1}-1}{a^{*T}a^*}a^{*T}x\right\}\\
\geq &K_{\nu}(\sqrt{2(a^*+\varepsilon\mathbf{1})^T(a^*+\varepsilon\mathbf{1})})\exp\left\{\frac{\sqrt{2a^{*T}a^*+1}-1}{a^{*T}a^*}a^{*T}a^*\right\}\\
\approx &\frac{1}{(2(a^*+\varepsilon\mathbf{1})^T(a^*+\varepsilon\mathbf{1}))^{1/4}}\exp\left\{\sqrt{2a^{*T}a^*+1}-1-\sqrt{2(a^*+\varepsilon\mathbf{1})^T(a^*+\varepsilon\mathbf{1})} \right\}.
\end{aligned}
$$
Here we use the fact that $K_{\nu}(x)\approx e^{-x}/\sqrt{x}$ as $x\to\infty$ where $\approx$ is up to constant factor. Since we have assumed that $a_i^*$ is polynomial in $\gamma$, we only need to show that $\exp\left\{\sqrt{2a^{*T}a^*+1}-1-\sqrt{2(a^*+\varepsilon\mathbf{1})^T(a^*+\varepsilon\mathbf{1})} \right\}$ does not decay exponentially fast in $\gamma$. In fact, we have that 
$$
\begin{aligned}
\sqrt{2a^{*T}a^*+1}-\sqrt{2(a^*+\varepsilon\mathbf{1})^T(a^*+\varepsilon\mathbf{1})}
&=\frac{1-4d\varepsilon^2-4\varepsilon a^{*T}\mathbf{1}}{\sqrt{2a^{*T}a^*+1}+\sqrt{2(a^*+\varepsilon\mathbf{1})^T(a^*+\varepsilon\mathbf{1})}}\\
&\geq \frac{1-4d\varepsilon^2-4\varepsilon \sqrt{d}\sqrt{a^{*T}a^*}}{\sqrt{2a^{*T}a^*+1}+\sqrt{2(a^*+\varepsilon\mathbf{1})^T(a^*+\varepsilon\mathbf{1})}}
\end{aligned}
$$
and the lower bound converges to a constant. Moreover, it is implied that the components of $s^*$ do not grow exponentially fast in $\gamma$. Therefore, all the required assumptions are satisfied.
\end{example}

\section{Conservativeness of Standard ERM and Lazy Learner}\label{app: conservativeness}

As in the setup of Theorem \ref{thm: set_ERM_modified}, suppose that Stage 1 samples are drawn from $q$. Suppose we use empirical risk minimization (ERM) to train $\hat g$, i.e., $\hat g:=\text{argmin}_{g\in\mathcal{G}}\{ {R}_{n_1}(g):=\frac{1}{n_1}\sum_{i=1}^{n_1}\ell(g(\tilde X_{i}),Y_{i})\}$
where $\ell$ is a loss function and $\mathcal G$ is the considered hypothesis class. Correspondingly, let $R(g):=E_{X\sim q}\ell(g(X),I(X\in\mathcal S_\gamma))$
be the true risk function and $g^{*}:=\arg\min_{g\in\mathcal{G}}R(g)$ its minimizer.
Also let $\kappa^{*}:=\min_{x\in\mathcal{S}_{\gamma}}g^{*}(x)$ be
the true threshold associated with $g^{*}$ in obtaining the smallest outer rare-event set approximation. Then we have the following result.

\begin{theorem}[Conservativeness of ERM-generated set approximation]\label{thm: set_ERM} 
Consider $\overline{\mathcal S}_\gamma^{\hat\kappa}$ obtained in Algorithm \ref{algo:stage1} where $\hat g$ is trained from an ERM. 
Suppose  the density $q$ has bounded 
support $K\subset[0,M]^{d}$ and $0<q_{l}\leq q(x)\leq q_{u}$ for any $x\in K$. Also suppose there exists
a function $h$ such that for any $g\in\mathcal{G}$, $g(x)\geq\kappa$ implies $\ell(g(x),0)\geq h(\kappa)>0$.
(e.g., if $\ell$ is the squared loss, then $h(\kappa)$ could be chosen 
as $h(\kappa)=\kappa^{2}$). Then, with probability at least $1-\delta$,
\begin{align*}
P_{X\sim q} &\left(X\in\bar{\mathcal{S}}_{\gamma}^{\hat{\kappa}}\setminus\mathcal{S}_{\gamma}\right)
\leq \frac{R(g^{*})+2\sup_{g\in\mathcal{G}}\left|R_{n_1}(g)-R(g)\right|}{h(\kappa^{*}-t(\delta,n_1)\sqrt{d} \text{Lip}(g^{*})-\left\Vert \hat{g}-g^{*}\right\Vert _{\infty})}.
\end{align*}
Here, $\text{Lip}(g^{*})$ is the Lipschitz parameter of 
$g^{*}$, and $t(\delta,n_1)=3\left(\frac{\log(n_1q_{l})+d\log M+\log\frac{1}{\delta}}{n_1q_{l}}\right)^{\frac{1}{d}}$.
\end{theorem}

A question is how to give a more refined bound for the false positive rate based on Theorem \ref{thm: set_ERM}, and Theorem \ref{thm: set_ERM_modified}, that depends on explicit constants of the classification model or training process. This would involve theoretical results for deep neural networks that are under active research. Let us examine the terms appearing in
Theorem \ref{thm: set_ERM} and give some related results. In machine learning theory, the term $\sup_{g_{\theta}\in\mathcal{G}}\left|R_{n_1}(g_{\theta})-R(g_{\theta})\right|$
is often bounded by the Rademacher complexity of the function class
(some results about the Rademacher complexity for neural networks are in \citealt{pmlr-v65-harvey17a,NIPS2019_9246}).
The convergence of $\left\Vert \hat{g}-g^{*}\right\Vert _{\infty}$
to 0 as $n_1\rightarrow\infty$ is implied by the convergence of the
parameters, which is in turn justified by
the empirical process theory \citep{WeakConvergence_VdV}. A bound for
$\text{Lip}(g^{*})$ could be potentially derived by adding norm constraints
to the parameters in the neural network \citep{pmlr-v97-anil19a}. On the other hand, if we let the network size
grow to infinity, the class of neural networks can approximate any continuous
function \citep{NIPS2017_7203}, and hence $R(g^{*})$ can be arbitrarily
small when the neural network is complex enough.  
However, if we restrict the choices of networks, for instance by the Lipschitz constant, then no results regarding the sufficiency of its expressive power for arbitrary functions are available in the literature to our knowledge, and thus it appears open how to simultaneously give bounds for $\text{Lip}(g^{*})$ and  $R(g^{*})$. Future investigations on the expressive power of restricted classes of neural networks would help refining our conservativeness results further.

We provide a corresponding result to quantify the conservativeness of the lazy-learner IS in terms of the false positive rate. Recall that the lazy learner constructs the outer approximation of the rare-event set using $\mathcal{H}({T_0})^c$, which is the complement of the orthogonal monotone hull of the set of all non-rare-event samples. The conservativeness is measured concretely by the set difference between $\mathcal{H}({T_0})^c$ and $\mathcal{S}_{\gamma}$, for which we have the following result:
\begin{theorem}[Conservativeness of lazy learner]
\label{thm: set_OMhull}Suppose that the density $q$ has bounded
support $K\subset[0,M]^{d}$, and $0<q_{l}\leq q(x)\leq q_{u}$ for any $x\in K$. 
Then, with probability at least $1-\delta$,
\begin{align*}
P_{X\sim q}(X\in\mathcal{H}(T_{0})^{c}\backslash\mathcal{S}_{\gamma})&\leq M^{d-1}q_{u}\left(\frac{\sqrt{d}}{2}\right)^{d-1}w_{d-1}t(\delta,n_1)\\
&=\sqrt{\frac{e}{\pi(d-1)}}\left(\frac{1}{2}\pi e\right)^{\frac{d-1}{2}}q_{u}t(\delta,n_1)(1+O(d^{-1})).
\end{align*}
Here  $t(\delta,n_1)=3\left(\frac{\log(n_1q_{l})+d\log{M}+\log\frac{1}{\delta}}{n_1q_{l}}\right)^{\frac{1}{d}}$,
$w_{d}$ is the volume of a $d-$dimensional Euclidean ball of radius
1, and the last $O(\cdot)$ is as $d$ increases. 

\end{theorem}

Theorems \ref{thm: set_ERM} and \ref{thm: set_OMhull} are stated with respect to $q$, the sampling distribution used in the first stage. We explain how to translate the false positive rate results to under the original distribution $p$. In the discussion below, we will consider Theorem \ref{thm: set_ERM} (and Theorem \ref{thm: set_OMhull} can be handled similarly). In this case, our target is to give an upper bound to $P_{X\sim p}(X\in\mathcal{S}_{\gamma}^{\hat{\kappa}}\backslash\mathcal{S}_{\gamma})$ based on the result of Theorem \ref{thm: set_ERM}.

If the true input distribution $p$ does not have a bounded
support, we can first choose $M$ to be large to make sure that $P_{X\sim p}(X\notin[0,M]^{d})$ is small compared to the probability of $\mathcal S_\gamma$. We argue that we do not need $M$ to be too large here. Indeed, if $p$ is light tail (e.g., a distribution with tail probability exponential in $M$), then the required $M$ grows at most polynomially
in $\gamma$. 

Having selected $M$, and with the freedom in selecting $q$ in Stage 1, we could make sure
that in $[0,M]^d $,  $q(x)$ is bounded away from 0 (e.g., we can choose $q$ to be the uniform distribution over $[0,M]^d$). Then, by Theorem \ref{thm: set_ERM} and a change of measure argument, we can give a bound for $P_{X\sim p}(X\in{[0,M]^d},X\in\mathcal{S}_{\gamma}^{\hat{\kappa}}\backslash\mathcal{S}_{\gamma})$.
Finally, we bound the false positive rate with respect to $p$ by  $P_{X\sim p}(X\in\mathcal{H}(T_{0})^{c}\backslash\mathcal{S}_{\gamma})\leq P_{X\sim p}(X\notin[0,M]^{d})+P_{X\sim p}(X\in{[0,M]^d},X\in\mathcal{H}(T_{0})^{c}\backslash\mathcal{S}_{\gamma})$.

\section{Lower-Bound Efficiency Certificate and Estimators}
\label{app:lowerbound}

In Section \ref{sec:Deep-PrAE}, we described an approach that gives an estimator for the rare-event probability with an upper-bound relaxed efficiency certificate. Here we present analogous definitions and results on the lower-bound relaxed efficiency certificate. This lower-bound estimator gives an estimation gap for the upper-bound estimator. Moreover, by combining both of them, we can obtain an interval for the target rare-event probability. 

The lower-bound relaxed efficiency certificate is defined as follows (compare with Definition \ref{relaxed certificate}):

\begin{definition}
We say an estimator $\hat\mu_n$ satisfies an lower-bound \emph{relaxed efficiency certificate} to estimate $\mu$ if
$P(\hat\mu_n-\mu>\epsilon\mu)\leq\delta$
with $n\geq\tilde O(\log(1/\mu))$, for given $0<\epsilon,\delta<1$. \label{def: LB_relaxed_certificate}
\end{definition}

This definition requires that, with high probability, $\hat{\mu}_n$ is a lower bound of $\mu$ up to an error of $\epsilon\mu$. We have the following analog to Proposition \ref{certificate prop simple}:

\begin{corollary}

Suppose $\hat\mu_n$ is downward biased, i.e., $\overline\mu:=E[\hat\mu_n]\leq\mu$. Moreover, suppose $\hat\mu_n$ takes the form of an average of $n$ i.i.d. simulation runs $Z_i$, with $RE=Var(Z_i)/\overline\mu^2=\tilde O(\log(1/\overline\mu))$. Then $\hat\mu_n$ possesses the lower-bound relaxed efficiency certificate. \label{certificate prop simple_LB}
\end{corollary}

This motivates us to learn an inner approximation of the rare-event set in Stage 1 and then in Stage 2, we use IS as in Theorem \ref{general IS} to estimate the probability of this inner approximation set. For the inner set approximation, like the outer approximation case, we use our Stage 1 samples $\{(\tilde X_i,  Y_i) \}_{i=1,\ldots,n_1}$ to construct an approximation set $\overline{\mathcal{S}}_{\gamma}$ that has zero false positive rate, i.e., 
\begin{equation}
P(X\in\overline{\mathcal{S}}_{\gamma},Y=0)=0.\label{eq: zero_FN}
\end{equation}
To make sure of (\ref{eq: zero_FN}), we again exploit the knowledge that the rare event set $\mathcal{S}_{\gamma}$ is orthogonally monotone.  Indeed, denote $T_1:=\{\tilde X_i : Y_i=1\}$ as the rare-event sampled points and for each point $x\in\mathbb{R}^d_+$, let $\mathcal Q(x):=\{x^{\prime} : x^{\prime}\geq x\}$. We construct $\mathcal J(T_1):=\cup_{x\in T_1}{\mathcal Q(x)}$ which serves as the ``upper orthogonal monotone hull" of $T_1$. The orthogonal monotonicity property of $\mathcal{S}_{\gamma}$ implies that $\mathcal J(T_1)\subset \mathcal{S}_{\gamma}$. Moreover, $\mathcal J(T_1)$ is the largest choice of $\overline{\mathcal{S}}_{\gamma}$ such that \eqref{eq: zero_FN} is guaranteed. Based on this observation, in parallel to Section \ref{sec:Deep-PrAE}, depending on how we construct the inner approximation to the rare-event set, we propose the following two approaches.

\textbf{Lazy-Learner IS (Lower Bound). }We now consider an estimator for $\mu$ where in Stage 1, we sample a constant $n_1$ i.i.d. random points from some density, say $q$. Then, we use the mixture IS depicted in Theorem 1 to estimate $P(X\in\mathcal J(T_1))$ in Stage 2. Since $\mathcal J(T_1)$ takes the form $\cup_{x\in T_1}{\mathcal Q(x)}$, it has a finite number of dominating points, which can be found by a sequential algorithm. But as explained in Section \ref{sec:Deep-PrAE}, this leads to a large number of mixture components that degrades the IS efficiency.

\textbf{Deep-Learning-Based IS (Lower Bound). }We train a neural network classifier, say $\hat g$, using all the Stage 1 samples $\{(\tilde X_i,Y_i)\}$, and obtain an approximate rare-event region $\overline{\mathcal S}_\gamma^\kappa=\{x:\hat g(x)\geq\kappa\}$, where $\kappa$ is say $1/2$. Then we adjust $\kappa$ minimally away from $1/2$, say to $\hat\kappa$, so that $\overline{\mathcal  S}_\gamma^{\hat\kappa}\subset\mathcal J(T_1)$, i.e., $\hat\kappa=\min\{\kappa\in\mathbb R: \overline{\mathcal  S}_\gamma^{\hat\kappa}\subset\mathcal J(T_1)\}$. Then $\overline{\mathcal  S}_\gamma^{\hat\kappa}$ is an inner approximation for $\mathcal S_\gamma$ (see Figure \ref{fig:illustration}(c), where $\hat\kappa=0.83$). Stage 1 in Algorithm \ref{algo: Deep-PrAE-LB} shows this procedure. With this, we can run mixture IS to estimate $P(X\in\overline{\mathcal  S}_\gamma^{\hat\kappa})$ in Stage 2.

\begin{algorithm}[htbp]
\KwIn{Black-box evaluator $I(\cdot\in\mathcal S_\gamma)$, initial Stage 1 samples $\{(\tilde X_i, Y_i) \}_{i=1,\ldots,n_1}$, Stage 2 sampling budget $n_2$, input density function $f(x)$.}
\KwOut{IS estimate $\hat\mu_n$.}

\nl \textbf{Stage 1 (Set Learning):}\\

\nl Train classifier with positive decision region $\overline{\mathcal  S}_\gamma^{\kappa}=\{x:\hat{g}(x) \geq\kappa\}$ using $\{(\tilde X_i, Y_i) \}_{i=1,\ldots,n_1}$;\\
\nl Replace $\kappa$ by $\hat\kappa=\min\{\kappa\in\mathbb R: \overline{\mathcal  S}_\gamma^{\hat\kappa}\subset \mathcal{J}(T_1)\}$;\\

\nl \textbf{Stage 2 (Mixture IS based on Searched dominating points):}\\
\nl The same as Stage 2 of Algorithm \ref{algo:stage1}.

    \caption{{\bf Deep-PrAE to estimate $\mu=P(X\in\mathcal S_\gamma)$ (lower bound).} 
    \label{algo: Deep-PrAE-LB}}
\end{algorithm}

As we can see, compared with Algorithm \ref{algo:stage1}, the only difference is how we adjust $\kappa$ in Stage 1. And similar to Theorem \ref{NN main}, we also have that Algorithm \ref{algo: Deep-PrAE-LB} attains the lower-bound relaxed efficiency certificate:

\begin{theorem}[Lower-bound relaxed efficiency certificate
]
Suppose $\mathcal S_\gamma$ is orthogonally monotone, and $\overline{\mathcal S}_\gamma^{\hat\kappa}$ satisfies the same conditions for $\mathcal S_{\gamma}$ in Theorem \ref{general IS}. Then Algorithm \ref{algo: Deep-PrAE-LB} attains the lower-bound relaxed efficiency certificate by using a constant number of Stage 1 samples.\label{thm: LB_achieved_by_algo}
\end{theorem}

Finally, we investigate the conservativeness of this bound, which is measured by the false negative rate $P(X\notin\overline{\mathcal{S}}_{\gamma}^{\hat{k}},Y=1)$.  Like in Section \ref{sec:Deep-PrAE}, we use ERM to train $\hat g$, i.e., $\hat g:=\text{argmin}_{g\in\mathcal{G}}\{ {R}_{n_1}(g):=\frac{1}{n_1}\sum_{i=1}^{n_1}\ell(g(\tilde X_{i}),Y_{i})\}$
where $\ell$ is a loss function and $\mathcal G$ is the considered hypothesis class. Let $g^*$ be the true risk minimizer as described in Section \ref{sec:Deep-PrAE}. For inner approximation, we let $\kappa^{*}:=\max_{x\in\mathcal{S}_{\gamma}^c}g^{*}(x)$ be
the true threshold associated with $g^{*}$ in obtaining the largest inner rare-event set approximation. Then we have the following result analogous to Theorem \ref{thm: set_ERM}.

\begin{theorem}[Lower-bound estimation conservativeness]\label{thm: inn_set_ERM} 
Consider $\overline{\mathcal S}_\gamma^{\hat\kappa}$ obtained in Algorithm \ref{algo: Deep-PrAE-LB} where $\hat g$ is trained from an ERM. 
Suppose  the density $q$ has bounded 
support $K\subset[0,M]^{d}$ and $0<q_{l}\leq q(x)\leq q_{u}$ for any $x\in K$. Also suppose there exists
a function $h$ such that for any $g\in\mathcal{G}$, $g(x)\leq\kappa$ implies $\ell(g(x),1)\geq h(\kappa)>0$.
(e.g., if $\ell$ is the squared loss, then $h(\kappa)$ could be chosen 
as $h(\kappa)=(1-\kappa)^{2}$). Then, with probability at least $1-\delta$,
\begin{equation*}
P_{X\sim q}\left(X\in\overline{\mathcal{S}}_{\gamma}^{\hat{\kappa}}\setminus\mathcal{S}_{\gamma}\right)
\leq \frac{R(g^{*})+2\sup_{g\in\mathcal{G}}\left|R_{n_1}(g)-R(g)\right|}{h(\kappa^{*}+t(\delta,n_1)\sqrt{d} \text{Lip}(g^{*})+\left\Vert \hat{g}-g^{*}\right\Vert _{\infty})}.
\end{equation*}
Here, $\text{Lip}(g^{*})$ is the Lipschitz parameter of 
$g^{*}$, and $t(\delta,n_1)=3\left(\frac{\log(n_1q_{l})+d\log M+\log\frac{1}{\delta}}{n_1q_{l}}\right)^{\frac{1}{d}}$.
\end{theorem}

\section{Proofs}\label{app:proofs}

\subsection{Proofs for the Dominating Point Methodologies}

\proof{Proof of Proposition \ref{naive Monte Carlo}}

Since $\mu$ is exponentially decaying in $\gamma$ while $n$ is polynomially growing in $\gamma$, we know that $\lim_{\gamma\rightarrow\infty}n\mu=0$. Since $n\hat{\mu}_n$ takes values in $\{0,1,\dots,n\}$, we get that $P(|\hat{\mu}_n-\mu|>\varepsilon\mu)=P(|n\hat{\mu}_n-n\mu|>\varepsilon n\mu)\rightarrow 1$ as $\gamma\rightarrow\infty$.

\proof{Proof of Lemma \ref{thm:properties_I}}
See \citet{rockafellar1970convex}.

\proof{Proof of Theorem \ref{general IS}}

For simplicity, we denote $s_j:=s_{a_j}$. We know that the likelihood ratio function 
$$
L(x)=\frac{1}{\sum_j \alpha_j\exp\{s_j^T x-\lambda(s_j)\}}\leq \frac{1}{\alpha_j\exp\{s_j^T x-\lambda(s_j)\}},\forall j.
$$
First, we develop an upper bound for the second moment of the IS estimator.
$$
\begin{aligned}
\tilde{E}[Z^2] &=\sum_j\tilde{E}[I(X\in\mathcal S_{\gamma}^j)L^2(X)]\\
&\leq \sum_j\tilde{E}[I(X\in \mathcal S_{\gamma}^j)\exp\{2\lambda(s_j)-2s_j^T X\}/\alpha_j^2]\\
&\leq \sum_j\frac{1}{\alpha_j^2}\exp\{2\lambda(s_j)-2s_j^T a_j\}\\
&=\sum_j\frac{1}{\alpha_j^2}\exp\{-2I(a_j)\}\\
&\leq \left(\sum_j\frac{1}{\alpha_j^2}\right)\exp\{-2I(a^*)\}.
\end{aligned}
$$
Next, we develop a lower bound for $\tilde{E}[Z]=P(X\in \mathcal S_{\gamma})$. Since $\mathcal S_{\gamma}$ is orthogonally monotone, we get that $\{x\in\mathbb{R}^d:x\geq a^*\}\subset \mathcal S_{\gamma}$
. We have also assumed that $f^*(x):=f(x)\exp\{s^{*T}x-\mu(s^*)\}\geq\varepsilon'(a^*)$ for any $a^*\leq x\leq a^*+\varepsilon\mathbf{1}$ where $\varepsilon>0$ is a constant and $\varepsilon'(a^*)>0$ at most decays polynomially in $\gamma$. Then we have that 
$$
\begin{aligned}
\tilde{E}[Z]&\geq P(a^*\leq X\leq a^*+\varepsilon\mathbf{1})\\
&=\int_{a^*\leq x\leq a^*+\varepsilon\mathbf{1}}\exp\{\mu(s^*)-s^{*T}x\}f^*(x)\mathrm{d} x\\
&\geq\exp\{\mu(s^*)-s^{*T}a^*\}\varepsilon'(a^*)\int_{a^*\leq x\leq a^*+\varepsilon\mathbf{1}}\exp\{-s^{*T}(x-a^*)\}\mathrm{d} x\\
&=\exp\{-I(a^*)\}\varepsilon'(a^*)\prod_{i=1}^d\frac{1-\exp\{-s_i^*\varepsilon\}}{s_i^*}.
\end{aligned}
$$
Again, the orthogonal monotonicity implies that $s_i^*\geq 0,\forall i=1,\dots,d$. If $s_i^*=0$, then we naturally use $\varepsilon$ to substitute $\frac{1-\exp\{-s_i^*\varepsilon\}}{s_i^*}$. Since $s_i^*$ at most grows polynomially in $\gamma$, finally we get that $\tilde{E}[Z^2]/\tilde{E}[Z]^2$ at most grows polynomially in $\gamma$. Thus we have proved the efficiency certificate.

\proof{Proof of Theorem \ref{counterexample}}
We know that $\tilde{E}[Z]=\bar{\Phi}(\gamma)+\bar{\Phi}(k\gamma)$. Moreover, 
    $$
    \tilde{E}[Z^2]=e^{\gamma^2}(\bar{\Phi}(2\gamma)+\bar{\Phi}((k-1)\gamma)).
    $$
    If $0<k\leq 1$, then $\tilde{E}[Z]=O\left( e^{-k^2\gamma^2/2}/\gamma\right)$ and $\tilde{E}[Z^2]=O\left( e^{\gamma^2}\right)$ as $\gamma\rightarrow\infty$. If $1<k<3$, then $\tilde{E}[Z]=O\left( e^{-\gamma^2/2}/\gamma\right)$ and $\tilde{E}[Z^2]=O\left( e^{(1-(k-1)^2/2)\gamma^2}/\gamma\right)$ as $\gamma\rightarrow\infty$. In both cases, we get that $\tilde{E}[Z^2]/\tilde{E}[Z]^2$ grows exponentially in $\gamma$. 
    
    On the other hand, we denote $Z_i'=I(X_i\geq\gamma)L(X_i)$ and we note that $\tilde{E}[Z']=\bar{\Phi}(\gamma),\tilde{E}[Z'^2]=e^{\gamma^2}\bar{\Phi}(2\gamma),\tilde{E}[Z'^4]=e^{6\gamma^2}\bar{\Phi}(4\gamma)$. We know that 
    $$
    \begin{aligned}
    &\tilde{P}\left(\left|\frac{1}{n}\sum_{i=1}^n Z_i-\bar{\Phi}(\gamma)\right|>\varepsilon\bar{\Phi}(\gamma)\right)\\
    \leq& \tilde{P}(\exists i:X_i\leq -k\gamma)+\tilde{P}\left(\left|\frac{1}{n}\sum_{i=1}^n Z_i'-\bar{\Phi}(\gamma)\right|>\varepsilon\bar{\Phi}(\gamma)\right).
    \end{aligned}
    $$
    Clearly $\tilde{P}(\exists i:X_i\leq -k\gamma)=1-(1-\bar{\Phi}((k+1)\gamma))^n=O\left( n\bar{\Phi}((k+1)\gamma)\right)$, which is exponentially decreasing in $\gamma$ as $n$ is polynomial in $\gamma$. Moreover, by Markov's inequality,
    $$
    \tilde{P}\left(\left|\frac{1}{n}\sum_{i=1}^n Z_i'-\bar{\Phi}(\gamma)\right|>\varepsilon\bar{\Phi}(\gamma)\right)
    \leq \frac{\tilde{E}[Z'^2]}{n\varepsilon^2\bar{\Phi}^2(\gamma)}=\frac{e^{\gamma^2}\bar{\Phi}(2\gamma)}{n\varepsilon^2\bar{\Phi}^2(\gamma)}=\Theta\left( \frac{\gamma}{n\varepsilon^2}\right).
    $$
    Thus $P(|\hat{\mu}_n-\bar{\Phi}(\gamma)|>\varepsilon\bar{\Phi}(\gamma))=O\left(\frac{\gamma}{n\varepsilon^2}\right)$. Similarly, we also know that 
    $$
    \begin{aligned}
    &\tilde{P}\left(\frac{\frac1n \sum_{i=1}^n Z_{i}^2}{\left(\frac1n\sum_{i=1}^n  Z_i\right)^2}\geq \frac{1+\varepsilon}{(1-\varepsilon)^2}\frac{e^{\gamma^2}\bar{\Phi}(2\gamma)}{\bar{\Phi}^2(\gamma)}\right)\\
    \leq &\tilde{P}(\exists i:X_i\leq -k\gamma)+\tilde{P}\left(\frac{\frac1n \sum_{i=1}^n Z_i'^2}{\left(\frac1n\sum_{i=1}^n  Z_i'\right)^2}> \frac{1+\varepsilon}{(1-\varepsilon)^2}\frac{e^{\gamma^2}\bar{\Phi}(2\gamma)}{\bar{\Phi}^2(\gamma)}\right)\\
    \leq &\tilde{P}(\exists i:X_i\leq -k\gamma)+\tilde{P}\left(\frac1n\sum_{i=1}^nZ_i'^2>(1+\varepsilon)e^{\gamma^2}\bar{\Phi}(2\gamma)\right)+\tilde{P}\left(\frac1n\sum_{i=1}^n Z_i'<(1-\varepsilon)\bar{\Phi}(\gamma)\right)\\
    =&\tilde{P}(\exists i:X_i\leq -k\gamma)+\tilde{P}\left(\left|\frac1n\sum_{i=1}^nZ_i'^2-e^{\gamma^2}\bar{\Phi}(2\gamma)\right|>\varepsilon e^{\gamma^2}\bar{\Phi}(2\gamma)\right)+\tilde{P}\left(\left|\frac1n\sum_{i=1}^nZ_i'-\bar{\Phi}(\gamma)\right|>\varepsilon\bar{\Phi}(\gamma)\right)\\
    \leq &\tilde{P}(\exists i:X_i\leq -k\gamma)+\frac{\tilde{E}[Z'^4]}{n\varepsilon^2e^{2\gamma^2}\bar{\Phi}^2(2\gamma)}+\frac{\tilde{E}[Z'^2]}{n\varepsilon^2\bar{\Phi}^2(\gamma)}\\
    =&\Theta\left(\frac{\gamma}{n\varepsilon^2}\right).
    \end{aligned}
    $$

\subsection{Proofs for the Relaxed Efficiency Certificate}

\proof{Proof of Proposition \ref{certificate prop simple}}
We have
$$P(\hat\mu_n-\mu<-\epsilon\mu)\leq P(\hat\mu_n-\overline\mu<-\epsilon\overline\mu)$$
since $\overline\mu\geq\mu$ and $1-\epsilon>0$. Note that the Markov inequality gives
$$P(\hat\mu_n-\overline\mu<-\epsilon\overline\mu)\leq\frac{\widetilde{Var}(Z_i)}{n\epsilon^2\overline\mu^2}$$
so that 
$$n\geq\frac{\widetilde{Var}(Z_i)}{\delta\epsilon^2\overline\mu^2}=\frac{RE}{\delta\epsilon^2}=\tilde O\left(\log\frac{1}{\overline\mu}\right)=\tilde O\left(\log\frac{1}{\mu}\right)$$
achieves the relaxed efficiency certificate.

\proof{Proof of Proposition \ref{prop:extend}}
The proof follows from that of Proposition \ref{certificate prop simple} with a conditioning on $D_{n_1}$. We have
\begin{align*}
P(\hat\mu_n-\mu<-\epsilon\mu|D_{n_1})&\leq P(\hat\mu_n-\overline\mu(D_{n_1})<-\epsilon\overline\mu(D_{n_1})|D_{n_1})
\end{align*}
since $\overline\mu(D_{n_1})\geq\mu$ almost surely and $1-\epsilon>0$. Note that the Markov inequality gives
$$P(\hat\mu_n-\overline\mu(D_{n_1})<-\epsilon\overline\mu(D_{n_1})|D_{n_1})\leq\frac{Var(Z_i|D_{n_1})}{n_2\epsilon^2\overline\mu(D_{n_1})^2}$$
so that 
$$n_2\geq\frac{Var(Z_i|D_{n_1})}{\delta\epsilon^2\overline\mu(D_{n_1})^2}=\frac{RE(D_{n_1})}{\delta\epsilon^2}=\tilde O\left(\log\left(\frac{1}{\overline\mu(D_{n_1})}\right)\right)=\tilde O\left(\log\frac{1}{\mu}\right)$$
almost surely. Thus,
$$n=n_1+n_2\geq\tilde O\left(\log\frac{1}{\mu}\right)$$
achieves the relaxed efficiency certificate.

\proof{Proof of Corollary \ref{relaxed prob}}
Follows directly from Proposition \ref{prop:extend}, since $\overline{\mathcal S}_\gamma\supset\mathcal S_\gamma$ implies $\overline\mu(D_{n_1})\geq\mu$ almost surely.

\proof{Proof of Theorem \ref{NN main}}
 We have assumed that $\overline{\mathcal S}_\gamma^{\hat\kappa}$ satisfies the assumptions for $\mathcal S_\gamma$ in Theorem \ref{general IS}. Then following the proof of Theorem \ref{general IS}, we obtain the efficiency certificate for the IS estimator in estimating its mean. Theorem \ref{NN main} is then proved by directly applying Corollary \ref{relaxed prob}.

\subsection{Proofs for Conservativeness}

Recall that $T_0=\{\tilde{X}_i:Y_i=0\}$ where the samples are generated as in Algorithm \ref{algo:stage1}. By some combinitorial argument, we can prove the following lemma which
says that with high probability, each point in  $\mathcal{S}_{\gamma}^c$ that has sufficient distance
to its boundary could be covered by $\mathcal{H}(T_{0})$. 
\begin{lemma} Suppose that the density $q$ has bounded support $K\subset[0,M]^{d}$,
and for any $x\in K$, suppose that $0<q_{l}\leq q(x)\leq q_{u}$.
Define $B_{t}:=\{x\in\mathcal{S}_{\gamma}^{c}:x+t\mathbf{1}_{d\times1}\in\mathcal{S}_{\gamma}^{c}\}$.
Then with probability at least $1-\delta$, we have that $B_{t(\delta,n_1)}\subset\mathcal{H}(T_{0})$.
Here $t(\delta,n_1)=3\left(\frac{\log(n_1q_{l})+d\log{M}+\log\frac{1}{\delta}}{n_1q_{l}}\right)^{\frac{1}{d}}.$\label{lem: distance}
\end{lemma}

\proof{Proof of Lemma \ref{lem: distance}} The basic idea is to construct a finite number of regions,
such that when there is at least one sample point in each of these
regions, we would have that $B_{t}\subset\mathcal{H}(T_{0})$. Then
we could give a lower bound to the probability of $B_{t}\subset\mathcal{H}(T_{0})$
in terms of the number of regions and the volume of each of these
regions.

By dividing the first $d-1$ coordinates into $\frac{M}{\delta}$
equal parts, we partition the region $[0,M]^{d}$ into rectangles,
each with side length $\delta$, except for the $d-$th dimension
(the $\delta$ here is not exactly the $\delta$ in the statement
of the lemma, since we will do a change of variable in the last step).
To be more precise, the rectangles are given by 
\[
Z_{j}=\left(\prod_{i=1}^{d-1}[(j_{i}-1)\delta,j_{i}\delta]\right)\times[0,M].
\]
Here $j\in J$ and $J$ is defined by 
\[
J:=\{j=(j_{1},\cdots,j_{d-1}),j_{i}=1,2,\cdots,\frac{M}{\delta}\}.
\]
Denote by $J_{0}$ the set which consists of $j\in J$ such that there
exist a point in $B_{2\delta}$ whose first $d-1$ coordinates are
$j_{1}\delta,j_{2}\delta,\cdots,j_{d-1}\delta$ respectively, i.e.,
$J_{0}=\left\{ j\in J:B_{2\delta}\cap\left(\left(\prod_{i=1}^{d-1}\{j_{i}\delta\}\right)\times[0,M]\right)\neq\emptyset\right\} .$
For all $j\in J_{0}$, let $p_{j}$ be the point such that

i) $p_{j}\in B_{\delta}$

ii) The first $d-1$ coordinates of $p_{j}$ are $j_{1}\delta,j_{2}\delta,\cdots,j_{d-1}\delta$
respectively

iii) $p_{j}$ has $d-$th coordinate larger than $-\delta+\sup_{p\text{ satisfies i),ii)}}\left(d\text{-th coordinate of \ensuremath{p}}\right)$.

From the definition of $J_{0}$ and the fact that $B_{\delta}\supset B_{2\delta}$,
$p_{j}$ is guaranteed to exist. We claim that $B_{2\delta}\cap Z_{j}\subset\mathcal{R}(p_{j})$,
where $\mathcal{R}({p_j})$ is the rectangle that contains 0 and $p_j$ as two of its corners. Clearly, from the definition of $Z_{j}$, for
any point $x\in B_{2\delta}\cap Z_{j}$, its first $d-1$ coordinates
are smaller than $j_{1}\delta,j_{2}\delta,\cdots,j_{d-1}\delta$ respectively.
For the $d-$th coordinate, suppose on the contrary that there exists
$x\in B_{2\delta}\cap Z_{j}$ with $d-$th coordinate greater than
the $d-$th coordinate of $p_{j}$. Since $x\in Z_{j}$, the first
$d-1$ coordinates of $x$ are at least $(j_{1}-1)\delta,(j_{2}-1)\delta,\cdots,(j_{d-1}-1)\delta$,
so we have that $x+\delta\mathbf{1}_{d\times1}\geq p_{j}+\delta e_{d}$.
Since $x\in B_{2\delta}$, we know that $x+2\delta\mathbf{1}_{d\times1}\in\mathcal{S}_{\gamma}^{c}$.
Hence by the previous inequality and the orthogonal monotonicity of
$\mathcal{S}_{\gamma}$, $p_{j}+\delta e_{d}+\delta\mathbf{1}_{d\times1}\in \mathcal{S}_{\gamma}^c$.
By definition of $B_{\delta}$, this implies $p_{j}+\delta e_{d}\in B_{\delta}$. This contradicts iii) in
the definition of $p_{j}$. By contradiction, we have shown that each
point in $B_{2\delta}\cap Z_{j}$ has $d-$th coordinate smaller than
the $d-$th coordinate of $p_{j}$. So the claim that $B_{2\delta}\cap Z_{j}\subset\mathcal{R}(p_{j})$
for any $j\in J_0$ is proved.

Then we consider those $j$ such that $j\in J-J_{0}$.
For any point $x\in Z_{j}$, the first $d-1$ coordinates
of $x+\delta\mathbf{1}_{d\times1}$ are at least $j_{1}\delta,j_{2}\delta,\cdots,j_{d-1}\delta$
respectively. Since $j\notin J_{0}$,  we have that $x+\delta\mathbf{1}_{d\times1}\notin B_{2\delta}$.
This implies $x+3\delta\mathbf{1}_{d\times1}\notin \mathcal{S}_{\gamma}^c$, or $x\notin B_{3\delta}$.
So we have shown that for any $j\notin J_{0}$, $B_{3\delta}\cap Z_{j}=\emptyset$.
This implies $B_{3\delta}$ has a partition given by $B_{3\delta}=\cup_{j\in J}\left(B_{3\delta}\cap Z_{j}\right)=\cup_{j\in J_{0}}\left(B_{3\delta}\cap Z_{j}\right)$.
Notice that $B_{3\delta}\subset B_{2\delta}$, from the result in
the preceding paragraph, we conclude that $B_{3\delta}\subset\cup_{j\in J_{0}}\mathcal{R}(p_{j})$.

For each $j\in J_{0}$ and the constructed $p_{j}$, consider the
region 
\[
G_{j}:=\{x\in S_{\gamma}^{c}:x\geq p_{j}\}.
\]
Observe that, if there exists a sample point in $T_0$ that lies in $G_{j}$,
then we have $p_{j}\subset\mathcal{H}(T_{0})$ which implies $\mathcal{R}(p_{j})\subset\mathcal{H}(T_{0})$.
Since $p_{j}\in B_{\delta}$ and $\mathcal{S}_{\gamma}$ is orthogonally monotone, we have that $G_j$ contains the rectangle which contains $p_j$ and $p_j+\delta \mathbf{1}_{d\times 1}$ as two of its corners, so $\text{Vol}(G_{j})\ge\delta^{d}$.
Hence the probability that $\mathcal{R}(p_{j})\subset\mathcal{H}(T_{0})$
has a lower bound given by 
\[
P(\mathcal{R}(p_{j})\subset\mathcal{H}(T_{0}))\geq P(T_0 \cap G_j \neq \emptyset)\geq1-\left(1-\delta^{d}q_{l}\right)^{n_1}\geq1-e^{-n_1q_{l}\delta^{d}}.
\]
Notice that $\left|J_{0}\right|\leq\left(\frac{M}{\delta}\right)^{d-1}$,
by union bound we have that 
\[
P(\cup_{j\in J_{0}}\mathcal{R}(p_{j})\subset\mathcal{H}(T_{0}))\geq1-\frac{M^{d-1}}{\delta^{d-1}}e^{-n_1q_{l}\delta^{d}}.
\]
Since we have shown that $B_{3\delta}\subset\cup_{j\in J_{0}}\mathcal{R}(p_{j})$,
this implies 
\[
P(B_{3\delta}\subset\mathcal{H}(T_{0}))\geq1-\frac{M^{d-1}}{\delta^{d-1}}e^{-n_1q_{l}\delta^{d}}.
\]
Based on this inequality, it is not hard to check that for $t(\delta,n_1)=3\left(\frac{\log(n_1q_{l})+d\log{M}+\log\frac{1}{\delta}}{n_1q_{l}}\right)^{\frac{1}{d}}$, we have that $P(B_{t(\delta)}\subset\mathcal{H}(T_0))\geq 1-\delta$. 

\proof{Proof of Theorem \ref{thm: set_OMhull}}
First, we show the inequality in the theorem, i.e., $P_{X\sim q}(X\in\mathcal{H}(T_0)^c\backslash\mathcal{S}_{\gamma})\leq M^{d-1}q_u\left(\frac{\sqrt{d}}{2}\right)^{d-1}w_{d-1}t(\delta,n_1)$. It suffices to show that with probability at least $1-\delta$, $\text{Vol}\left(\mathcal{H}(T_{0})^{c}\backslash\mathcal{S}_{\gamma}\right)\leq M^{d-1}\left(\frac{\sqrt{d}}{2}\right)^{d-1}w_{d-1}t(\delta,n_1)$,
or equivalently $\text{Vol}(\mathcal{S}_{\gamma}^{c}\backslash\mathcal{H}(T_{0}))\leq M^{d-1}\left(\frac{\sqrt{d}}{2}\right)^{d-1}w_{d-1}t(\delta,n_1)$.
Since by lemma \ref{lem: distance} we have that $B_{t(\delta,n_1)}\subset\mathcal{H}(T_{0})$
with probability at least $1-\delta$, it suffices to show that $\text{Vol}(\mathcal{S}_{\gamma}^{c}\backslash B_{t(\delta,n_1)})\leq M^{d-1}\left(\frac{\sqrt{d}}{2}\right)^{d-1}w_{d-1}t(\delta,n_1)$.
This latter inequality actually follows from the definition of $B_{t(\delta,n_1)}$
and some geometric argument. Indeed, by definition of $B_{t(\delta,n_1)}$,
for each $x\in\mathcal{S}_{\gamma}^{c}\backslash B_{t(\delta,n_1)}$,
$x$ belongs to the area which is obtained by moving the boundary
of $\mathcal{S}_{\gamma}$ in direction $-\frac{\mathbf{1}_{d\times1}}{\sqrt{d}}$
for a distance of $t(\delta,n_1)\sqrt{d}$. So the volume of $\mathcal{S}_{\gamma}^{c}\backslash B_{t(\delta,n_1)}$
is bounded by 
\begin{align*}
 & t(\delta,n_1)\sqrt{d}\times\text{Vol}_{d-1}(\text{projection of the boundary of \ensuremath{S_{0}} in direction}\ \mathbf{1}_{d\times1})\\
\leq & t(\delta,n_1)\sqrt{d}\times\text{Vol}_{d-1}(\text{projection of \ensuremath{[0,M]^{d}} in direction}\ \mathbf{1}_{d\times1})
\end{align*}
Here $\text{Vol}_{d-1}$ means computing volume in the $d-1$ dimensional
space. Notice that $[0,M]^{d}$ is contained in a ball with radius
$\frac{M\sqrt{d}}{2}$, we have that 
\[
\text{Vol}_{d-1}(\text{projection of \ensuremath{[0,M]^{d}} in direction}\ \mathbf{1}_{d\times1})\leq M^{d-1}\left(\frac{\sqrt{d}}{2}\right)^{d-1}w_{d-1}.
\]
Combining the preceding two inequalities, we have proved the inequality in the theorem. Next we show the equality in the theorem. Indeed, when $d$ is large, we have asymptotic formula
$w_{d}=\frac{1}{\sqrt{d\pi}}\left(\frac{2\pi e}{d}\right)^{\frac{d}{2}}(1+O(d^{-1}))$.
Plugging this into the RHS above, we will obtain the asymptotic bound
as stated in the theorem. 
 
\proof{Proof of Theorem \ref{thm: set_ERM}}
By Markov inequality and the definition of $h$,$\bar{\mathcal{S}}_{\gamma}^{\hat{\kappa}}$, we
know that 
\begin{equation}
P_{X\sim q}\left(X\in\bar{\mathcal{S}}_{\gamma}^{\hat{\kappa}},X\in\mathcal{S}_{\gamma}^{c}\right)= P_{X\sim q}(\hat{g}(X)\geq\hat\kappa,X\in{\mathcal{S}_{\gamma}^c})\leq\frac{R(\hat{g})}{h(\hat{\kappa})}.\label{eq: Markov}
\end{equation}
We will compare the numerator and denominator of the RHS of (\ref{eq: Markov}) with their counterparts for the true minimizer $g^*$. For the numerator, since $\hat{g}$ is the empirical risk minimizer, we have that 
\begin{align}
R(\hat{g}) & \leq R_{n_1}(\hat{g})+\sup_{g_{\theta}\in\mathcal{G}}\left|R_{n_1}(g_{\theta})-R(g_{\theta})\right|\leq R_{n_1}(g^{*})+\sup_{g_{\theta}\in\mathcal{G}}\left|R_{n_1}(g_{\theta})-R(g_{\theta})\right|\nonumber \\
 & \leq R(g^{*})+2\sup_{g_{\theta}\in\mathcal{G}}\left|R_{n_1}(g_{\theta})-R(g_{\theta})\right|.\label{eq: R_g_difference}
\end{align}
For the denominator, from the definition of $\bar{\mathcal{S}}_{\gamma}^{\hat{\kappa}}$, it is not hard to verify that, in Algorithm \ref{algo:stage1}, our choice of $\hat{\kappa}$ is given by $\hat{\kappa}=\min\{\hat{g}(x):x\in\mathcal{H}(T_0)^c\}$. By lemma \ref{lem: distance}, we have that with probability at least
$1-\delta$, $B_{t(\delta,n_1)}\subset\mathcal{H}(T_{0})$, which implies
that with probability at least $1-\delta$, 
\begin{align*}
\hat{\kappa} & \geq\min\{\hat{g}(x):x\in B_{t(\delta,n_1)}^{c}\}\geq\min\{g^{*}(x):x\in B_{t(\delta,n_1)}^{c}\}-\left\Vert \hat{g}-g^{*}\right\Vert _{\infty}\\
 & \geq\min\{g^{*}(x):x\in \mathcal{S}_{\gamma }\}-t(\delta,n_1)\sqrt{d}\text{Lip}(g^{*})-\left\Vert \hat{g}-g^{*}\right\Vert _{\infty}\\
 & =\kappa^{*}-t(\delta,n_1)\sqrt{d}\text{Lip}(g^{*})-\left\Vert \hat{g}-g^{*}\right\Vert _{\infty}.
\end{align*}
Putting the preceding two inequalities into the Markov inequality
(\ref{eq: Markov}), and notice that $h$ is non decreasing by its definition, the theorem is proved.

\proof{Proof of Theorem \ref{thm: set_ERM_modified}}

The proof goes the same way as the proof of Theorem \ref{thm: set_ERM}. We only need to replace $g$ and $g_{\theta}$ with $f$ and $f_{\theta}$ in \eqref{eq: R_g_difference} (Note that \eqref{eq: Markov} still holds because of our assumption that $g_{\theta}(x)\geq\kappa\Rightarrow \ell(f_{\theta}(x),0)\geq h(\kappa)$).

\end{document}